\documentclass{aa}  
\pdfoutput=1
\usepackage{graphicx}
\usepackage{dblfloatfix}
\usepackage{txfonts}
\usepackage{hyperref}
\usepackage{siunitx}
\usepackage{multirow}
\usepackage{multicol}
\usepackage{float}
\usepackage{bm}
\usepackage[section]{placeins}
\usepackage[super]{nth}
\usepackage{natbib}
\bibpunct{(}{)}{;}{a}{}{,} 

\begin{document}

\title{Effects of neighbouring planets on the formation\\
of resonant dust rings in the inner Solar System}


\author{M.~Sommer
	\inst{1}
	\and
	H.~Yano
	\inst{2,3}
	\and
	R.~Srama
	\inst{1}
	}

\institute{Institute of Space Systems, University of Stuttgart, Germany\\
           \email{sommer@irs.uni-stuttgart.de}
         \and
           Institute of Space and Astronautical Science, Japan Aerospace Exploration Agency (JAXA)
           , Sagamihara, Japan\\
           \email{yano.hajime@jaxa.jp}
         \and
           Graduate University for Advanced Studies (SOKENDAI), Japan\\
           }

\date{Received September 11, 2019; accepted January 10, 2020}
  
 
  \abstract
   {Findings by the \textit{Helios} and \textit{STEREO} mission have indicated the 
   presence of a resonant circumsolar ring of dust associated with Venus. 
   Attempts to model this phenomenon as an analogue to the resonant ring of Earth -- as a result of migrating dust trapped in external mean-motion 
   resonances (MMRs) -- have so far been unable to reproduce the observed dust feature.
   Other theories of origin have recently been put forward.
   However, the reason for the low trapping efficiency of Venus's external MMRs remains unclear.}
   {Here we look into the nature of the dust trapping resonant phenomena that 
   arise from the multi-planet configuration of the inner Solar System, 
   aiming to add to the existent understanding of resonant dust rings in single planet systems.}
   {We numerically modelled resonant dust features associated with the inner planets
   and specifically looked into the dependency of these structures and the trapping efficiency of particular 
   resonances on the configuration of planets.}
   {Besides Mercury showing no resonant interaction with the migrating dust cloud, 
   we find Venus, Earth, and Mars to considerably interfere with each other's resonances, influencing their ability to form circumsolar rings. 
   We find that the single most important reason for the weakness of Venus’s external MMR ring is 
   the perturbing influence of its outer neighbour -- Earth.
   In addition, we find Mercury and Mars to produce crescent-shaped density features,
   caused by a directed apsidal precession occurring in particles traversing their orbital region.}
   {}


  \maketitle
%

  \section{Introduction}

   The presence of a circumsolar density enhancement of the zodiacal dust cloud near 
   the orbit of Venus has first been indicated by photometry data obtained by the Venus-orbit 
   crossing \textit{Helios} mission \citep{2007A&A...472..335L}. 
   These findings have been supported by imagery data from the \textit{STEREO} 
   spacecraft (\citealt{Jones960}; \citealt{2017Icar..288..172J}).
   In addition, in-situ data from the large-area dust impact detector aboard the solar sail 
   demonstration mission \textit{IKAROS} show a dust flux variation that may be connected
   to the existence of a Venusian ring \citep{2013LPI....44.2743Y, 2014cosp...40E3705Y}.
   This phenomenon has been thought to be akin to the resonant dust ring of Earth, 
   which is the result of dust particles becoming trapped in the planet's external 
   mean-motion resonances (MMRs) as they migrate inwards from the asteroid belt 
   \citep{1989Natur.337..629J, 1994Natur.369..719D}. 
   However, attempts to model the Venus resonant ring as a result of migrating dust could 
   not reproduce a meaningful enhancement due to an overall low trapping probability 
   \citep{2012LPICo1667.6201J}.
      
   While also deeming migrating dust in external resonances to be the unlikely
   cause for the observed ring, \citet{2019ApJ...873L..16P} have recently found that dust stemming from 
   a hypothetical population of Venus co-orbital asteroids can produce a distinct ring structure that 
   matches the \textit{STEREO} observations. 
   Dust particles ejected from these co-orbital asteroids would remain in the 1:1 MMR, 
   bypassing the low trapping probability and forming a co-orbital resonant dust ring. 
   \citet{2019ApJ...873L..16P} show that it is possible for a fraction of a primordial population of 
   Venus co-orbitals to remain stable for the lifetime of the Solar System, substantiating 
   the hypothesis of the co-orbital ring. 
   
   However, it is still unclear as to what causes the proclaimed absence of Venus's external MMR ring, 
   which Earth so significantly has been shown to entail \citep{1994Natur.369..719D, 1995Natur.374..521R, REACH2010848}. 
   \citet{2017Icar..288..172J} note that the lower over-density of Venus's ring is in qualitative 
   agreement with the expectation that higher Keplerian speeds closer to the Sun would lead to a lower 
   probability of resonant capture. 
   Similarly, \citet{1994Natur.369..719D} argued that the higher Poynting-Robertson drag rates 
   at Venus would reduce trapping efficiency to some extent. 
   Indeed, the debris disc catalogue by \citet{2008ApJ...686..637S}, who analysed the dependency 
   of the dust disc structure on the parameters of a single planet, shows a slight decrease of 
   resonant ring overdensity with a planet's semi-major axis at a fixed dust particle size. 
   On the other hand, \citet{2012LPICo1667.6201J} as well as \citet{2019ApJ...873L..16P} report no 
   meaningful dust enhancement due to migrating dust at Venus, which is in contrast to the expectation 
   of a less prominent version of Earth's resonant ring. 
   This discrepancy indicates the existence of an additional effect diminishing the trapping 
   efficiency of Venus's external MMRs specifically. 
   An inhibition of MMRs by neighbouring planets, as first suggested by 
   \citet{1989Natur.337..629J} and as \citet{Liou99} found to occur 
   in migrating Edgeworth-Kuiper belt dust around the giant planets of the 
   outer Solar System, seems likely. 
   
   To further our understanding of resonant dust around the inner planets, 
   we have performed a set of simulations to model the dust features associated 
   with them. 
   Similar to \citet{2008ApJ...686..637S}, we produce a series of dust discs for dust of 
   different particle sizes. 
   However, instead of considering a single planet, we look into the disc structure and resonant 
   behaviour that arise from the interaction with our multi-planetary system. 
   Finally, we try to isolate the effect of neighbouring planets by changing the planetary 
   configuration altogether. 

  \section{Methods}
 
\subsection{Dynamical model}
	In order to model the dynamics of dust particles migrating through the inner Solar System, 
	we integrate their equations of motion in the N-body system using a customised version of 
	\textit{Mercury6} \citep{1999MNRAS.304..793C}. 
	This is a hybrid symplectic integrator that extends a symplectic integrator with the ability
	to compute close encounters by switching the integration method (e.g. to a Bulirsch-Stoer 
	algorithm, as used in our model). 
	To account for all relevant forces, we have modified \textit{Mercury6}  to include the effects of 
	radiation pressure, Poynting-Robertson drag, and solar wind drag \citep{Burns19791}.
	Solar wind drag is treated as a multiplying factor of PR drag. 
	For the ratio of solar wind drag to PR drag, we assume a constant value of 0.35, 
	following \citet{1994AREPS..22..553G}. 
	The magnitude of these effects is given by the dimensionless parameter $\beta$, which is 
	the ratio of the force resulting from radiation pressure to the solar gravitational force 
	and is dependent on particle properties. 
	The implementation of radiation forces was verified by comparing simulated particle 
	evolution to the analytic solutions for particles under the effect of radiation in the 
	two-body problem, derived by \citet{1950ApJ...111..134W}. 
	Furthermore, we include the gravity of the Sun and its eight planets (unless otherwise specified), 
	all of which are treated as point masses. The initial positions of the planets are taken 
	from the ephemerides DE430 \citep{folkner2014planetary} at the epoch J2000.\hspace{-.2mm}0.
	
	All our simulations use an integration time step of 1/\nth{40} of a year or about 1/\nth{24} of 
	a Venus year. 
	As argued by \citet{KORTENKAMP2011669}, the traditional range of 1/\nth{5} to 1/\nth{10} of the 
	orbital period of a host planet used for symplectic integration can inhibit trapping of 
	particles in the co-orbital resonance. 
	To avoid this effect, a time step of less than 1/\nth{20} of the host planet's orbital period 
	is required. 
	Our time step was chosen to satisfy this criterion in the case of Venus.
	
\subsection{Particle starting populations}
	Our initial dust populations for each simulated disc consist of 50,000 test particles. 
	We modelled discs for five values of $\beta=$~0.1, 0.05, 0.03, 0.02, 0.01, 0.005\hspace{.2mm}. 
	These correspond to particle diameters of roughly $D=$~5~\si{\micro\meter}, 
	10~\si{\micro\meter}, 17~\si{\micro\meter}, 25~\si{\micro\meter}, 50~\si{\micro\meter}, 
	100~\si{\micro\meter} assuming perfectly absorbing, spherical particles with a density of 
	$\rho=$~2.3~\si{\gram\per\cubic\centi\metre}. 
	The initial particle distribution is described by ranges of semi-major axis, 
	eccentricity and inclination. 
	The particles are released on orbits with semi-major axes uniformly distributed 
	between 2.2~\si{\astronomicalunit} and 3.2~\si{\astronomicalunit}. 
	Initial eccentricities and initial inclinations are uniformly distributed from 0 to 0.2 
	and \ang{0} to \ang{15}, respectively. 
	These ranges were chosen to roughly span the orbital elements of the majority of bodies 
	in the asteroid belt \citep{MPC_inc}. 
	The initial angles for the argument of pericentre, longitude of the ascending node, and mean 
	anomaly are uniformly distributed from 0 to 2$\pi$. 
	
	The dust disc that we intend to model with this initial distribution can therefore be 
	considered of asteroidal origin only. 
	We neglect dynamically hot dust, mainly produced by comets, as it is unlikely to become 
	trapped in resonances of the inner planets \citep{Borderies1984, 2011MNRAS.413..554M}. 
	Cometary dust therefore mainly contributes a smooth background distribution, which 
	reduces the contrast of the resonant features modelled here.
	\begin{table}
	\caption{Initial particle populations}
	\label{table:populations}
	\centering
	\begin{tabular}{l l} 
	\hline\hline\\[-5pt]
	\multicolumn{2}{l}{Six simulated particle sizes of 50,000 particles each:}\\[4pt]
	\hline\\[-5pt]
	$\beta$ &  0.1\hspace{2.3mm} 0.05\hspace{2.4mm} 0.03\hspace{2.4mm} 0.02\hspace{2.4mm} 0.01\hspace{2.4mm} 0.005\\
	D (2.3~\si{\gram\per\cubic\centi\metre})& 5\hspace{0.4mm}\si{\micro\meter} 10\hspace{0.4mm}\si{\micro\meter} 17\hspace{0.4mm}\si{\micro\meter} 25\hspace{0.4mm}\si{\micro\meter} 
	50\hspace{0.4mm}\si{\micro\meter} 100\hspace{0.4mm}\si{\micro\meter}\\[4pt]
	\hline\hline\\[-5pt]
	\multicolumn{2}{l}{With initial orbital elements uniformally distributed in:}\\[4pt]
	\hline\\[-5pt]
	   semi-major axis & 2.2~\si{\astronomicalunit} -- 3.2~\si{\astronomicalunit} \\
	   eccentricity & 0 -- 0.2\\
	   inclination & \ang{0} -- \ang{15}\\
	\hline
	\end{tabular}
	\end{table}
	
\subsection{Particle recording} \label{subsect:particle-recording}
	In order to produce the steady state dust disc from the integration of particle trajectories, 
	we employed the technique of superimposing states of the evolution of the initial cloud, 
	simulating continuous emission of dust. 
	That is, we record the states of all particles of a simulation at regular intervals in time, 
	transform their coordinates to a frame co-rotating with a planet and stack their positions 
	in a 2D-histogram. 
	This method has been widely used by dust modellers to obtain the equilibrium state of dust 
	discs whose features are dominated by a single planet 
	\citep{Liou99, 2002AJ....124.2305M, 0004-637X-625-1-398, 2008ApJ...686..637S}.
	In our multi-planetary case, however, this approach is problematic as it azimuthally blurs out 
	resonant structures that revolve with an angular speed different from that of the observed 
	planet, that is, resonant structures that are associated with a different planet. 
	Therefore, we can only resolve the azimuthal structure of the resonant features of one planet 
	at a time.
	
	Furthermore, in order to obtain the equilibrium state of the disc the simulation would need 
	to run until all test particles have been destroyed (or ejected). 
	This cannot be easily satisfied, as the dynamical lifetime of some particles can be unusually 
	high as they get trapped in stable, long-lasting resonances. 
	\citet{2002AJ....124.2305M} and \citet{0004-637X-625-1-398} showed, however, that the disc 
	structure is created quickly and that it does not strongly depend on the contribution from 
	these overly long-living particles. 
	Here, we use a fixed simulation time that is larger than the nominal PR-lifetime by a factor, 
	such that the fraction of particles still alive at the end is less than 1\% of the starting 
	population in any case. Particles are considered to be destroyed once they reach a heliocentric 
	distance of 0.2~\si{\astronomicalunit}.
	
	The produced steady state distributions are collisionless discs, meaning we neglect mutual 
	particle collisions. 
	\citet{1985Icar...62..244G} calculated that particle evolution 
	becomes dominated by their collisional lifetimes instead of their PR-lifetimes only at a particle 
	size in the order of \textgreater100~\si{\micro\meter}. 
	However, more recent efforts to model the sporadic meteoroid complex indicate an
	inconsistency of the \citet{1985Icar...62..244G} timescales with the dynamics of larger 
	grains derived from meteor radars, which suggest orders of magnitude higher collisional 
	lifetimes for particles larger than several 100~\si{\micro\meter} 
	\citep{Nesvorn__2010, Nesvorn__2011, Pokorn__2014, SojaR_2019_IMEM2}.
	In section~\ref{Ch4}, we discuss the implications grain-grain collisions 
	might have for the simulated discs at the larger end of our particle size range.

  \section{Results}

\subsection{All planets present}	\label{Ch31}
	Since the formation of resonant rings in single planet systems has been extensively discussed
	in previous works \citep[amongst others]{2008ApJ...686..637S,10.1093/mnras/stv045}, we are going
	to start by looking into the dust disc that arises under the presence of all eight planets as the 
	nominal case. We will compare the emerging rings by their overall geometry as well as the peak
	overdensity of their trailing blob $\sigma_{TB,max}$ and the peak overdensity of their azimuthally 
	averaged surface density $\sigma_{AA,max}$.
	In addition, we look at the semi-major axis histograms of the dust discs, which give insight into
	the underlying resonant behaviour.

	\begin{figure}[!t]
		\centering
		\includegraphics[width=.96\hsize,trim={0 7mm 0 0},clip]{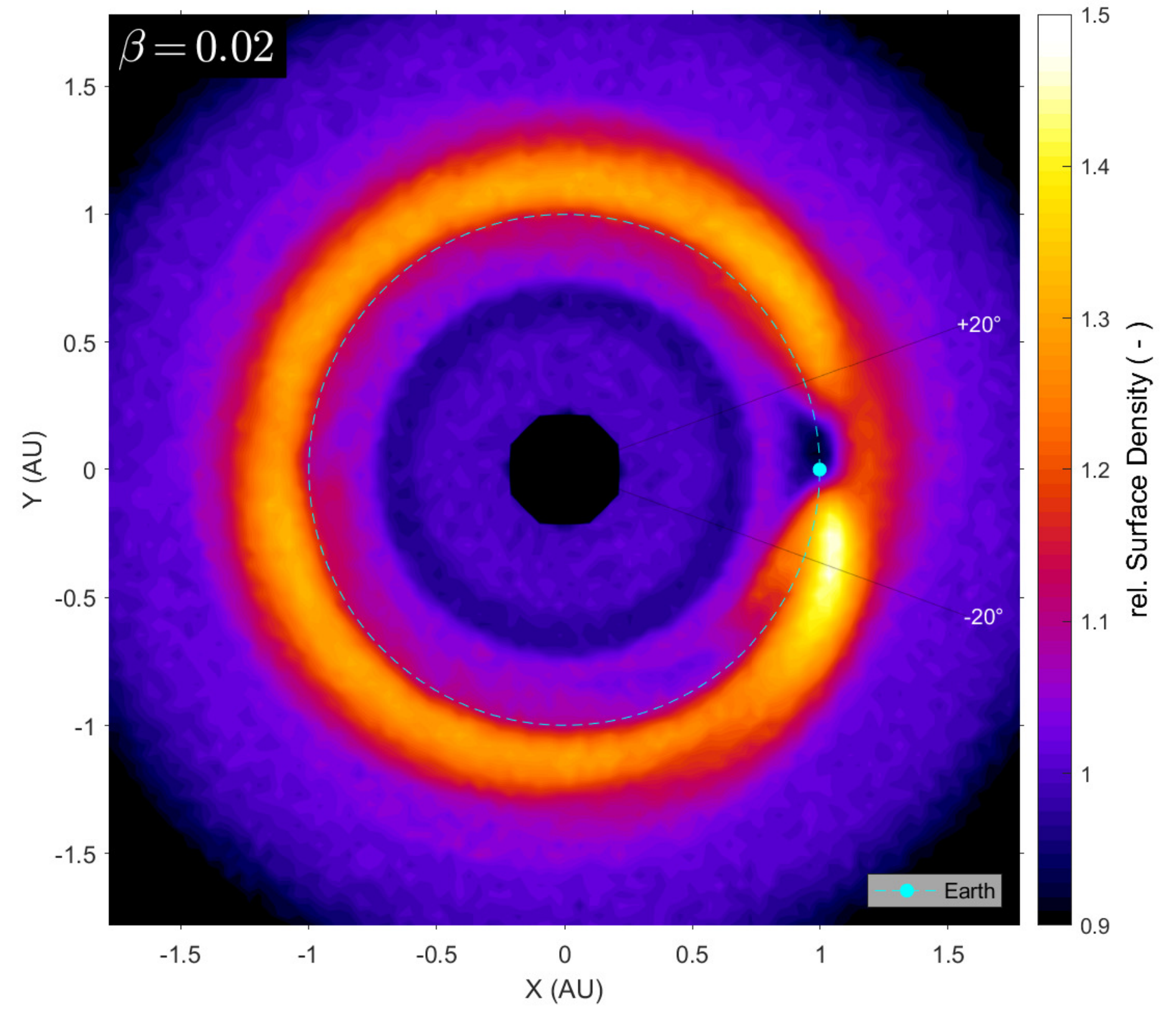}
		\includegraphics[width=.96\hsize,trim={0 0 0 1mm},clip]{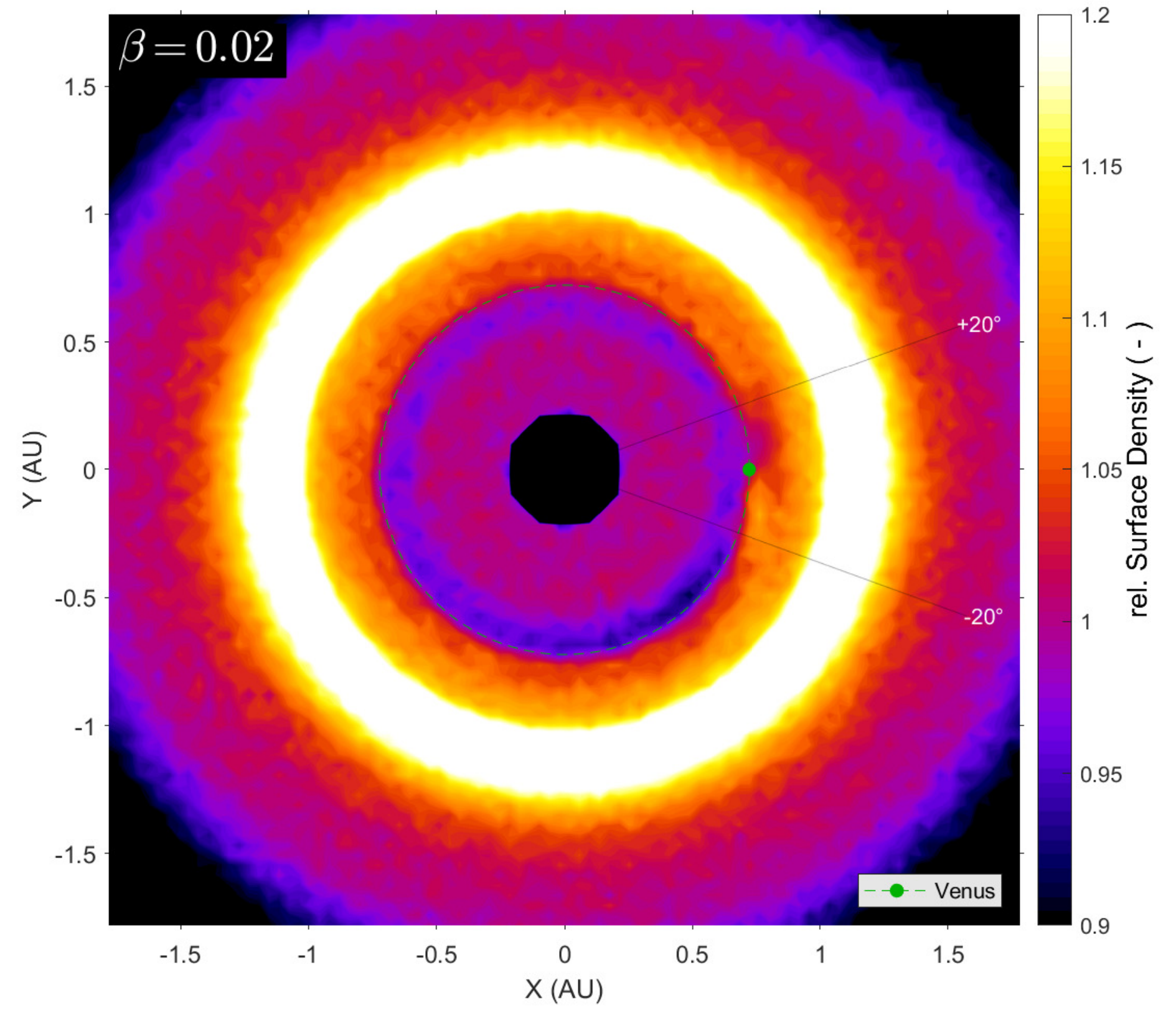}
		\caption{Surface density of simulated particles for the case of $\beta=$~0.02 (all planets present). 
		The relative surface density is a 2D-histogram of the particle distribution, 
		recorded in a frame co-rotating with Earth \textit{(top)} and Venus \textit{(bottom)}, 
		that is then normalised to a background density that arises under the absence
		of the inner planets. 
		Histogram bin size is 0.02~\si{\astronomicalunit}~$\times$~0.02~\si{\astronomicalunit}.
		Structures revolving with an orbital period different from the observed planet are washed out 
		azimuthally. 
		The density distributions for Earth and Venus are thus presented separately.
		In the case of Venus \textit{(bottom)} the colour scale has been changed to bring 
		out the fainter structure.
		Other particle $\beta$ are given in Figs.~\ref{fig:A_XY_AP_1} \&~\ref{fig:A_XY_AP_2}.
		}
		\label{fig:b02_XY_AP}
	\end{figure}
	\setcounter{figure}{3}
	\begin{figure*}[b]
		\includegraphics[width=\textwidth]{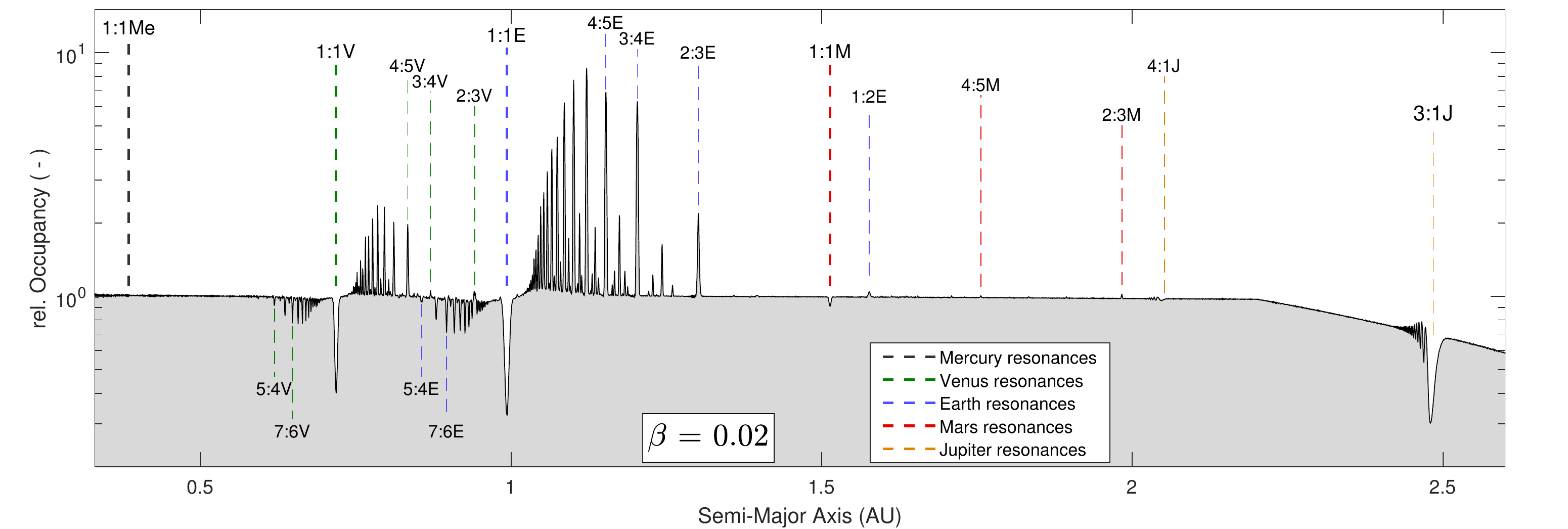}
      	\caption{Semi-major axis histogram under the presence of all planets for 
      	$\beta=$~0.02\hspace{0.2mm}.
      	Occupancy is the number of particles per 0.0005~\si{\astronomicalunit} bin divided 
      	by solar distance of that bin and then normalised to background count 
      	(i.e. under the absence of resonances).
      	Vertical dashed lines indicate the locations of several first order as well as the 
      	co-orbital MMRs (including a shifting factor to account for radiation pressure).
      	Mercury and Mars show no and virtually no effect on dust migration rate at this particle size.
      	Trapping and displacing by Venus and Earth MMRs is significant, while the trapping by Earth MMRs is 
      	much more pronounced (considering the logarithmic scale).
      	Internal Jupiter MMRs cause displacement analogous to Kirkwood gaps, but shifted in location
      	due to radiation pressure. 
      	Other particle $\beta$ are given in Fig.~\ref{fig:A_SMA_AP}.
      	}
        \label{fig:b02_SMA_AP}
	\end{figure*}
	\setcounter{figure}{1}
	\begin{figure}[]
		\centering
		\includegraphics[width=\hsize, trim=0 7mm 0 0, clip]{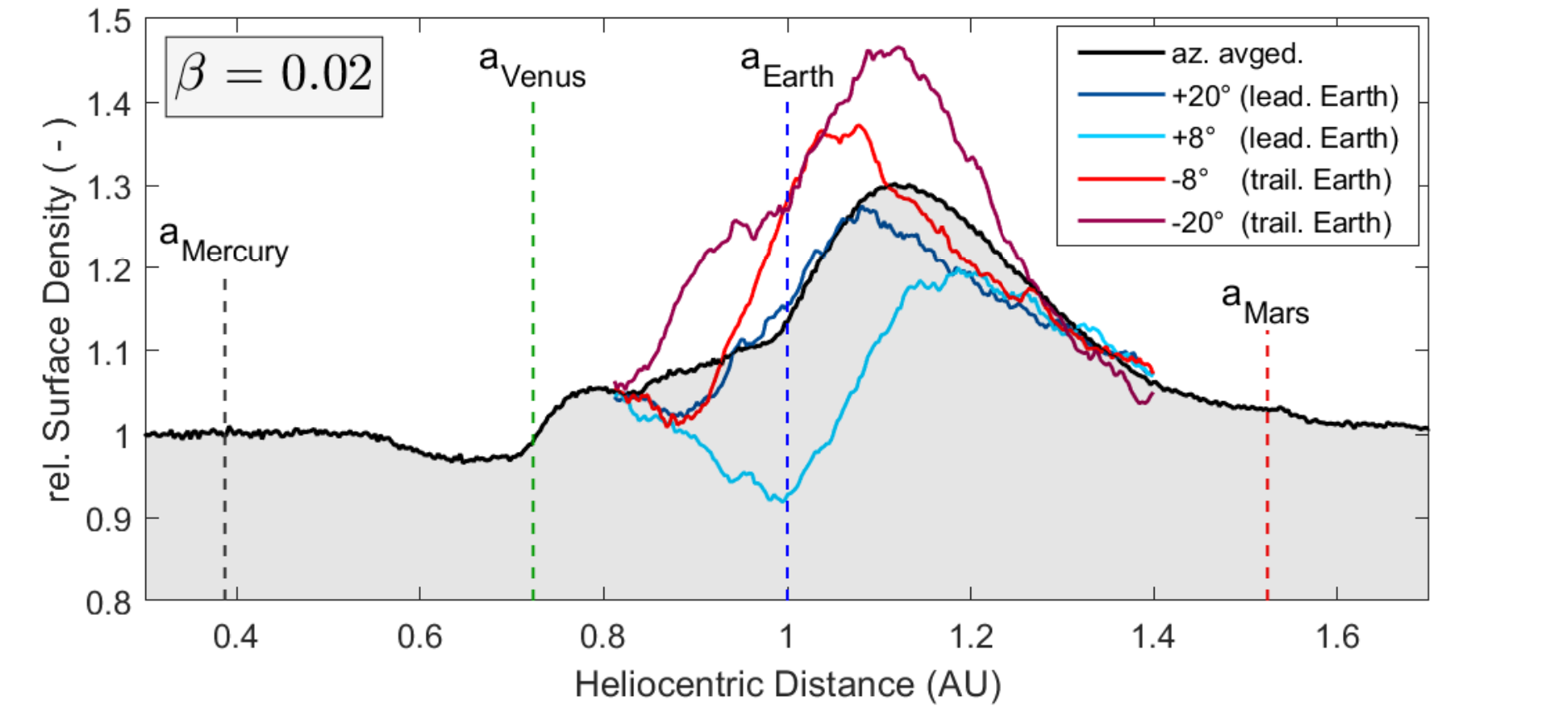}
		\includegraphics[width=\hsize]{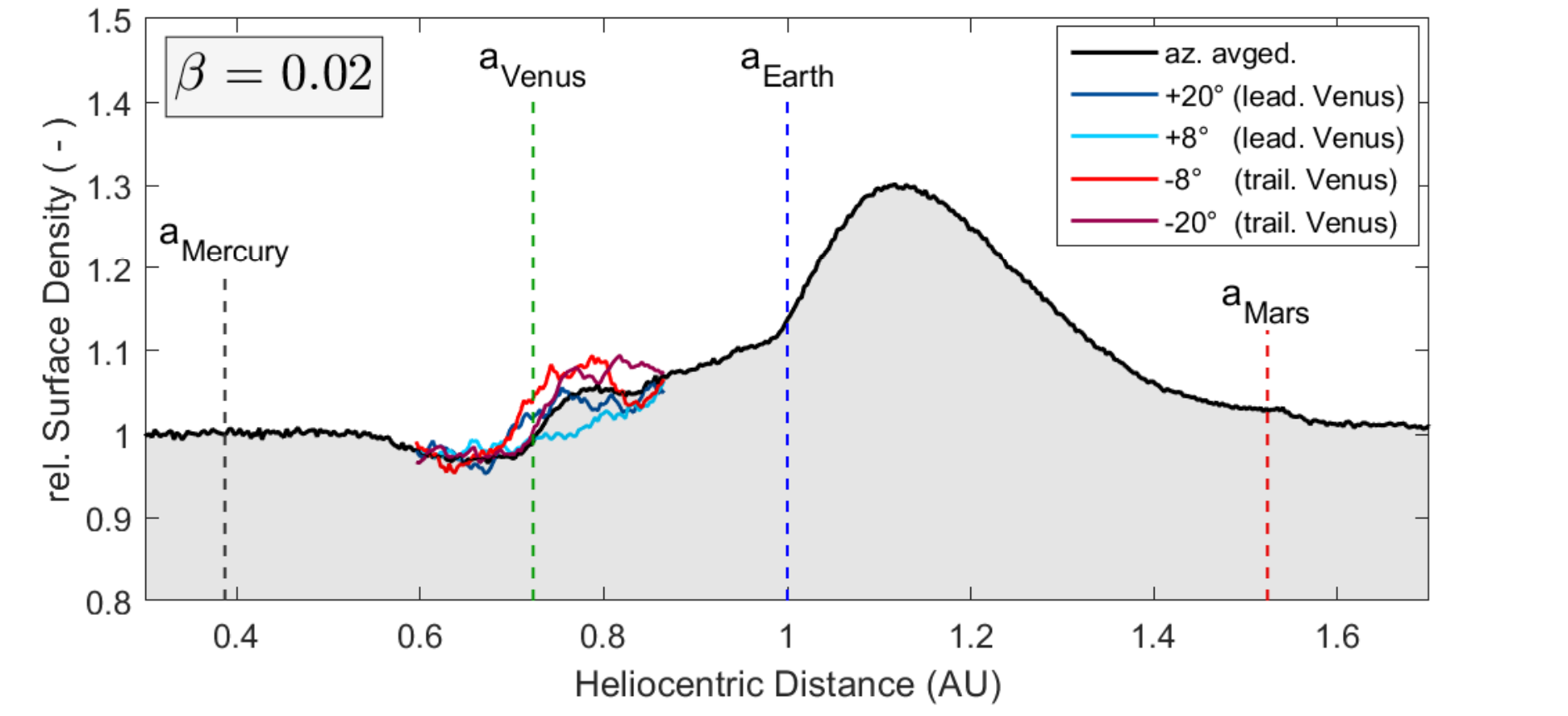}
		\caption{Radial profiles of the dust disc surface density seen in Fig.~\ref{fig:b02_XY_AP}. 
		Radial profiles are given for the azimuthal average (indicated
		by the grey area, equivalent in both plots), as well as at cuts leading and trailing Earth 
		\textit{(top)} and Venus \textit{(bottom)}.
		The radial cuts located at azimuths of $\pm\ang{8}$ and $\pm\ang{20}$ from the planets illustrate
		the azimuthal asymmetry present in the rings (leading gap and trailing enhancement).
		Also shown are the locations of the semi-major axes of the inner planets.
		Other particle $\beta$ are given in Figs.~\ref{fig:A_XY_AP_1} \&~\ref{fig:A_XY_AP_2}.
		}
		\label{fig:b02_RAD_AP}
	\end{figure}

	Figure~\ref{fig:b02_XY_AP} shows the surface density distribution resulting near Earth and Venus
	for $\beta=0.02$\hspace{0.2mm}.
	The corresponding radial profiles of the surface density is given in Fig. \ref{fig:b02_RAD_AP}.
	In agreement with previous ring modellers, we find a prominent circumsolar ring at Earth orbit with
	an enhancement trailing Earth in its orbit in addition to a leading gap (as first described by
	\citet{1994Natur.369..719D}.
	On the contrary, however, we do not see a sharp inner edge of the ring \citep[e.g. described 
	by][]{2008ApJ...686..637S},	but rather a gradual decline of density until smoothly transitioning 
	into the resonant ring of Venus.
	Here, the ring of Venus exhibits a qualitatively similar, yet generally fainter set of features.
	Since the crossover distance from the Venus ring to that of Earth cannot be defined precisely, we
	choose $\sigma_{AA,max,Venus}$ as the first maximum on the azimuthally averaged density profile
	beyond the semi-major axis of Venus and $\sigma_{TB,max,Venus}$ as the maximum surface density
	within a heliocentric distance of 0.82~\si{\astronomicalunit}.
	The difference in ring overdensity between Earth and Venus is about a factor 5 at $\beta=0.02$, with
	$\sigma_{AA,max,Venus}/\sigma_{AA,max,Earth}=0.2$ 
	and $\sigma_{TB,max,Venus}/\sigma_{TB,max,Earth}=0.23$.
	
	The density distributions resulting at Venus and Earth for the other modelled values of $\beta$ are 
	shown in the appendix in Figs.~\ref{fig:A_XY_AP_1}~\&~\ref{fig:A_XY_AP_2}.
	The corresponding development of their ring overdensities as $\beta$ decreases is given in 
	Fig.~\ref{fig:overdensity_AP}. For Earth we see the general trend of increasing ring contrast
	from $\sigma_{AA,max,Earth}=6\%$ at $\beta=0.1$ to $\sigma_{AA,max,Earth}=38\%$ at $\beta=0.01$,
	followed by a slight retraction going to $\beta=0.005$. 
	An increase in ring density with decreasing $\beta$ agrees with the expectation that slower 
	migration times
	give resonances more opportunity to exert their effect on particles migrating their effective range, 
	thus increasing trapping probability.
	However the reversal of this trend at $\beta\leq0.01$ disagrees with this consideration, which also 
	poses a deviation from the results of single-planet models 
	(e.g.~\citet{2008ApJ...686..637S}).
	Furthermore, we do not see that trend in the resonant structures of Venus. With decreasing $\beta$ 
	the ring density is stagnating and falling off even earlier, virtually disappearing at 
	$\beta\leq0.01$	with an overdensity of less than 1/\nth{10} of that of Earth. 
	This is in disagreement with the early prediction made by \citet{1994Natur.369..719D}, that the
	resonant ring of Venus would only be shifted towards a higher particle size range, as the closer
	distance to the Sun would favour the capture of low $\beta$ particles.
	An additional notable feature in the density distribution is a toroidal depletion inside the orbit 
	of Venus that becomes apparent at $\beta \leq 0.01$, reaching a density reduction of  
	$\sim\!10\%$ at 0.62~\si{\astronomicalunit} -- about 0.85 of the semi-major axis of Venus.
	
	\setcounter{figure}{2}
	\begin{figure}[]
		\centering
		\includegraphics[width=\hsize, trim=0 1mm 0 0, clip]{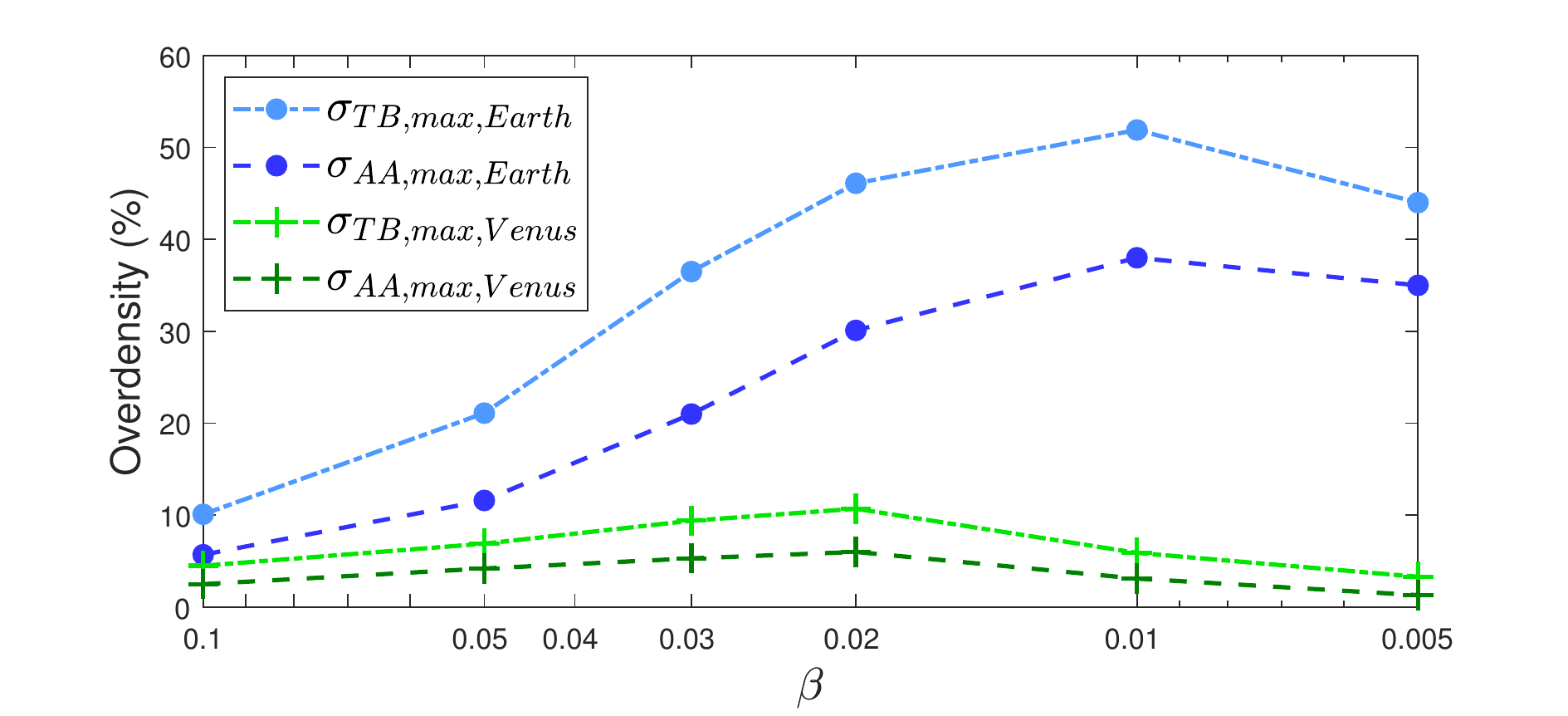}
		\caption{Dependency of ring overdensity on $\beta$.
		Overdensity is defined as the relative surface density subtracted by one.
		Maximum overdensity of the azimuthal average ($\sigma_{AA,max}$) and maximum overdensity
		of the trailing enhancement ($\sigma_{max}$) increase with decreasing $\beta$ in the case
		of Earth but remain low for Venus. Mercury and Mars do not produce meaningful enhancements
		over the modelled $\beta$-range.
		}
		\label{fig:overdensity_AP}
	\end{figure}
	
	To connect the presence and absence of these structures to the efficacy of different resonances
	we can look at the semi-major axis distribution of the dusk discs, shown in Fig.~\ref{fig:b02_SMA_AP}
	for $\beta=0.02$. Other particle $\beta$ are given in Fig.~\ref{fig:A_SMA_AP}.
	Here, MMRs appear as peaks and troughs according to their ability to trap or displace dust particles. 
	This is a useful indicator when trying to understand the planets' gravitational interaction 
	with the dust disc. 
	In Fig.~\ref{fig:b02_SMA_AP} we can see the general pattern of resonances produced by both Earth and 
	Venus. 
	Internal MMRs as well as the co-orbital resonance cause a net displacement of migrating particles. 
	External MMRs on the other hand hold particle migration, causing their 
	accumulation outside the planet's orbit and resulting in the distinctive ring.
	
	When comparing the semi-major axis distribution at different $\beta$ 
	(Fig.~\ref{fig:A_SMA_AP}), we see an increase of resonance occupancy with particle size 
	in the case of external Earth resonances, which is in accordance with the noted increasing ring 
	overdensity. 
	In~particular, resonances with a revolving period closer to that of Earth, as well as higher order 
	resonances (being located at semi-major axes between those of first order MMRs) 
	are increasingly occupied with decreasing $\beta$. 
	External MMRs of Venus, however, stop increasing in occupancy for
	$\beta \leq 0.03$. 
	If trapping efficiency would only be limited by PR drag, their occupancy should further 
	increase with particles size, as it is the case for Earth. 
	This indicates the existence of another limiting factor for the external MMRs of Venus. 
	A lowered trapping probability and/or a destabilisation of resonating particles through Earth's 
	gravitational influence must be presumed.
	This is also supported by the fact that internal MMRs of Venus, being less affected by the influences 
	of Earth, do indeed continue to grow in efficacy with decreasing PR drag.
	Another way to visualise this, would be to look at the orbital evolution of resonating particles.
	Dust particles trapped in external resonances typically have their eccentricity gradually increased 
	until reaching a limiting eccentricity that is characterised by the final equilibrium solution of
	that resonance~\citep{BEAUGE1994239}.
	As a result, particles resonating with Venus have their aphelion steadily raised towards 
	1~\si{\astronomicalunit}, where they become more and more susceptible to the gravitational perturbations 
	from Earth. 
	Eventually, Earth's influence causes the particles to fall out of the resonances.
	The thus reduced dwell time in these resonances decreases the overall number of particles resonating
	with Venus at any given time, reducing the intensity of any resonant structure.
	In addition, the reduced final eccentricities that particles falling out of a specific resonance exhibit,
	necessarily raise their final perihelia and lower their final aphelia.
	Premature termination of resonances should therefore also show itself in a confinement of the ring's radial 
	extent likewise in both directions. 
	This is indeed indicated by the azimuthally averaged radial profile of the dust disc, particularly
	at higher masses (see $\beta=0.02$, $0.01$ Fig.~\ref{fig:A_XY_AP_2}), 
	showing the ring of Venus not extending inside Venus's orbit, contrary to the ring of Earth.
	
	A similar argument can reasonably be made for dust particles trapped in MMRs with Earth.
	Here, the absence of a sharp inner edge of Earth's ring also hints at a premature termination 
	of resonances.
	As argued by \citet{2008ApJ...686..637S}, the formation of a distinct inner edge 
	at $\sim\!0.83 a_p$  in single planet systems 
	(where $a_p$ is the semi-major axis of the observed planet) is caused by a culmination 
	of limiting pericentres of the dominant resonances at a similar heliocentric distance -- 
	limiting pericentre being the pericentre that particles assume when 
	approaching the equilibrium solution of a resonance, regardless of $\beta$.
	The cause of particles resonating with Earth not reaching or remaining near this equilibrium 
	solution -- presumably the presence of Venus or Mars --
	is investigated in Sect.~\ref{Ch32}.
	
	Furthermore, we can explain the toroidal depletion inside the orbit of Venus at the lower 
	$\beta$-range end with the growing troughs at internal MMRs of Venus, which peak 
	around the same heliocentric distance that the radial density profile reaches its minimum 
	($\sim$~0.64~\si{\astronomicalunit}; 
	cp. Figs.~\ref{fig:A_XY_AP_2} \& \ref{fig:A_SMA_AP} for $\beta\hspace{-1mm}=\hspace{-1mm}0.01$). 
	An analogue toroidal depletion associated with Earth is not apparent in the density distribution,
	for one because the ring of Venus would superpose any such depletion. 
	Additionally, these displacing internal Earth MMRs appear to 
	be impaired, similar to the way external Venus MMRs are impaired by Earth.
	The troughs they cause in the semi-major axis distribution 	
	do not increase when reducing $\beta$ from 0.02 to 0.01, 
	as is notably the case for the internal resonances of Venus.
	This again suggests a resonance weakening effect by a neighbouring planet, in this case
	on the displacing internal MMRs of Earth caused by the presence of Venus.
	
	Neither Mercury nor Mars produced a meaningful density enhancement over the examined
	$\beta$-range, using the described stacking method.
	One has to note however, that due to their orbital eccentricity, any resonant pattern must be 
	presumed to change with the planet's orbital phase. 
	Since the presence of multiple planets introduces variations into their orbital elements, 
	in particular an apsidal precession and eccentricity fluctuation, recording particle 
	positions in a frame rotating with the planet at fixed time intervals inevitably superimposes distributions
	at different orbital phases.
	This approach may therefore dilute density variations associated with these 
	eccentric planets.
	The case of Mars is aggravated by the fact that its resonant features might be overshadowed by 
	the outer region of the ring of Earth. 	
	Nevertheless, we see that an absence of resonant features associated with Mercury is in 
	accordance with the semi-major axis distribution, showing no meaningful resonant interaction
	even at the lowest $\beta$ (Fig.~\ref{fig:b02_SMA_AP} \& \ref{fig:A_SMA_AP}).
	Likewise, while there are minor peaks at Mars resonances in the semi-major axis 
	distribution at $\beta\leq0.01$, their extent stays well below even the far out 1:2~MMR of Earth,
	suggesting a non-significant impact of Martian mean-motion resonances.

\subsection{Changing the planetary configuration} \label{Ch32}

	\setcounter{figure}{4}
	\begin{figure*}[t]
		\includegraphics[width=\textwidth,trim={0 5.5mm 0 0mm},clip]{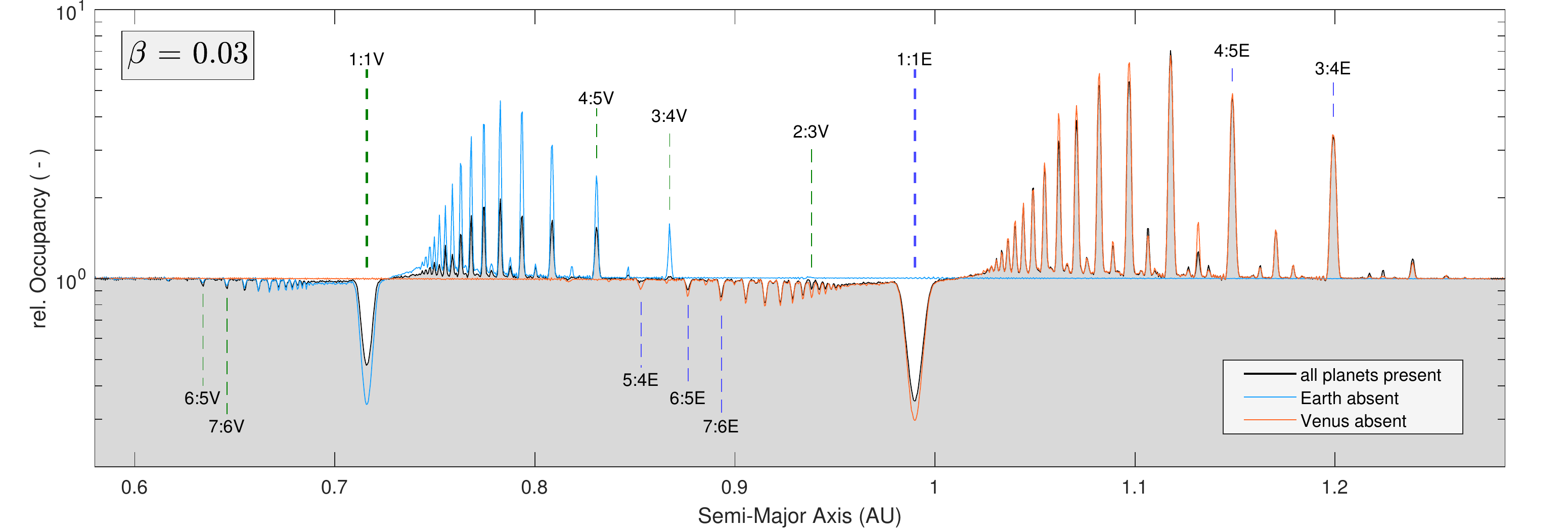}
		\includegraphics[width=\textwidth,trim={0 0mm 0 2mm},clip]{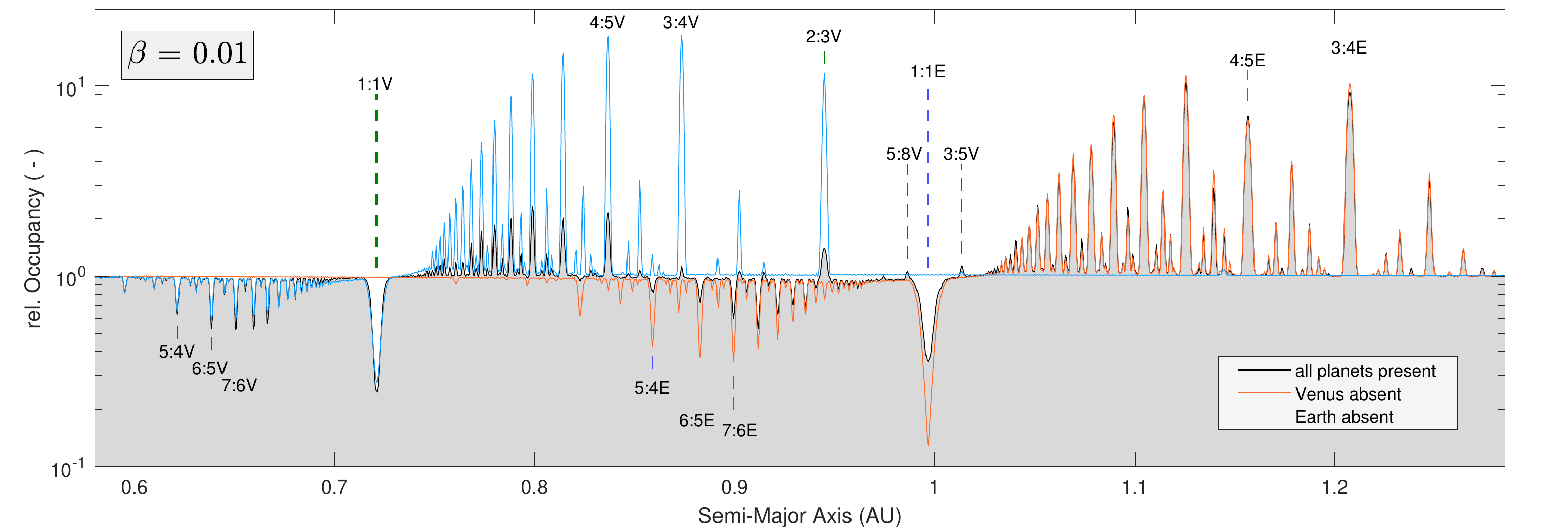}
      	\caption{Semi-major axis distribution under the interaction of Venus and Earth. 
      	The semi-major axis histogram is given for the cases of all planets present (also indicated
      	by the grey area) and either Earth or Venus absent for $\beta=$~0.03 \textit{(top)} 
      	and $\beta=$~0.01 \textit{(bottom)}.
      	Vertical dashed lines indicate the locations of several first order as well as the 
      	co-orbital MMRs (including a shifting factor to account for radiation pressure).
      	A significant impact on the efficacy of resonances in the region between Earth and Venus
      	is evident, becoming more pronounced at a lower $\beta$.}
        \label{fig:SMA_NENV}
	\end{figure*}
	\begin{figure*}[!b]
		\centering
		\begin{tabular}{ccc}
		\hspace{-2mm}\includegraphics[width=60mm,trim={3mm 0 4mm 0},clip]{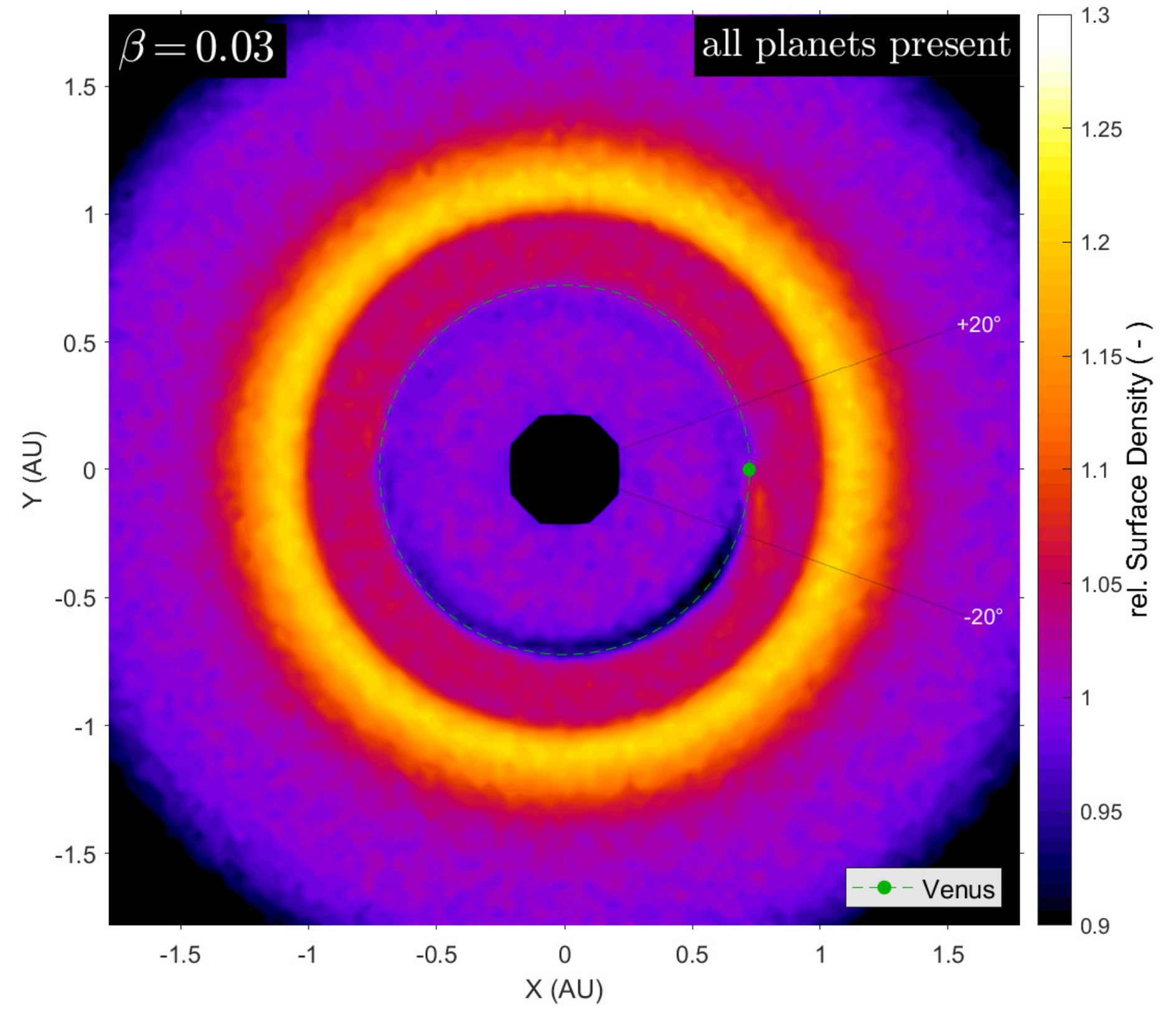} &
		\hspace{-2mm}\includegraphics[width=60mm,trim={3mm 0 4mm 0},clip]{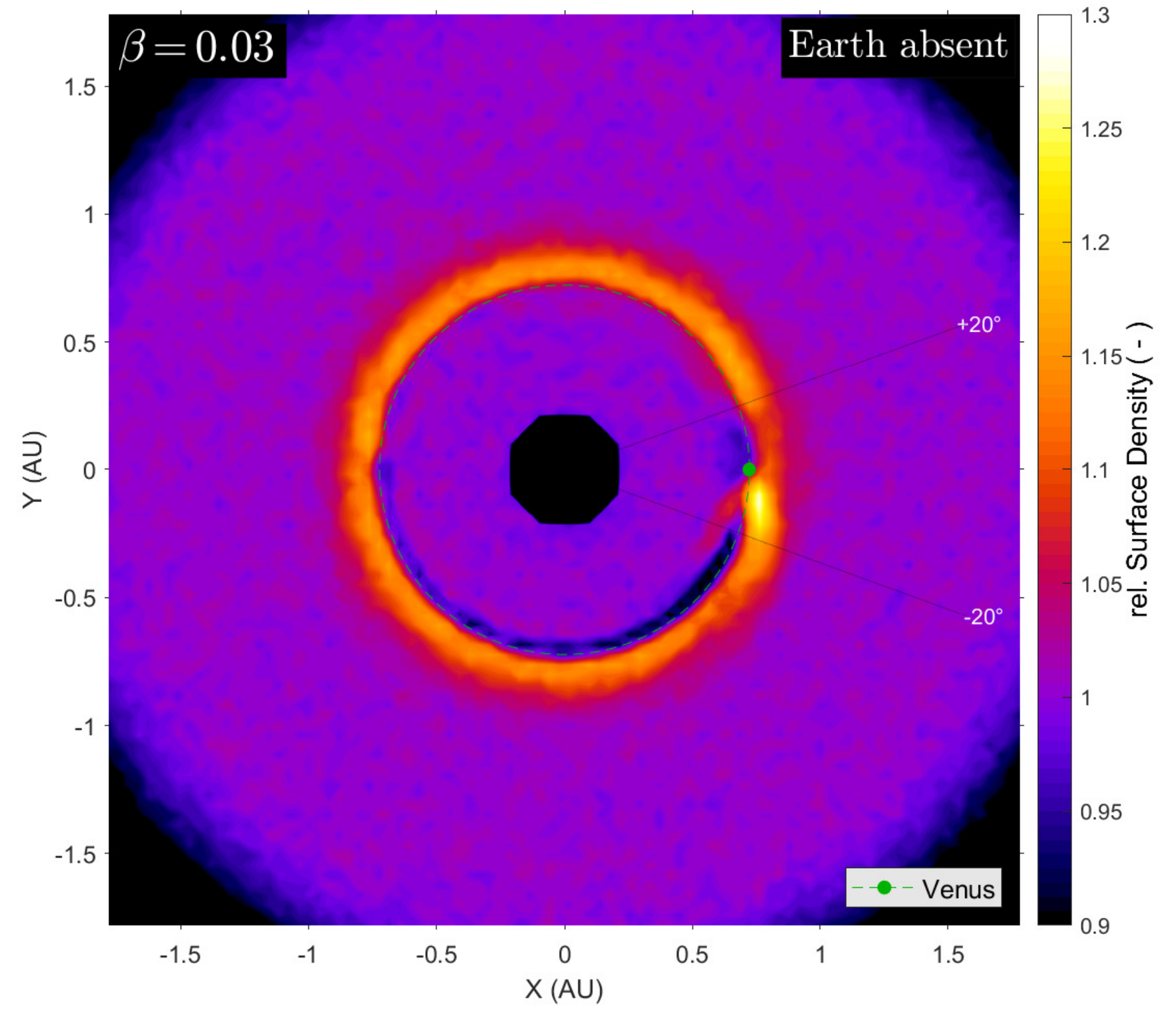} &
		\hspace{-2mm}\raisebox{11.6mm}{\includegraphics[width=60mm,trim={4mm 0 8mm 0},clip]{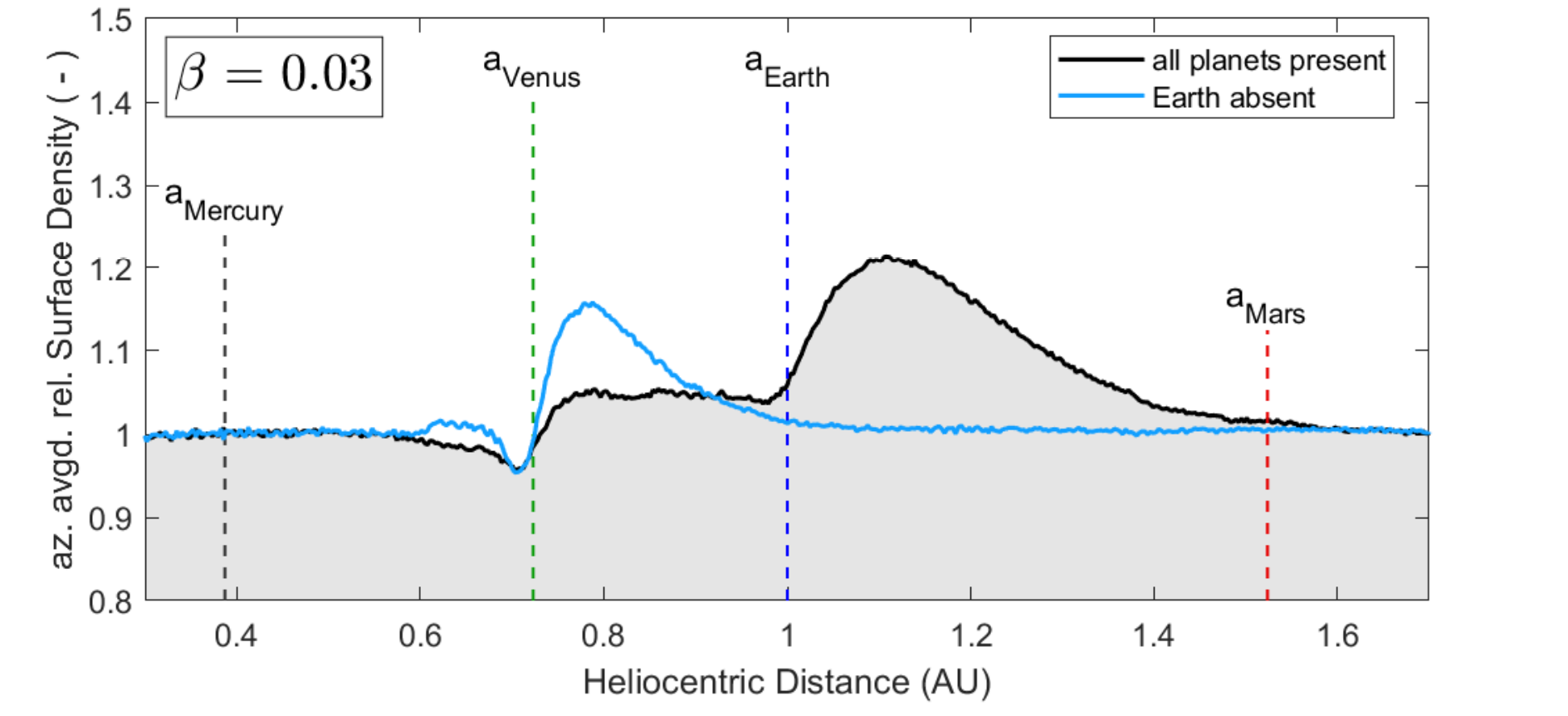}}\\[-2ex]
		\hspace{-2mm}\includegraphics[width=60mm,trim={3mm 0 4mm 0},clip]{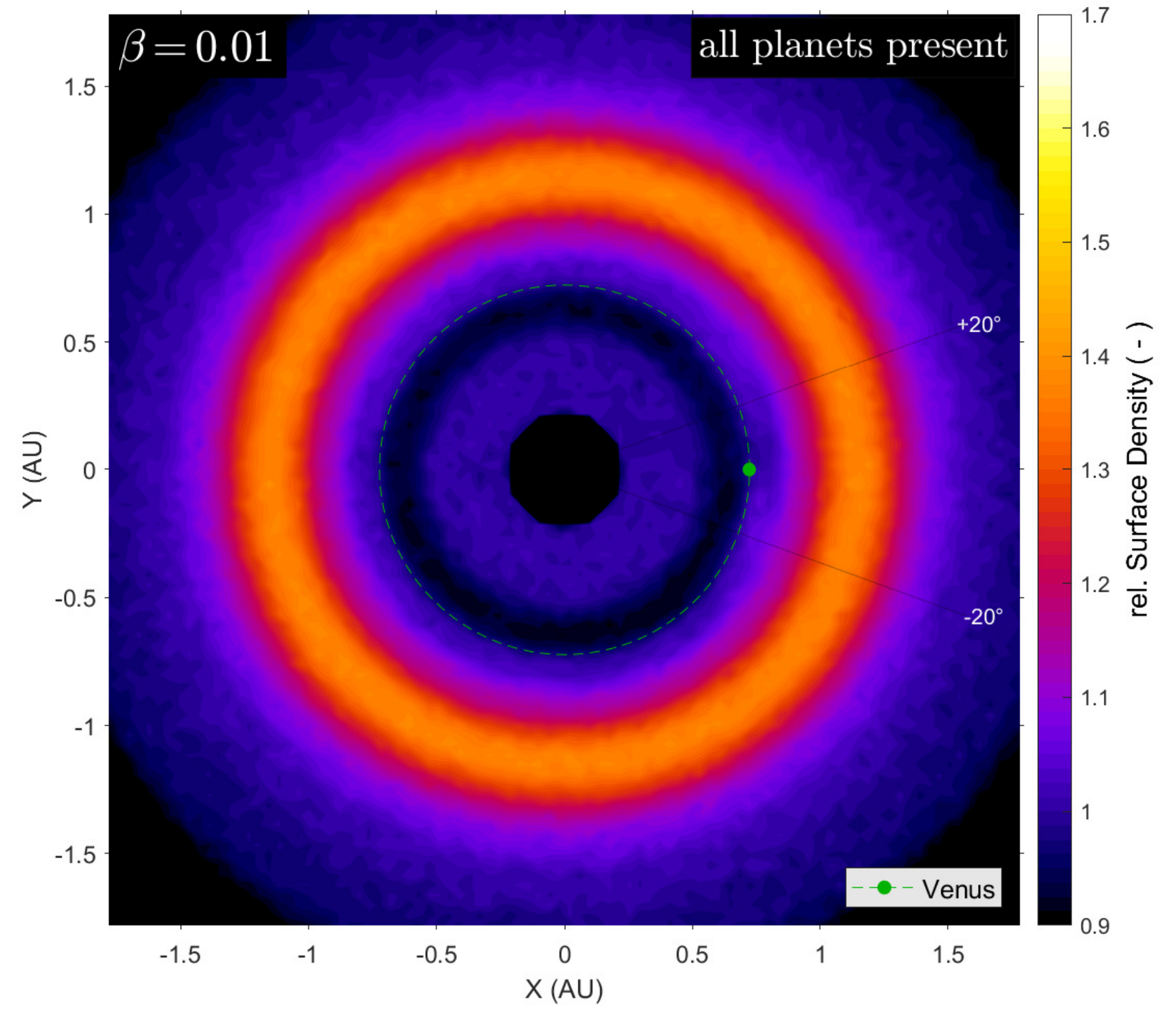} &
		\hspace{-2mm}\includegraphics[width=60mm,trim={3mm 0 4mm 0},clip]{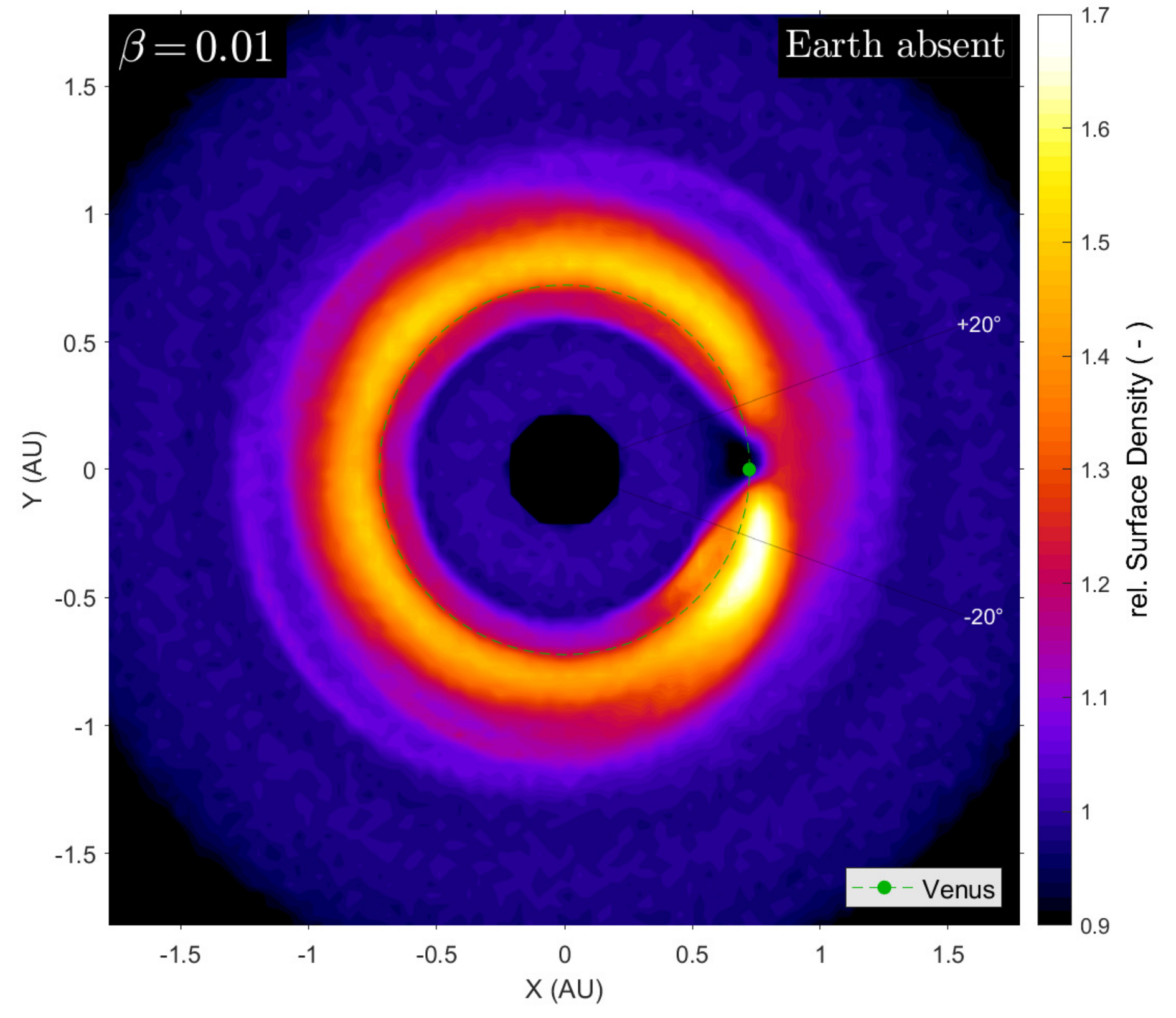} &
		\hspace{-2mm}\raisebox{11.6mm}{\includegraphics[width=60mm,trim={4mm 0 8mm 0},clip]{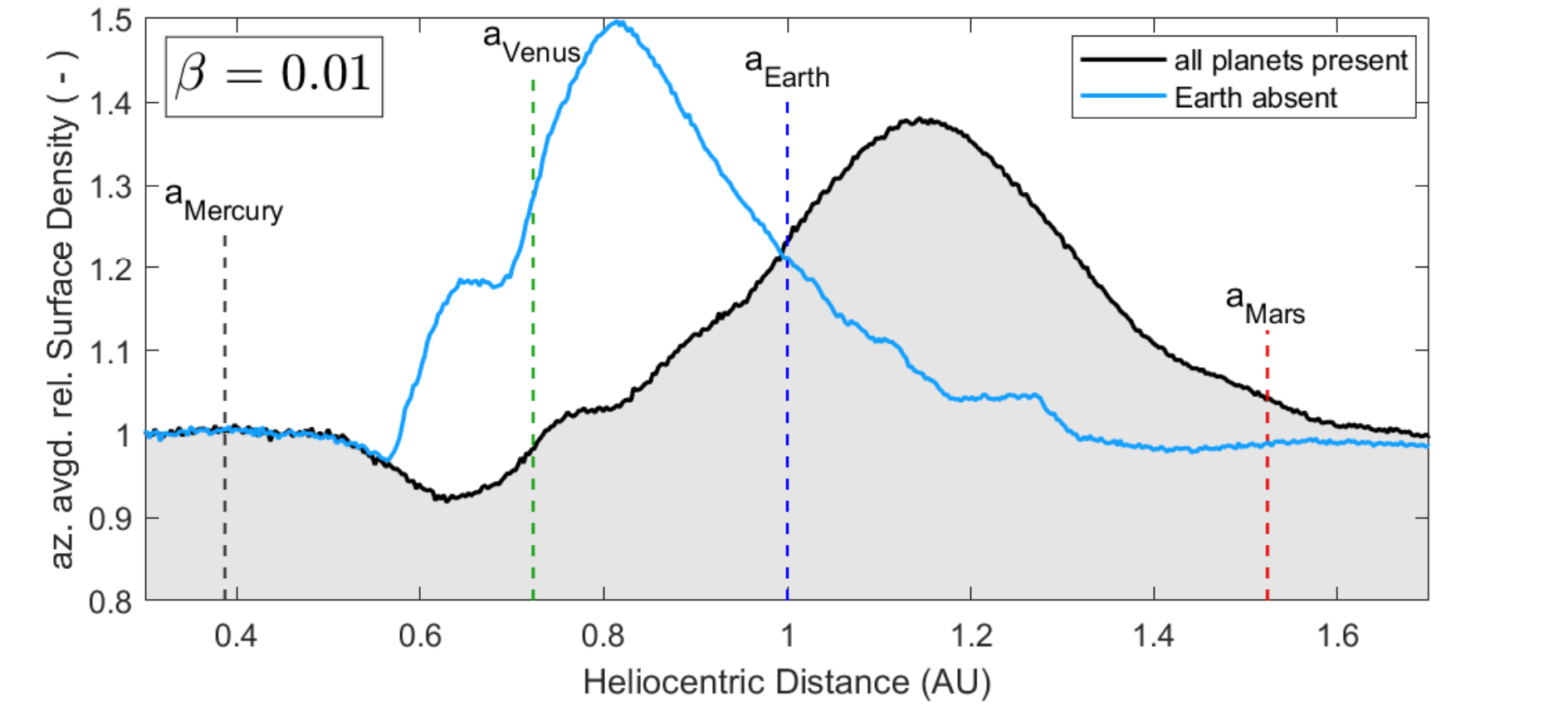}}
		\end{tabular}
      	\caption{Density distribution of Venus's resonant ring in a frame co-rotating with the planet 
      	under the presence \textit{(left)} and absence \textit{(middle)} of Earth for a $\beta$ of 0.03 
      	\textit{(top)} and 0.01 \textit{(bottom)}, as well as
      	the corresponding azimuthally averaged radial profiles \textit{(right)}. 
      	Removing Earth increases $\sigma_{AA,max,Venus}$ from 6\% to 19\% at $\beta=0.03$ and from 
      	4\% to 49\% at $\beta=0.01$, which is even higher than $\sigma_{AA,max,Earth}$ with 38\% in the case of
      	all planets present.}
        \label{fig:XY_Venus_NE}
	\end{figure*}

	In section~\ref{Ch31} we argued that the gravitational perturbations of neighbouring planets
	is a decisive factor in the formation of resonant rings in the inner Solar System.
	To confirm this conclusion, we conducted a series of additional simulations with changing planetary
	configurations. By removing planets from the simulations one by one, we can expose how 
	each of them affects the resonance trapping of another. All other simulation parameters 
	were left unchanged.
	
	First, we are going to look at how Venus and Earth interfere with each others' capability 
	of manipulating dust migration. 
	Figure~\ref{fig:SMA_NENV} compares the semi-major axis distribution resulting from a Solar System 
	missing either Earth or Venus with the nominal case (i.e. all planets present), 
	for two different particle sizes. 
	At $\beta=0.03$, we see a significant change in the semi-major axis occupancy at external Venus 
	MMRs when removing Earth from the simulation. 
	Here, first order resonances have their peak occupancy increased by about a factor of 2.
	The displacing of particles around the co-orbital resonance also becomes slightly aggravated.
	On the other hand, there is only a marginal impact on the occupancy of Earth resonances 
	when removing Venus at this particle size.
	Going to $\beta=0.01$, however, the deviation from the nominal case becomes much more apparent.
	While internal Venus resonances as well as external Earth resonances remain largely unaffected by the 
	respective other planet, resonances in the region between Venus and Earth are evidently
	impaired.
	Here, the removal of Earth increases peak occupancy of external first order Venus MMRs 
	by a factor in the order of 10, especially notable for those closer to Earth~(2:3V,
	3:4V, and 4:5V).
	Second order resonances also become occupied subsequent to virtually no activity under the presence
	of Earth, now reaching around 1/\nth{5} of the concentration of adjacent first order MMRs. 
	Interestingly, there are certain Venus resonances that, albeit weakly, aggregate particles in the nominal
	case, that disappear under the absence of Earth (5:8V and 3:5V).
	We revisit their occurrence in appendix~\ref{ChFR}.
	
	Looking at the internal Earth MMRs, we see a qualitatively mirrored albeit weaker deviation when 
	removing Venus from the model. 
	Since the internal resonances cause a net displacements of particles, the increase in 
	resonance efficacy manifests itself in an amplification of the troughs.
	This also contributes to the missing of the toroidal depletion inside Earth's orbit, 
	as we have argued in the previous section.
	In addition, we see a notable increase in the displacing ability of Earth's co-orbital
	resonance under the absence of Venus.
	
	With increasing particles size, the mutual interference of Earth and Venus evidently becomes the decisive factor for the
	strength of resonances in the region between them.
	This relation indicates that the effectiveness of a gravitational interference is dependent
	on particle migration time. 
	Especially trapping resonances, where migration is essentially halted, are most susceptible
	to gravitational termination by another planet. 
	In other words, gravitational perturbation becomes more likely to be the resonance terminating
	cause with increasing resonating time. 
	At higher $\beta$, where the resonating time of particles is much more limited by PR drag,
	gravitational perturbations by other planets are only secondary in affecting resonance occupation.
	
	\begin{figure*}[t]
		\includegraphics[width=\textwidth,trim={0 5mm 0 0mm},clip]{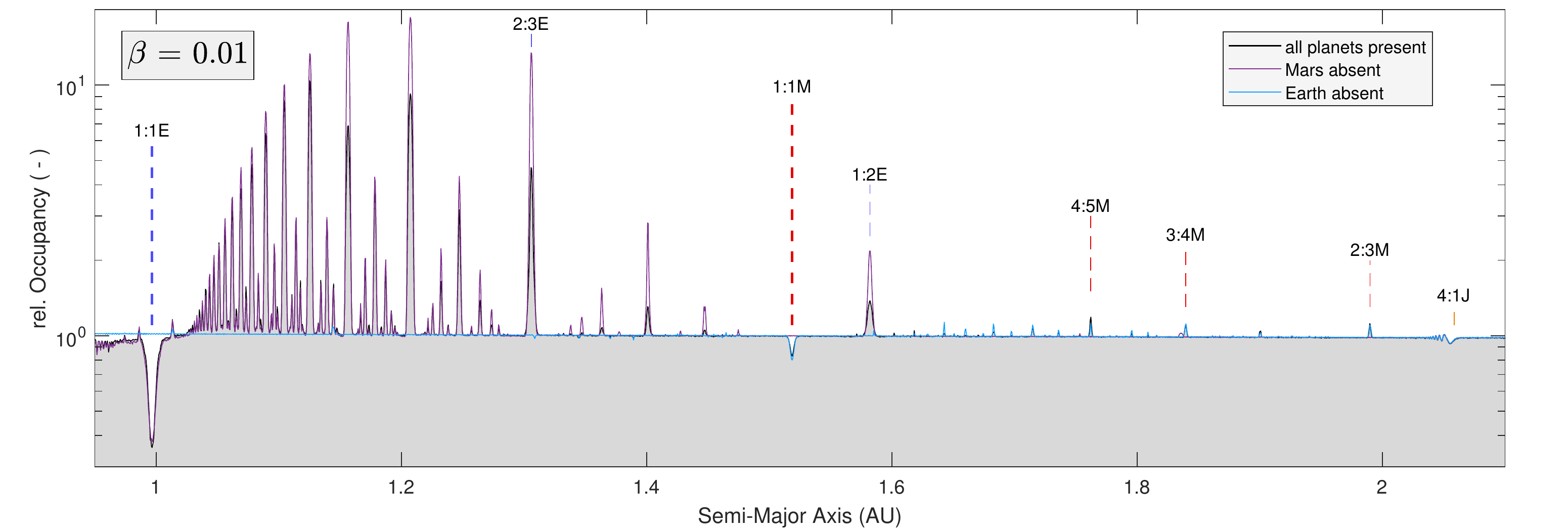}
		\includegraphics[width=\textwidth,trim={0 0mm 0 0mm},clip]{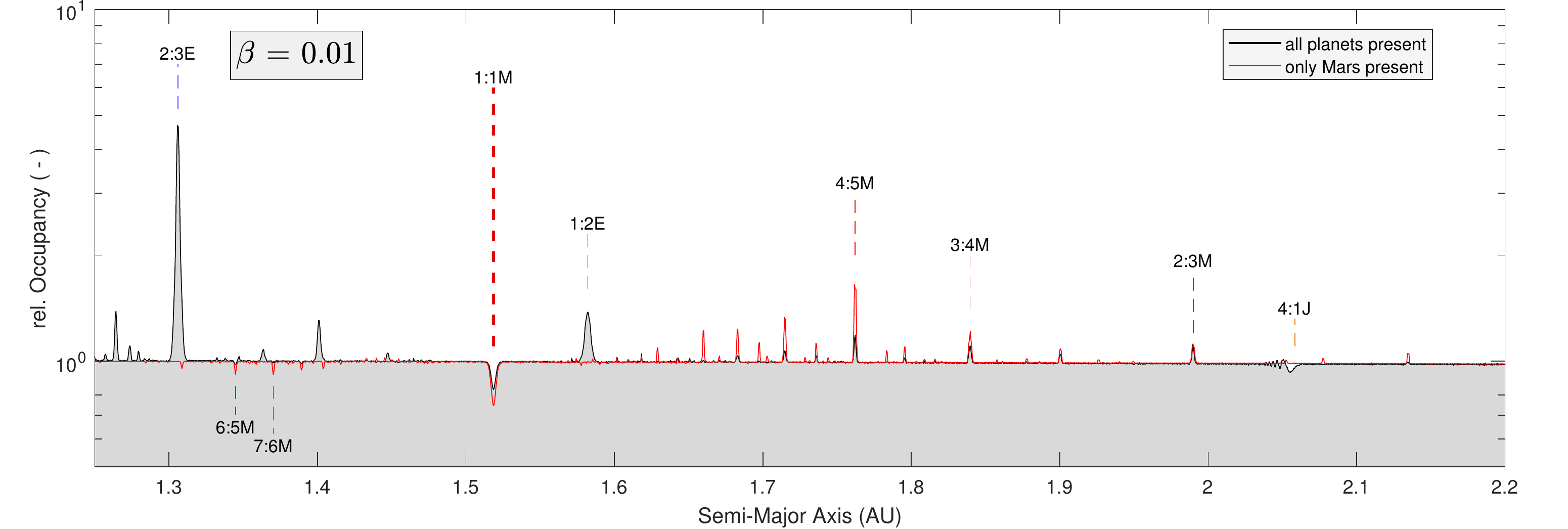}
      	\caption{Semi-major axis distribution under the interaction of Earth and Mars. 
      	The semi-major axis histogram compares the case of all planets present (also indicated
      	by the grey area) with either Earth or Mars absent \textit{(top)} or with Mars as the only 
      	present planet \textit{(bottom)} for $\beta=$~0.01\hspace{0.2mm}.
      	Vertical dashed lines indicate the locations of several first order as well as the 
      	co-orbital MMRs (including a shifting factor to account for radiation pressure).
      	Removing Earth has no significant effect on Mars MMRs, while the impact of Mars on 
      	Earth MMRs is more notable. Removing all other planets increases occupancy of external
      	Mars MMRs slightly, although remaining low compared to MMRs of Earth.}
        \label{fig:SMA_NENM}
	\end{figure*}

	Figure~\ref{fig:XY_Venus_NE} shows the density distribution of the ring of Venus that emerges 
	under the absence of Earth.
	Compared to the nominal case, $\sigma_{AA,max,Venus}$ increases by a factor of 3 at $\beta=0.03$
	and by  a factor of 12 at $\beta=0.01$\hspace{.2mm}. 
	This puts the resonant ring at a contrast similar to that of Earth, even surpassing it 
	in the case of $\beta=0.01$\hspace{.2mm}. 
	The explanation for the latter being that Earth's resonant ring is also suppressed to some extent
	by the presence of Mars, which we examine next.

	The interference of Mars on resonances of Earth must arguably be smaller than that of Earth on 
	Venus, given that Mars only has about 1/\nth{10} the mass of Earth and it being slightly more 
	distant to Earth, than Earth is from Venus ($a_{Mars} = 1.52 \times a_{Earth}$ compared to 
	$a_{Earth} = 1.37 \times a_{Venus}$).
	For these reasons, their interaction is only of consequence in the larger particle size range
	($\beta\hspace{-1mm}<\hspace{-1mm}0.02$).
	Figure~\ref{fig:SMA_NENM} (\textit{top}) shows the semi-major axis distribution at 
	$\beta\hspace{-.6mm}=\hspace{-.6mm}0.01$ for either Earth or Mars absent. 
	Removing Mars causes a prominent enhancement of Earth resonances, becoming 
	most visible at the outer first order resonances (4:5E, 3:4E, and 2:3E), whose peak occupancies
	more than double.
	Removing Earth on the other hand, does not produce a meaningful enhancement of Mars resonances.
	Even the most notable Mars MMRs stay well below the far-out 1:2E resonance.
	
\begin{figure*}[!pht]
		\centering
		\begin{tabular}{ccc}
		\hspace{-2mm}\includegraphics[width=60mm,trim={3mm 0 4mm 0},clip]{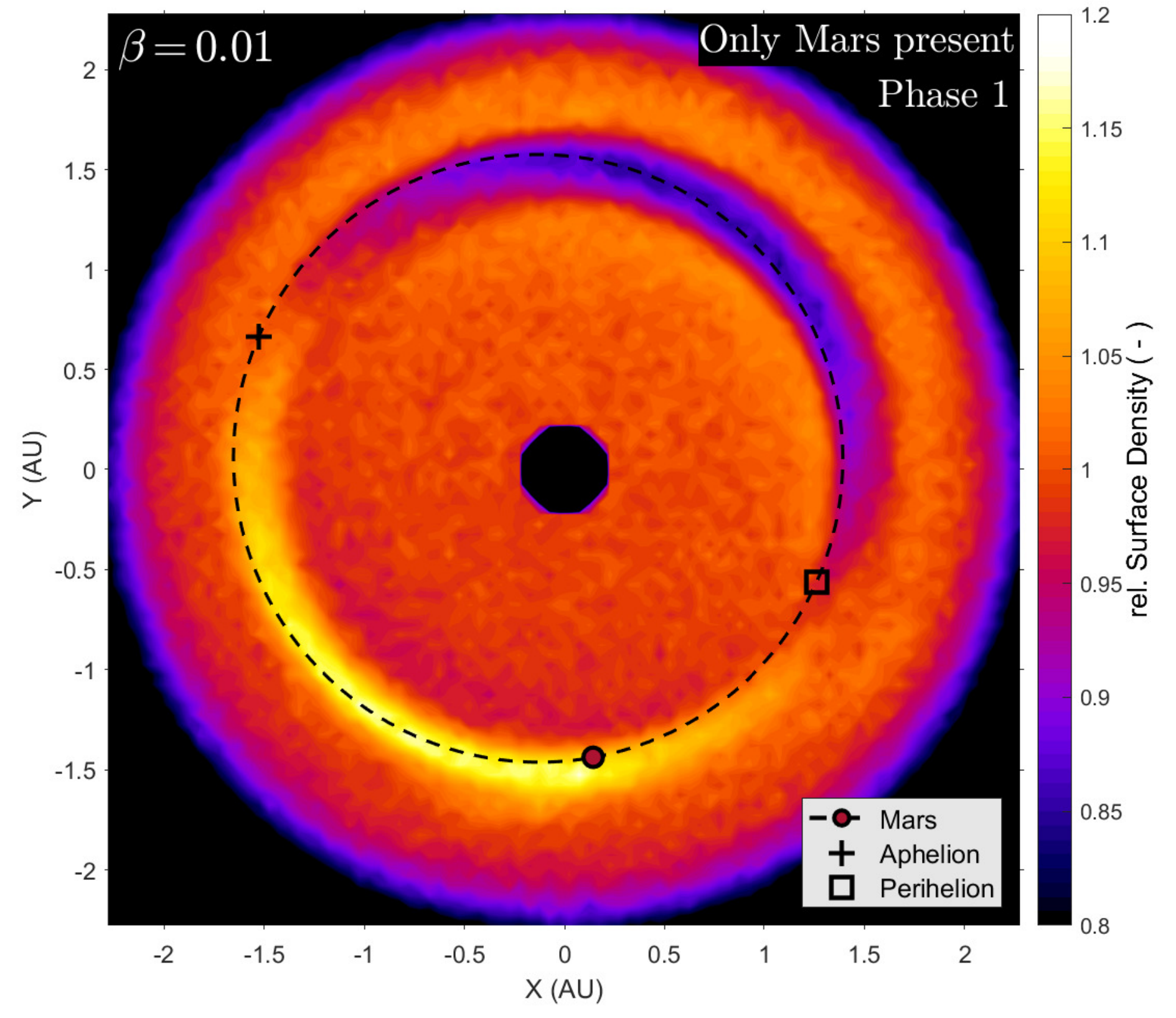} &
		\hspace{-2mm}\includegraphics[width=60mm,trim={3mm 0 4mm 0},clip]{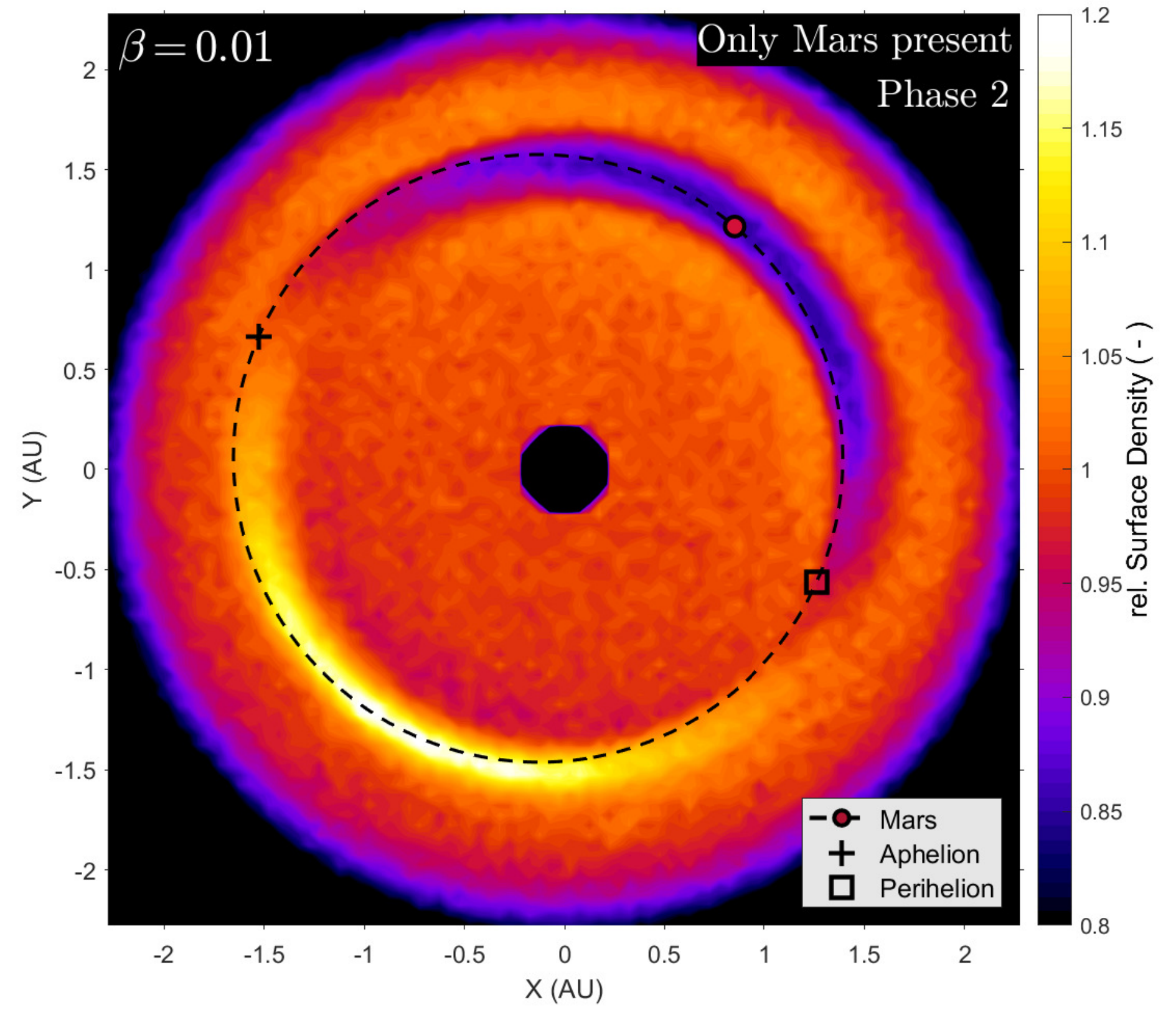} &
		\hspace{-2mm}\includegraphics[width=60mm,trim={3mm 0 4mm 0},clip]{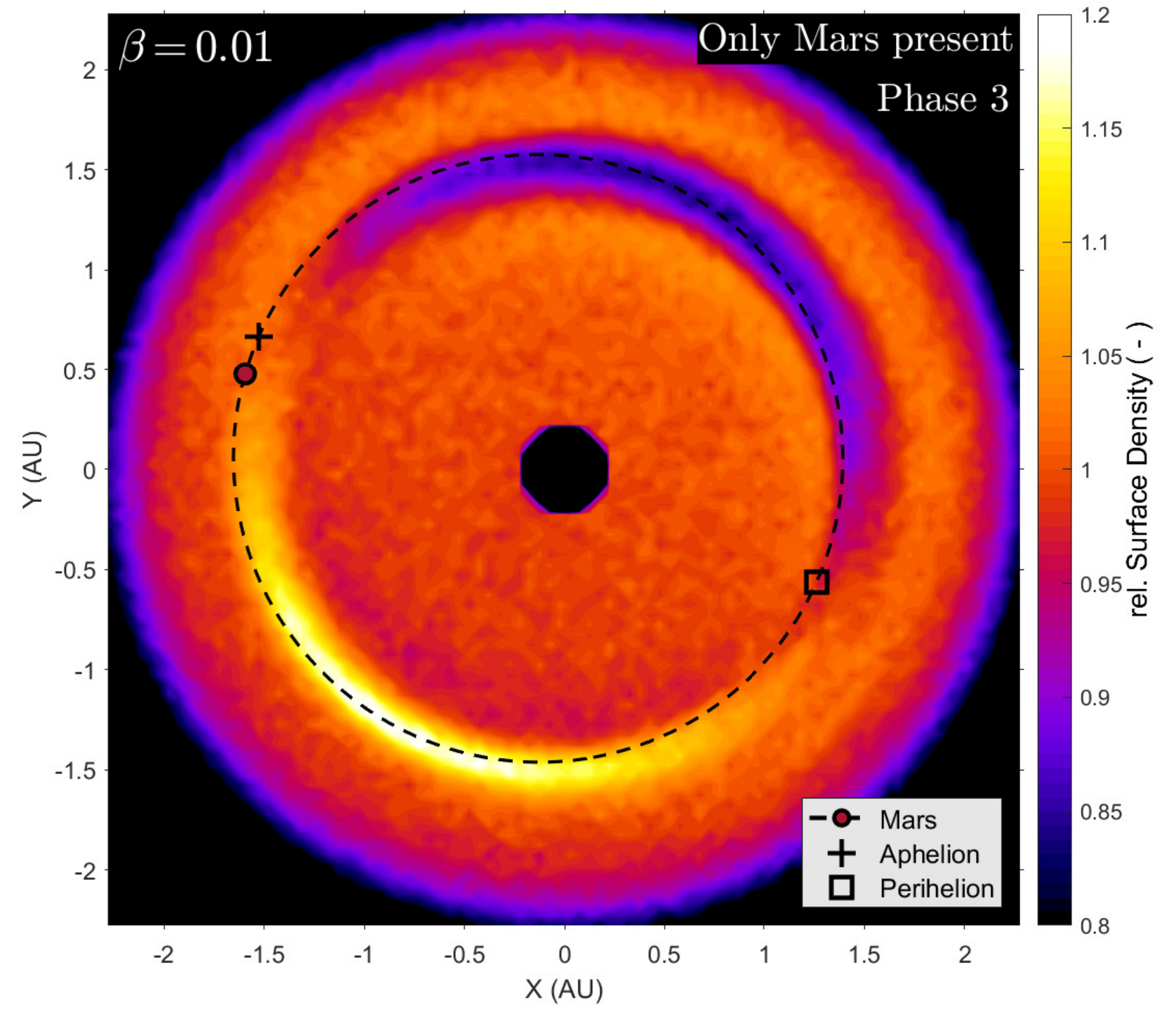}
		\end{tabular}
      	\caption{Density distribution in a Martian single planet system in a fixed reference frame at 
      	three different planetary phases for $\beta=0.01$\hspace{.2mm}. 
      	Instead of a pattern rotating with the planet, Mars generates a double-crescent-shaped density feature that remains fixed
      	with respect to the alignment of Mars's orbit and varies in contrast only marginally with Mars's 
      	orbital phase.
      	The pattern is mirrored roughly at Mars's line of apsides.}
        \label{fig:XY_OM}
	\end{figure*}

\begin{figure*}[]
		\centering
		\begin{tabular}{ccc}
		\hspace{-2mm}\includegraphics[width=60mm,trim={3mm 0 4mm 0},clip]{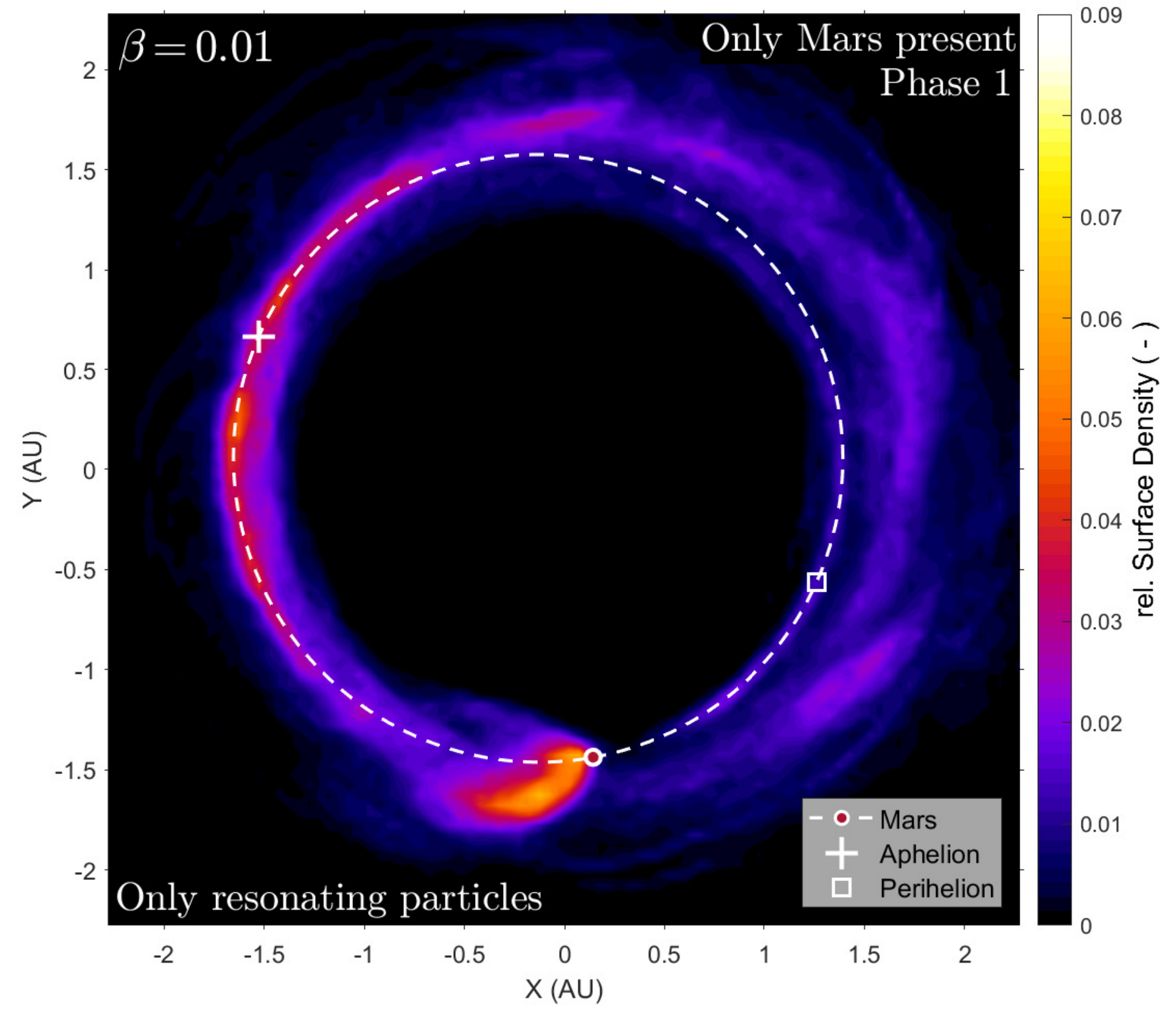} &
		\hspace{-2mm}\includegraphics[width=60mm,trim={3mm 0 4mm 0},clip]{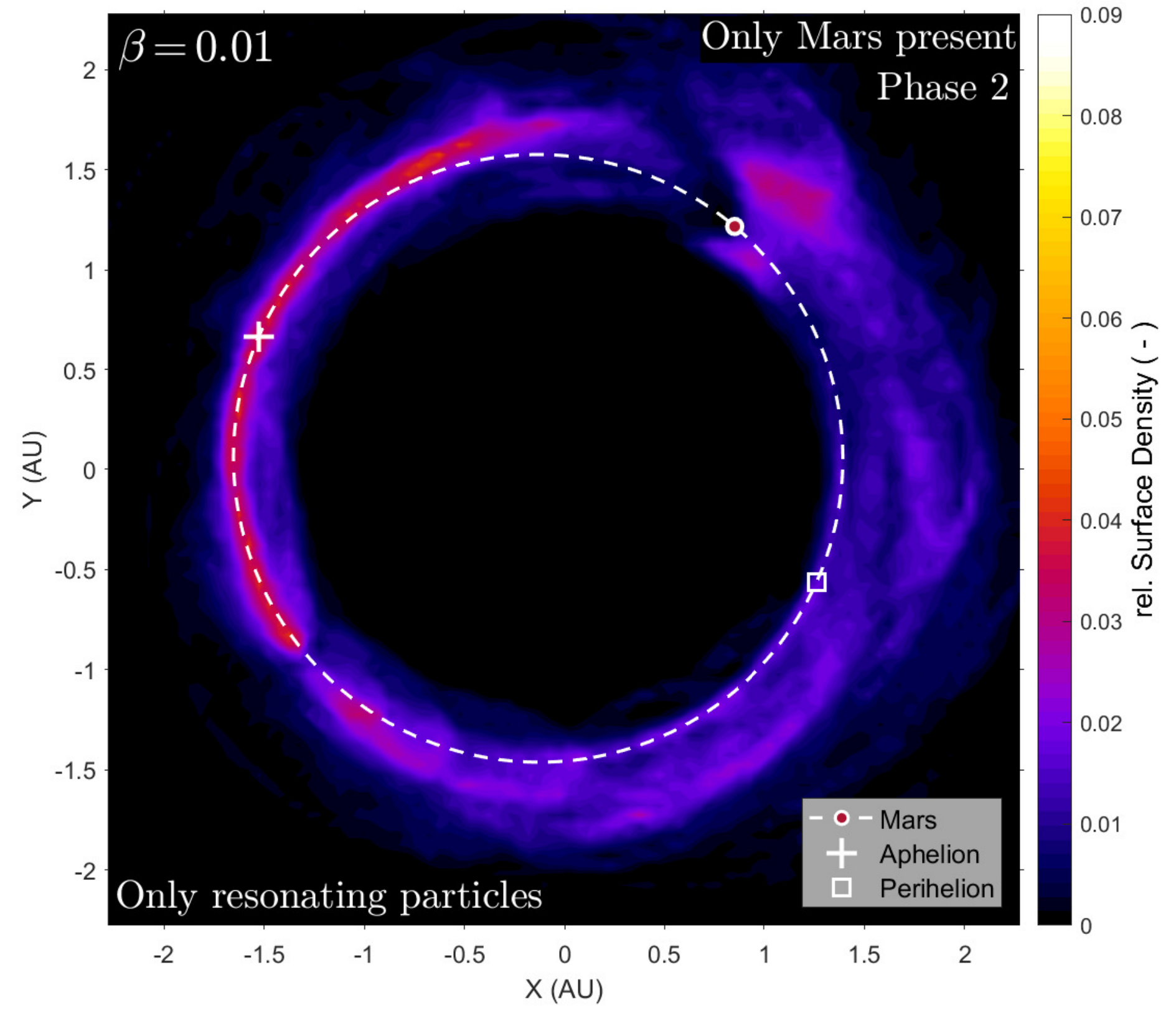} &
		\hspace{-2mm}\includegraphics[width=60mm,trim={3mm 0 4mm 0},clip]{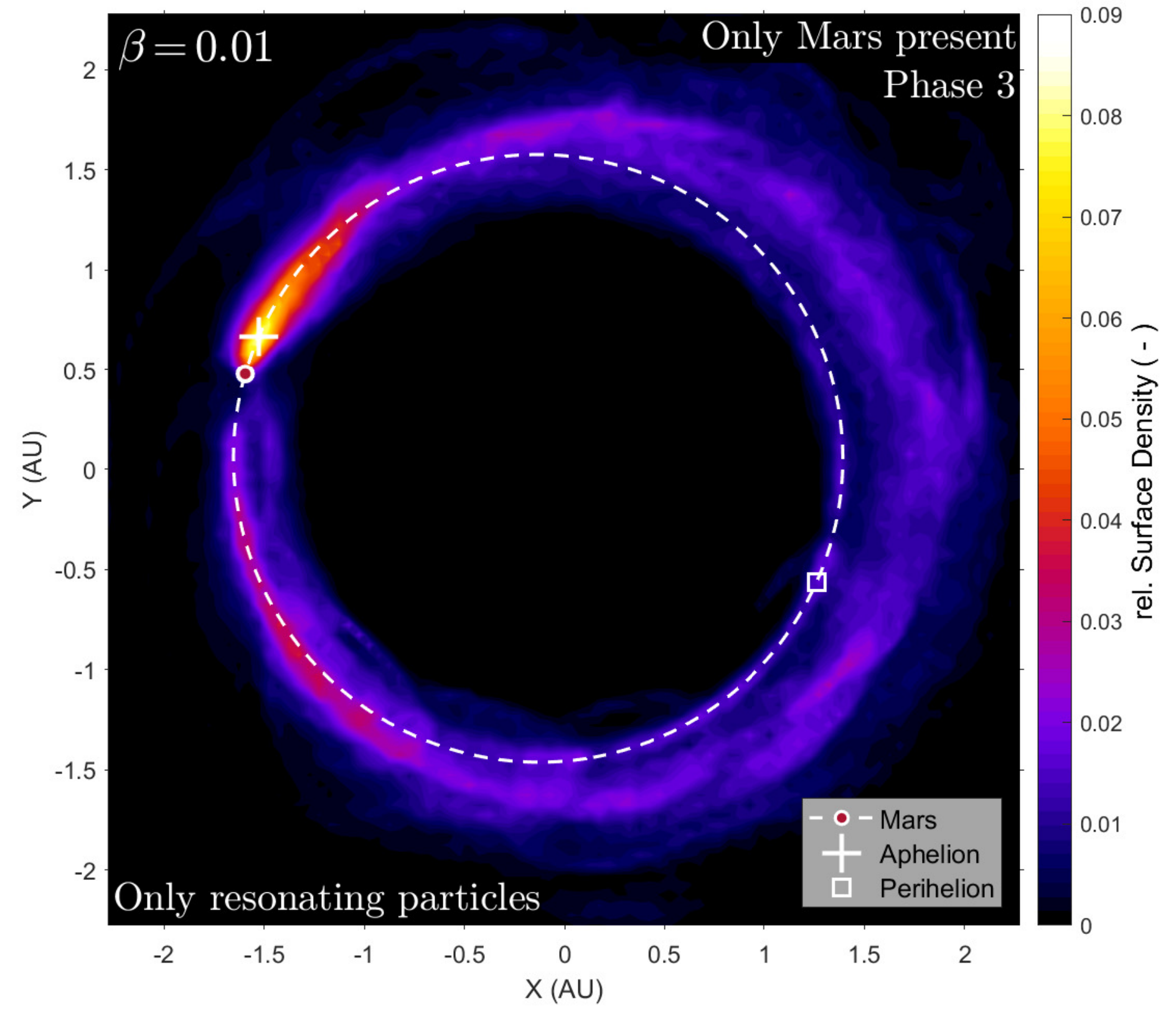}
		\end{tabular}
      	\caption{Density distribution considering only particles trapped in 
      	resonances with Mars in a Martian single planet system in a fixed reference frame at 
      	three different planetary phases for $\beta=0.01$\hspace{.2mm}.
      	To identify trapped particles, their semi-major axis evolution was analysed. Only when the 
      	semi-major axis hovered about a constant value a particle was considered trapped and
      	only then it was accounted for in the density distribution.
      	As with the previous density plots, the density is normalised to the surface density arising under 
      	the absence of planets.
      	The low density as well as the different geometry of these resonant features makes them inadequate
      	as to solely account for the overall density variations produced by Mars 
      	(Fig.~\ref{fig:XY_OM}), indicating a formation mechanism other than resonant trapping.
      	}
        \label{fig:XY_OM_OR}
	\end{figure*}

	Suspecting Jupiter to be an additional MMR weakening factor here, we conducted an additional 
	simulation considering Mars as the only planet, to gauge its ability to trap 
	dust in the absence of other planets (see Fig.~\ref{fig:SMA_NENM}, \textit{bottom}).
	At $\beta=0.01$, this yields particle concentrations at some external Mars MMRs that are 
	comparable to that at 1:2E.
	To plot the resulting density distribution, however, we can not resort to the stacking approach 
	we have used so far, as stated in Sect.~\ref{Ch31}:
	Due to the higher eccentricity of the orbit of Mars, we can expect a density pattern that 
	changes with respect to the planet as it revolves around the Sun, making 
	merely recording particles in a reference frame co-rotating with the planet inappropriate.
	In a single planet system, however, the planet's orbit remains unaltered by perturbations. 
	By adjusting the saving interval, the stacking method can therefore be modified 
	such that particle recording takes place in a fixed	reference frame, 
	but with the planet always in the same orbital phase,	 
	as adopted by~\citet{0004-637X-625-1-398}. 
	Figure~\ref{fig:XY_OM} shows the resulting density distribution for three different phases along the
	orbit of Mars.
	We note that the density pattern generated by Mars alone differs substantially from the resonant 
	rings of Venus and Earth. 
	Instead of co-rotating with the planet, the distribution appears stationary
	in a fixed reference frame and aligns with the orientation of the orbit of Mars.
	The emerging pattern can be described as two regions in the form of narrow crescents, mirrored 
	at the line of apsides and confined in radial extent roughly by Mars's peri- and aphelion distances. 
	The region spanning from Mars's aphelion to perihelion is characterised by a density enhancement, whereas
	the region spanning peri- to aphelion exhibits a depletion, comparable in magnitude.
	The overdensity of the enhancement varies slightly between the three examined phases, 
	from $\sigma_{TB,max,Mars}=17\%$ at phase 1 to $\sigma_{TB,max,Mars}=20\%$ at 
	phase 2 \& 3.
	The shape of these features is unaffected by changing $\beta$, while their contrast is not.
	At $\beta=0.1$ the density variation nearly vanishes reaching only around 3\%,
	whereas $\beta=0.005$ produces an enhancement of 40\% and a trough reaching a reduction of 27\%.
	The dissimilarity between this double crescent pattern and the resonant rings modelled 
	so far creates doubt about	a common formation mechanism.
	Investigating this discrepancy, we looked at how resonating particles contributed to the discs 
	seen in Fig.~\ref{fig:XY_OM}.
	Figure~\ref{fig:XY_OM_OR} shows the density distribution in the Martian single planet system exclusively of 
	those particles that are trapped in mean-motion resonances with Mars. 
	This was achieved by only allowing particles to contribute when their semi-major axis hovered about 
	a constant value.
	The result differs substantially from the double crescent pattern, showing a diffuse ring with a trailing 
	enhancement varying in geometry and contrast with planetary phase.
	It is evident that the low density as well as the different geometry of these resonant features makes 
	them inadequate as to solely account for the distinct double crescent pattern.
	In addition, the crescents are discernible throughout the particle size range, while the semi-major
	axis distribution shows no meaningful trapping in Mars resonances for $\beta\geq 0.02$ 
	(see Figs.~\ref{fig:b02_SMA_AP}~\&~\ref{fig:A_SMA_AP}).
	An effect other than resonance trapping must therefore be presumed as the underlying cause.
	We investigate these non-resonant features in Sect.~\ref{Ch33}.
	
	Examining the interaction of Earth and Mars, revealed a notable impact on external Earth MMRs at 
	$\beta=0.01$, even though Mars holding only 10\% of its mass (cf.~Fig.~\ref{fig:SMA_NENM}).
	This gives us an idea of the sensitivity of these resonances to even small gravitational 
	perturbations.
	A non-negligible influence of yet more distant planets must be presumed. 
	Therefore, we complemented our set of simulations by additionally removing Jupiter and ultimately
	removing all planets but Earth.	
	
\begin{figure*}
		\centering
		\begin{tabular}{cc}
		\includegraphics[width=68mm,trim={3mm 0 4mm 0},clip]{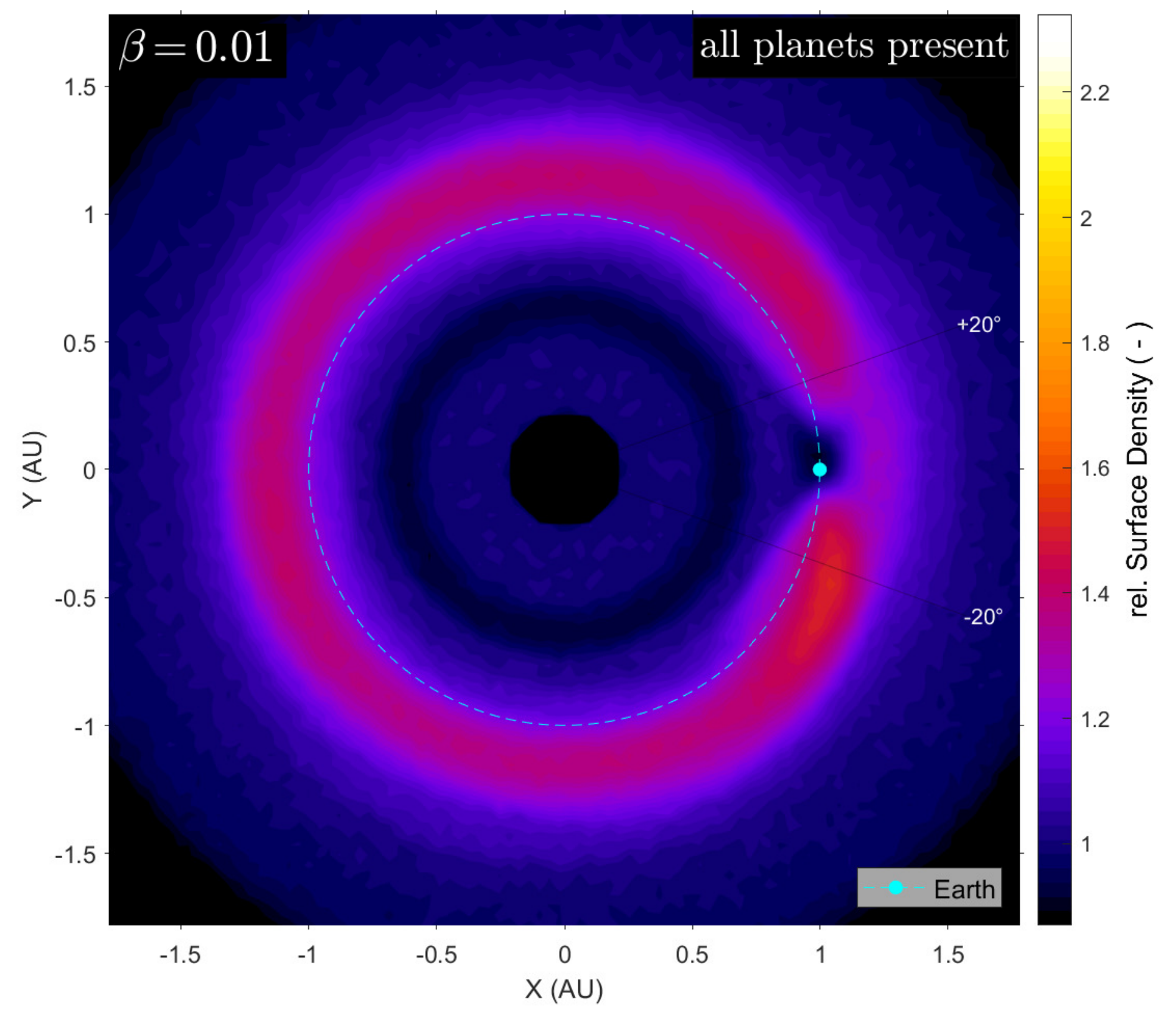} &
		\includegraphics[width=68mm,trim={3mm 0 4mm 0},clip]{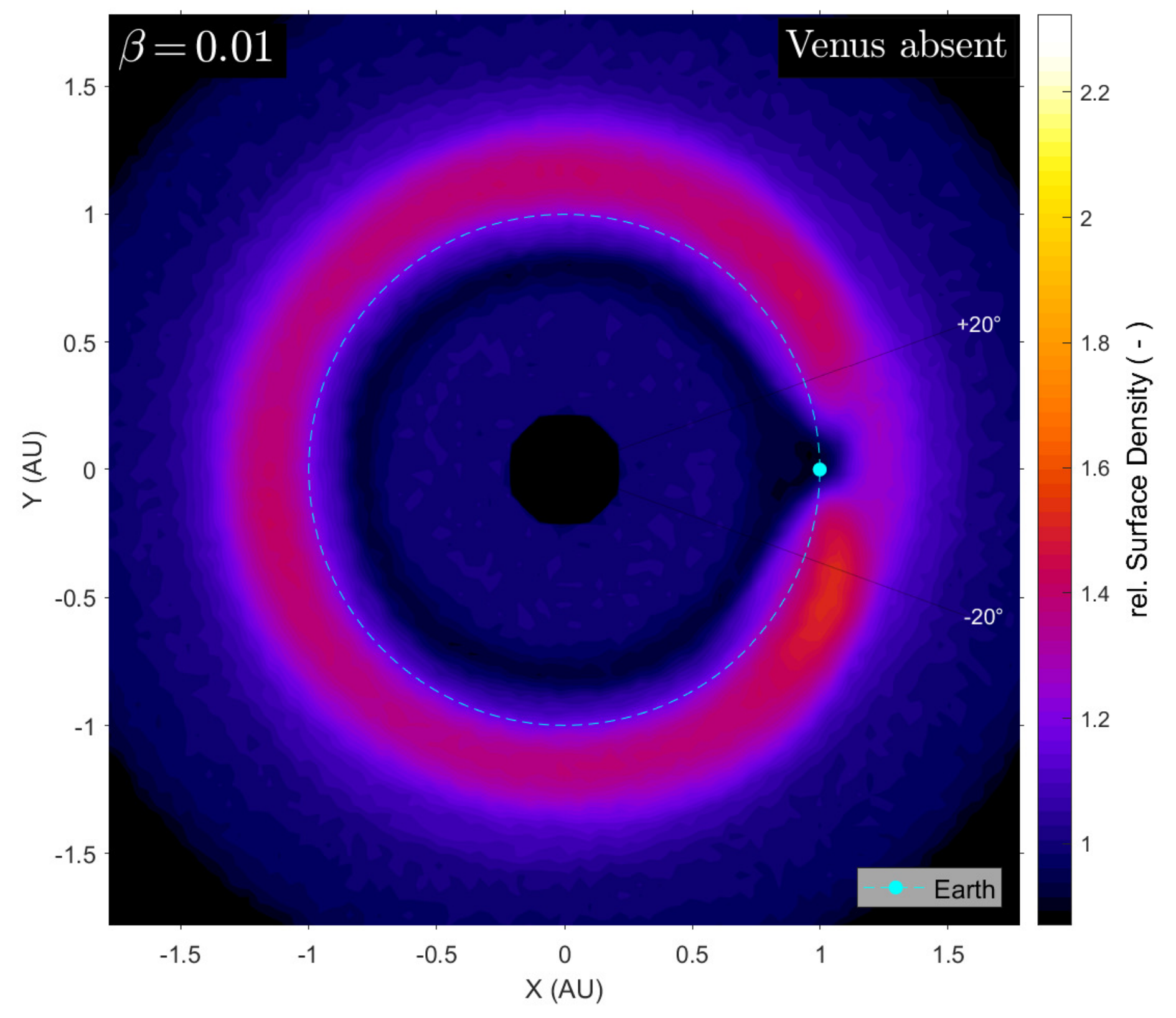} \\[-2.5ex]
		\includegraphics[width=68mm,trim={3mm 0 4mm 0},clip]{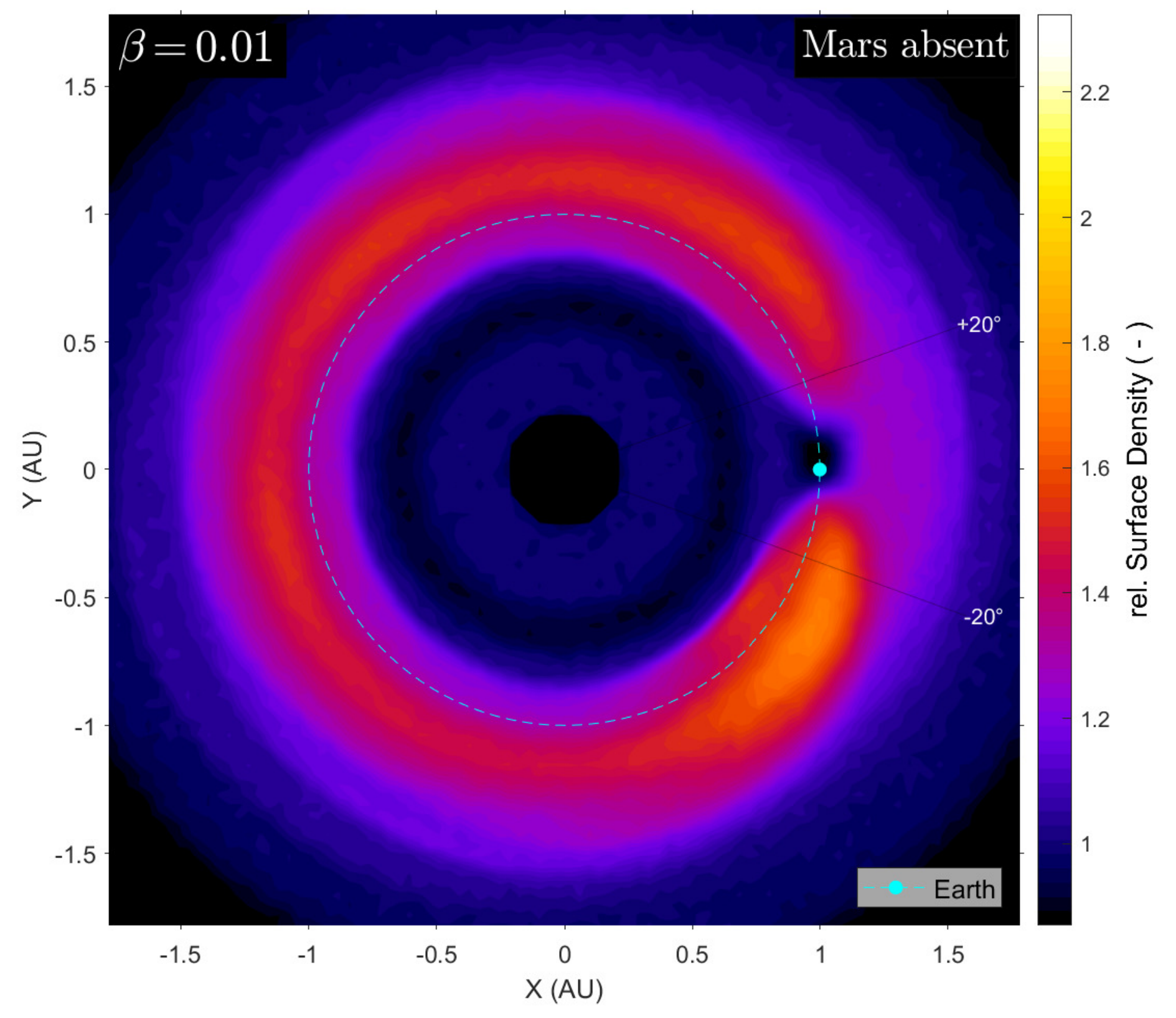} &
		\includegraphics[width=68mm,trim={3mm 0 4mm 0},clip]{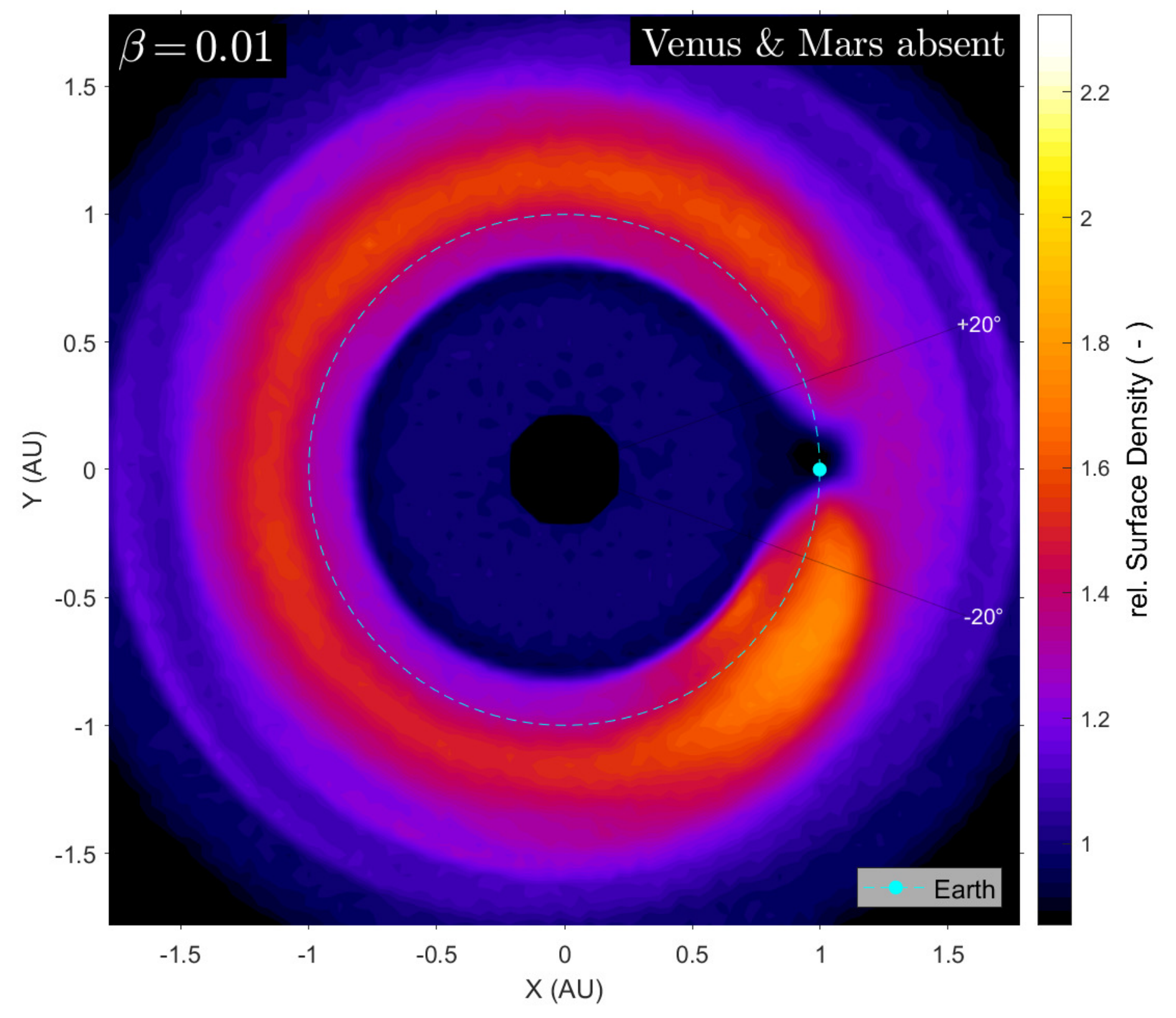} \\[-2.5ex]
		\includegraphics[width=68mm,trim={3mm 0 4mm 0},clip]{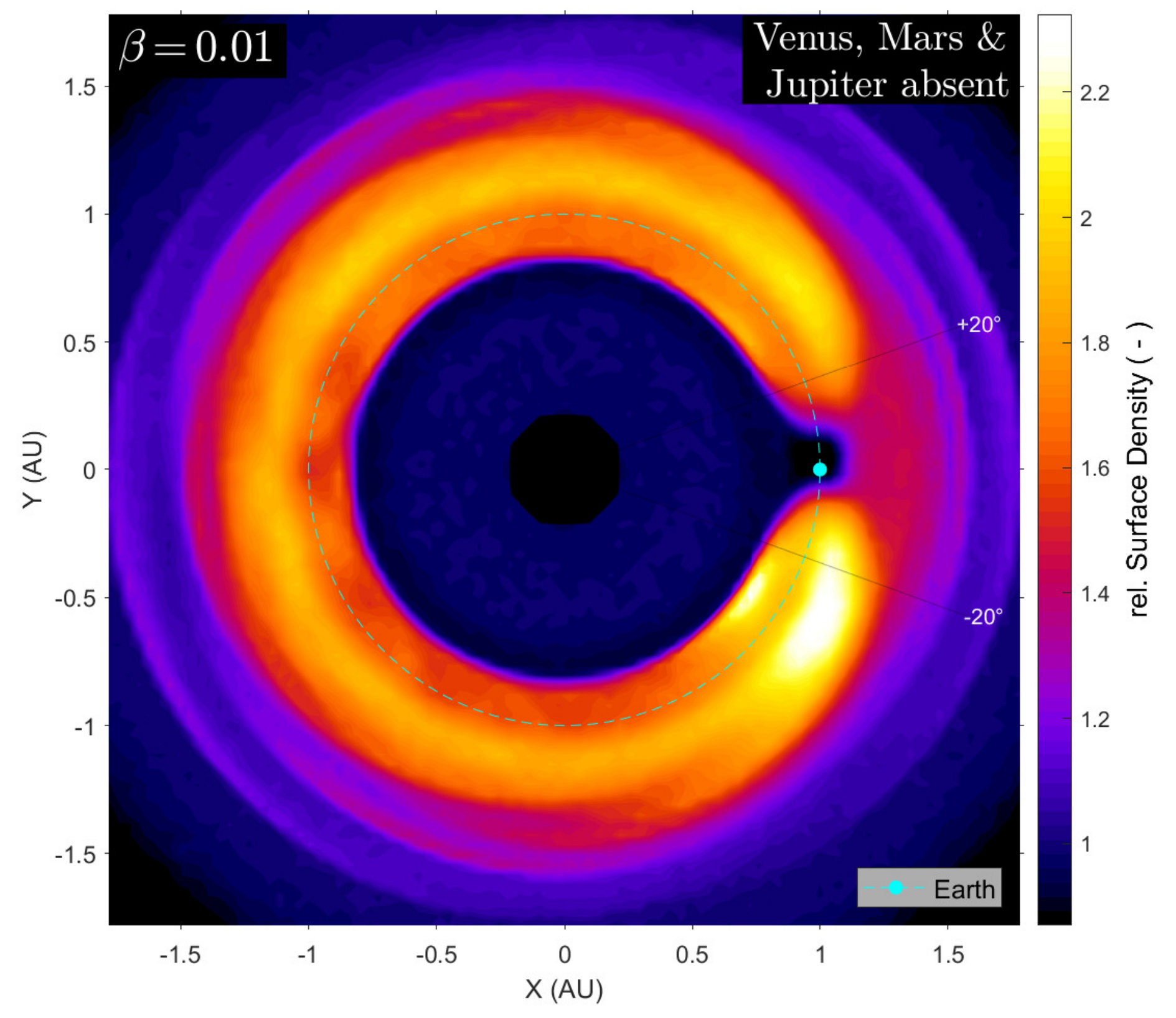} &
		\includegraphics[width=68mm,trim={3mm 0 4mm 0},clip]{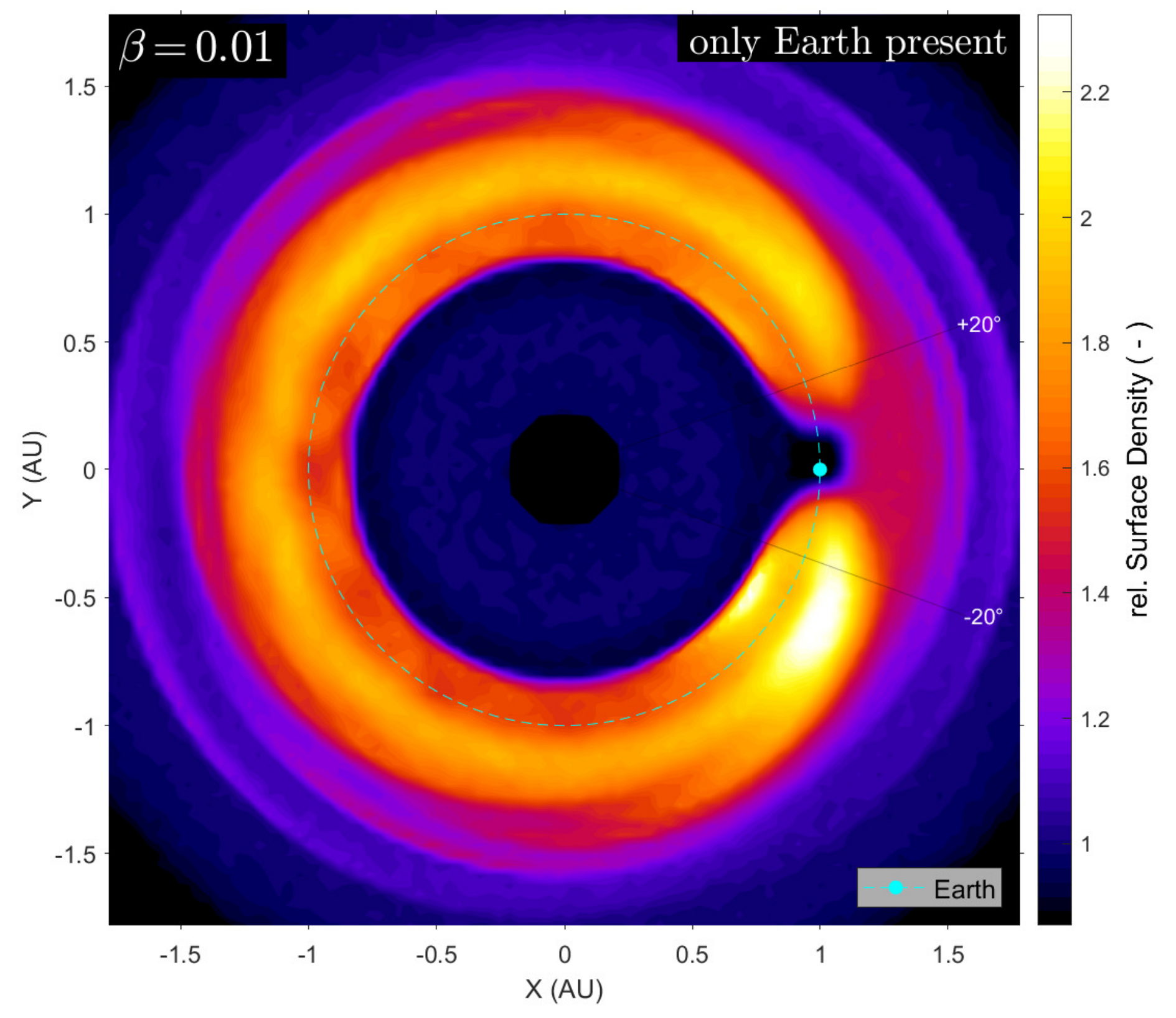} \\[-2.5ex]
		\end{tabular}\par
		(a)\\[3ex]
		\begin{tabular}{cc}
		\includegraphics[width=85mm,trim={0 0mm 0 0mm},clip]{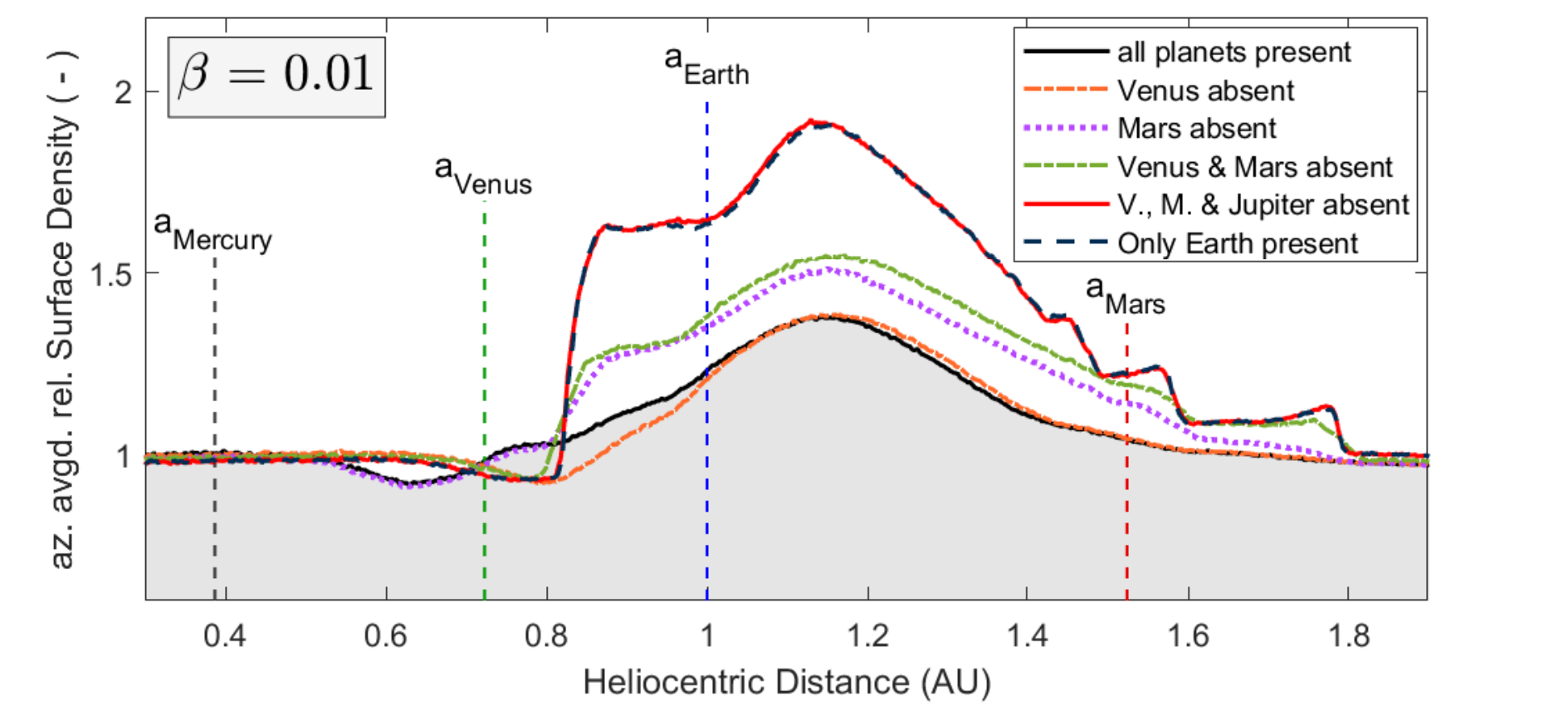} & 
		\includegraphics[width=85mm,trim={0 0mm 0 0mm},clip]{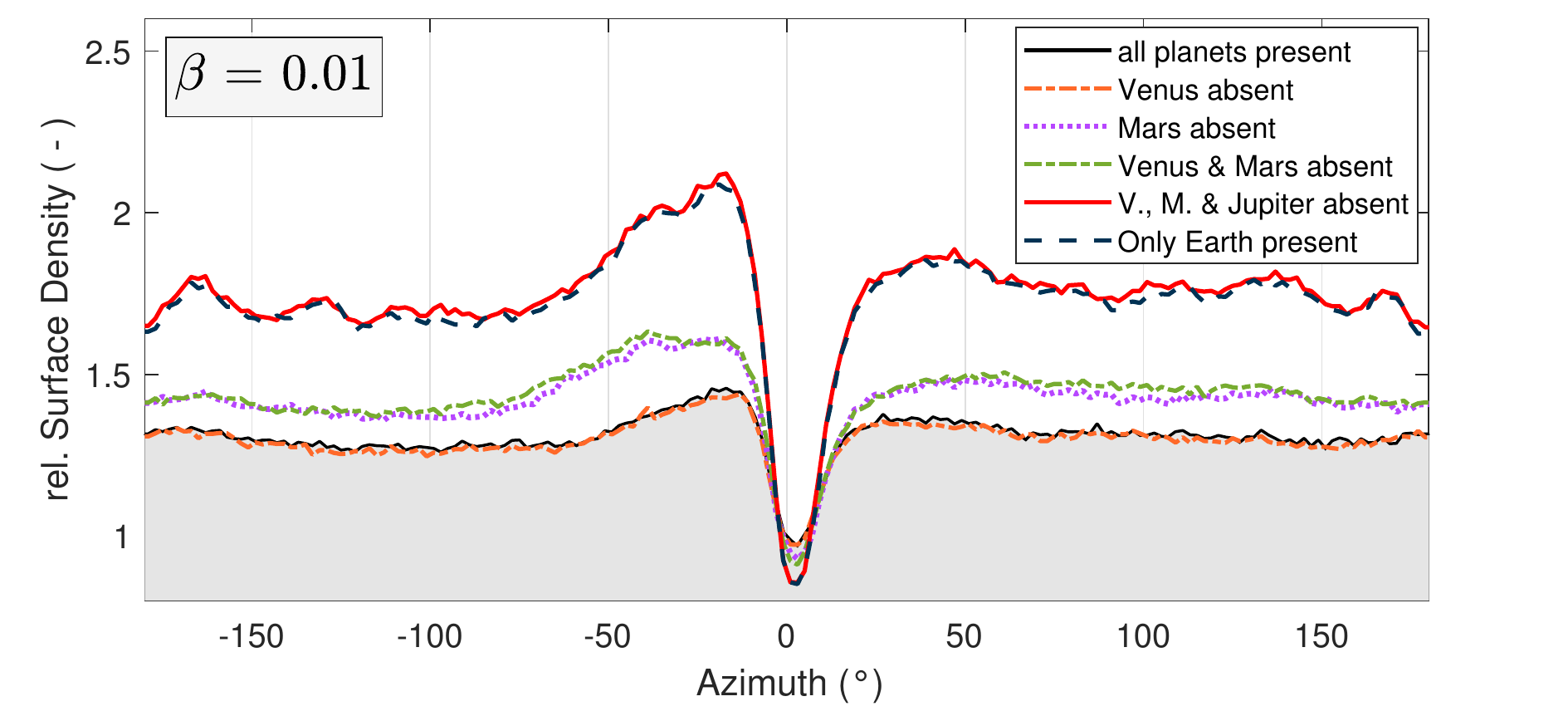}\\[-.5ex]
		(b) & (c)
		\end{tabular}
      	\caption{Density distribution of Earth's resonant ring under varying planetary configurations 
      	for $\beta=0.01$ (a), 
      	as well as corresponding azimuthally averaged radial profiles (b), 
      	and azimuthal profiles at a heliocentric distance of 1~\si{\astronomicalunit} (c).
      	The incremental increase of ring density with the removal of other planets is apparent.
      	Counterintuitively, Venus shows only little influence on the inner edge of Earth's resonant ring.
    	Mars and Jupiter on the other hand suppress both the inner and outer edge of the ring confining 
      	its overall width, while also significantly reducing the peak density of the azimuthal average.}
        \label{fig:EarthRing_1}
	\end{figure*}

	Figure~\ref{fig:EarthRing_1}a shows the density distribution of Earth's resonant ring that emerges 
	under these	different planetary configurations for $\beta=0.01$\hspace{.2mm}. 
	The result is a progression of ring contrast and extent as interferences of other planets
	are incrementally eliminated. 
	While the impact of Venus is largely limited to the undoing of the toroidal depletion inside
	Earth's orbit that results from the impairment of internal Earth MMRs, 
	the effects of Mars and Jupiter are much more severe. 
	Removing Mars not only causes an increase of $\sigma_{AA,max,Earth}$ from 38\% to 50\%, 
	but also allows for the formation of the distinct inner edge.
	As discussed in section~\ref{Ch31}, this edge forms as a result of resonating particles	having 
	their dwell time increased long enough to approach the equilibrium solution of their resonance.
	Counterintuitively, it is not Venus that causes the inner edge of Earth's resonant ring
	to disappear. 
	However, since the particles constituting the inner edge assume a high value of
	eccentricity as they approach the equilibrium solution, it is logical to presume
	their stability being most sensitive to an outer neighbour planet. 
	It is near aphelion where these high eccentricity particles spend most of their time
	and are thus most susceptible to gravitational perturbation.
	
	The simultaneous absence of Venus and Mars further increases $\sigma_{AA,max,Earth}$ 
	to 54\% and sharpens the edges of the ring. 
	Especially the outer features of the ring become more distinct, which are generated
	by trapping in the remote, yet effective 2:3E and 3:4E resonance.
	This again hints at more particles being able to approach the equilibrium solution 
	in these resonances. 
	With a high limiting eccentricity leading these particles deep inside the orbit 
	of Earth, a further increase of these resonances with the removal of Venus is to be expected 
	(limiting pericentre of the 2:3E and 3:4E MMRs are 0.823~\si{\astronomicalunit} 
	and 0.835~\si{\astronomicalunit}, derived from \citet{BEAUGE1994239}).
	However, the fact that this increase only occurs under the simultaneous absence 
	of Venus and Mars, indicates that this effect is only secondary to termination
	of resonances through an outer neighbour planet. 
	On the radial density profile (Fig.~\ref{fig:EarthRing_1}b), these manifest as a stair-like pattern between 
	1.4~\si{\astronomicalunit} and 1.8~\si{\astronomicalunit}. 
	
	Finally with the absence of Jupiter, the overall density as well as the edges
	are again substantially enhanced. 
	Compared to the nominal case, $\sigma_{AA,max,Earth}$ more than doubles to
	around 92\%. 
	With regard to density, the more distant Jupiter must therefore be considered even more 
	effective in suppressing the Earth's resonant ring than Mars.
	Removal of the remaining planets does not lead to a meaningful change in the density
	distribution, implying that the influence of Mercury and the outer three gas giants on 
	MMRs of Earth is negligible at this particle size.
	
\begin{figure}[]
		\centering
		\includegraphics[width=\hsize, trim=0 6mm 0 0, clip]{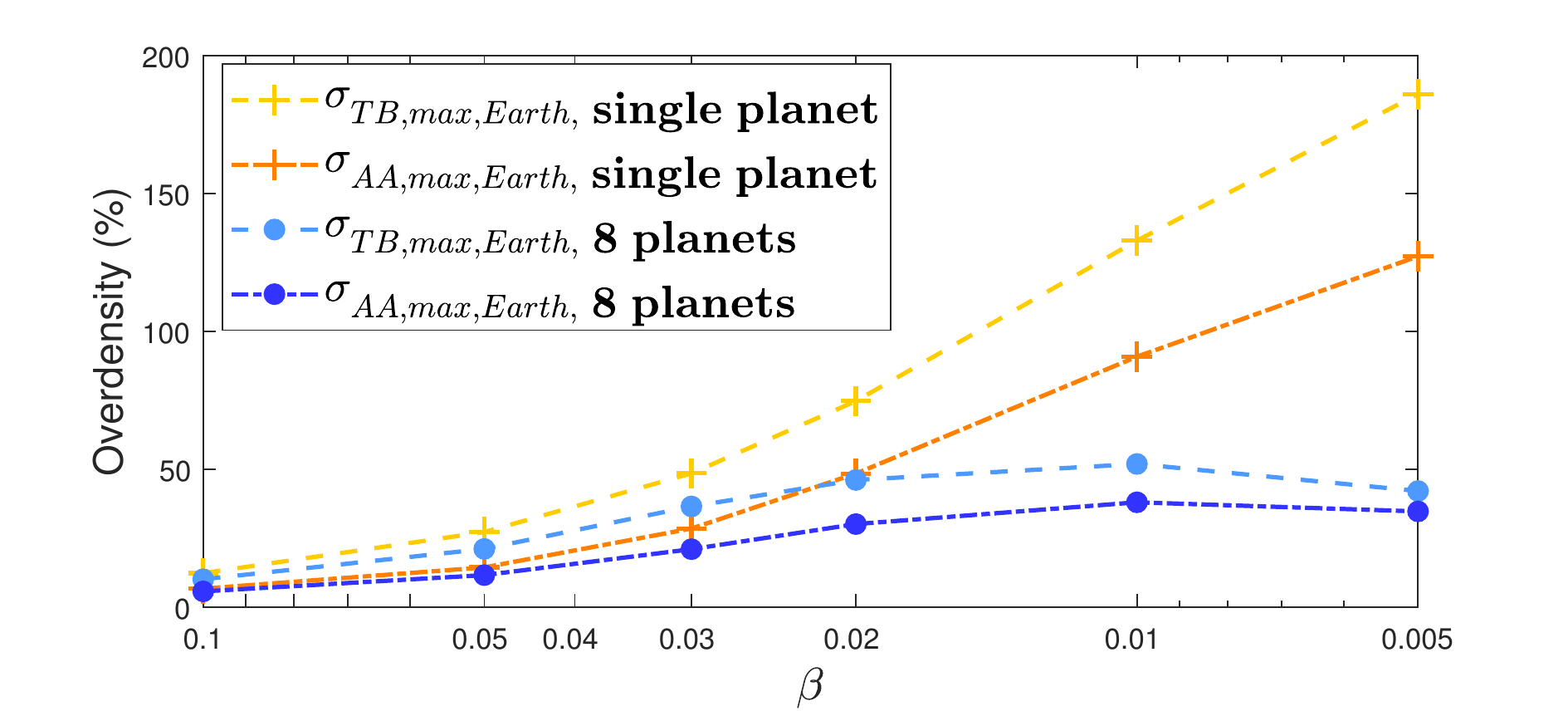} \\
		\includegraphics[width=\hsize, trim=0 1mm 0 5.6mm, clip]{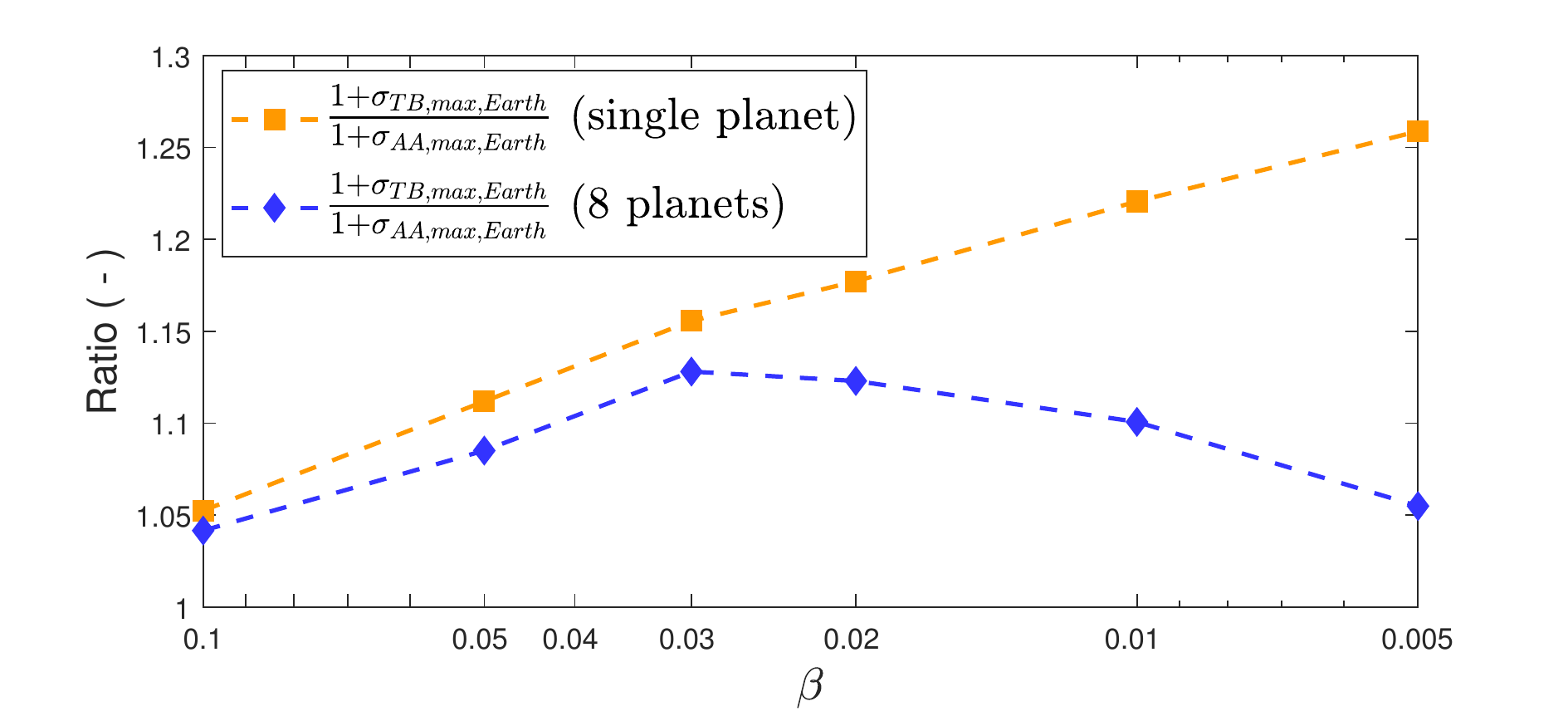}
		\caption{Severity of the weakening of Earth's resonant ring through other planets, 
		with respect to $\beta$.
		\textit{Top:} Maximum overdensity of the azimuthal average ($\sigma_{AA,max}$) and maximum 
		overdensity of the trailing enhancement ($\sigma_{TB,max}$) of the resonant ring of Earth are given
		for the nominal case (all 8 planets present) and for Earth as a single planet.
		The impact of the other planets of the Solar System increases with $\beta$, reaching a
		reduction of overdensities of more than 50\% at $\beta=0.01$.
		\textit{Bottom:} Ratio of maximum density of the trailing enhancement to maximum density of the 
		azimuthal average. 
		A decline towards lower $\beta$ in the nominal case indicates a reduction of the ring's 
		leading--trailing asymmetry.}
		\label{fig:overdensity_APOE}
	\end{figure}

	Figure~\ref{fig:overdensity_APOE}~(\textit{top}) shows the overdensity metrics for Earth's ring for the 
	case of all	planets present as well as for Earth as the single planet, with respect to $\beta$. 
	The increasing deviation again illustrates that the consequences of resonance impairment 
	through gravitational interference become more severe with decreasing migration rate.
	Furthermore we see a convergence of $\sigma_{TB,max}$ and $\sigma_{AA,max}$ towards lower $\beta$
	in the nominal case, which is apparent in Fig.~\ref{fig:overdensity_APOE}~(\textit{bottom})
	as a decline of the ratio $\left(1+\sigma_{TB,max}\right)/\left(1+\sigma_{AA,max}\right)$ 
	for $\beta\leq0.02$.
	This implies a weakening of the ring's leading--trailing asymmetry following the introduction
	of other planets, in addition to the lowered ring contrast.
	
	Lastly we note that Mercury did not show any meaningful trapping of migrating dust in the examined
	$\beta$-range, even under the absence of its neighbour Venus.
	To rule out a possible inhibition of resonances due to integration step length, we lowered the time 
	step to 1/\nth{100} of a year (1/\nth{24} of a Mercury year) specifically when looking into Mercury.
	Nevertheless, this did not lead to discernible trapping in its mean-motion resonances.
	
	
	\FloatBarrier
	\subsection{Non-resonant density features} \label{Ch33}
\begin{figure}[]
		\centering
		\includegraphics[height=75mm,trim={3mm 2mm 4mm 11mm},clip]{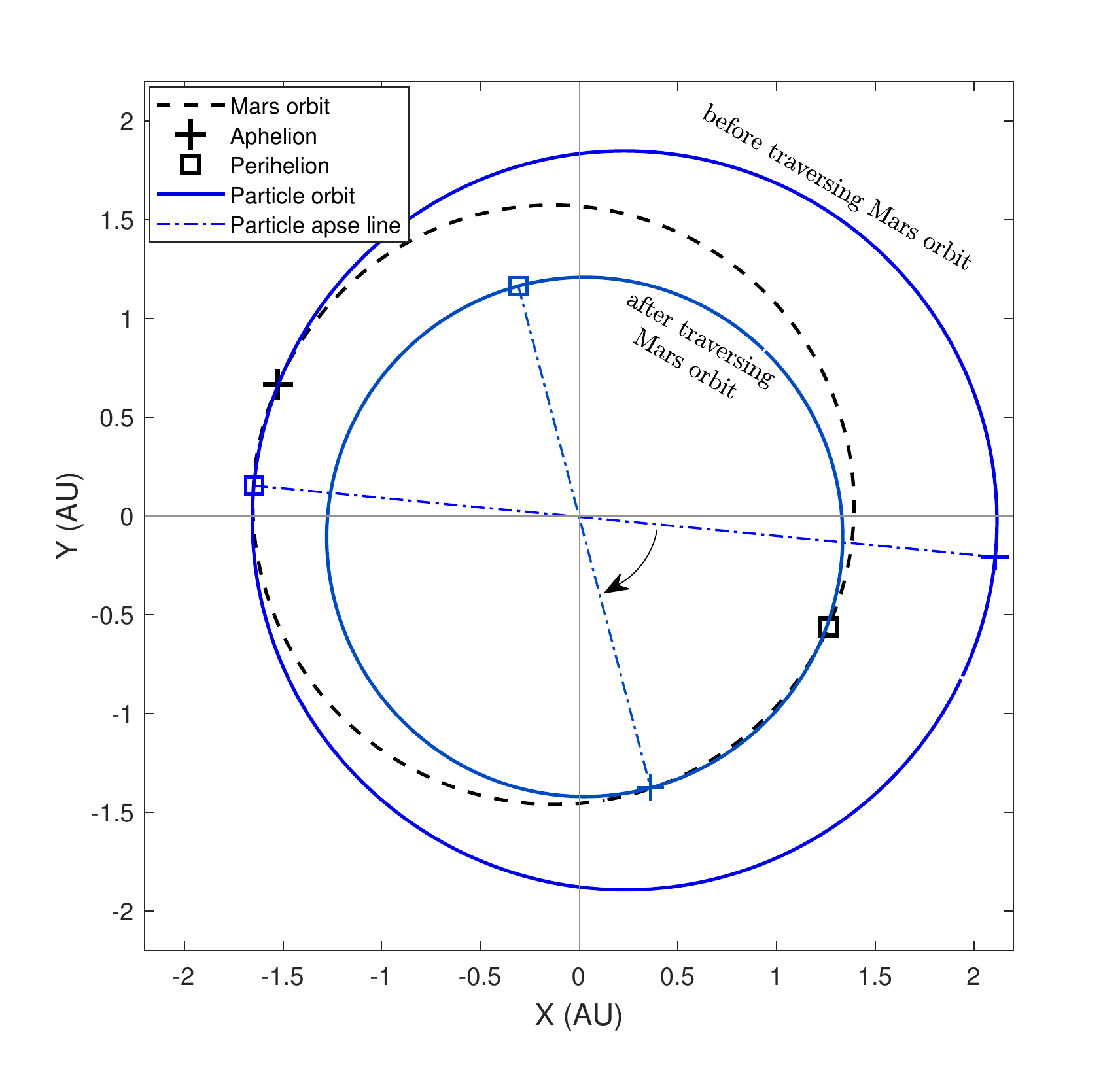}
      	\caption{Apsidal precession of migrating particles as the cause for the double-crescent density pattern 
      	generated by Mars (no other planets present, $\beta=0.005$).
      	The figure shows the orbit of an exemplary particle before and after traversing the orbit of Mars.
      	As the particle's perihelion drops inside the orbit of Mars under PR drag, it starts to experience a 
      	negative apsidal drift that is maintained until it revolves entirely within the orbit of Mars.
      	The apsidal drift is causing one side of the trajectory to fall off faster than the other, 
      	apparent in the varying gap width between the two orbits.
      	All particles experiencing a drift with preferred final apse line orientation with respect
      	to that of Mars, thus generate a mirrored depletion--enhancement pattern.
      	}
        \label{fig:XY_PERI_DRIFT_exemplary}
	\end{figure}

\begin{figure}[]
		\centering
		\includegraphics[height=78mm,trim={3.5mm 0 4mm 5mm},clip]{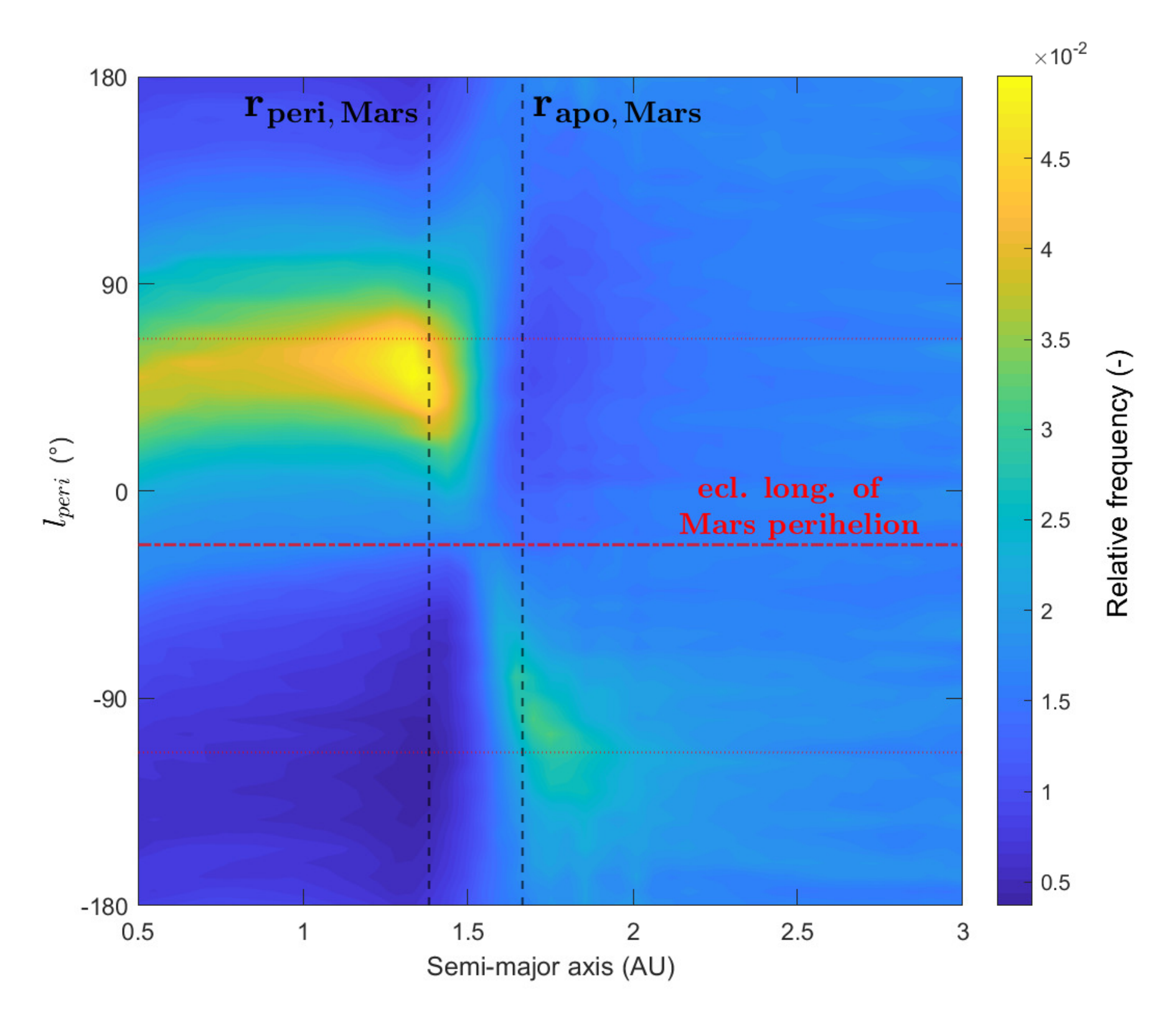}
      	\caption{Relative frequency of the ecliptic longitude of particle perihelia \boldsymbol{$(l_{peri})$} recorded at semi-major axis bins 
      	of 0.05~\si{\astronomicalunit} (no planets but Mars present, $\beta=0.005$). 
      	The relative frequency is given for 60 bins in \boldsymbol{$l_{peri}$}, thus a frequency of
      	$1.67\times10^{-2}$ in each bin implies a uniform distribution. 
      	The ecl. long. of Mars's perihelion \boldsymbol{$(l_{peri,\, Mars})$} is indicated by the red line, 
      	as are $l_{peri,\, Mars} \pm90$\si{\degree}.
      	Mars's perihelion and aphelion distances are indicated by vertical lines. 
      	Particle perihelia are equally distributed when in a semi-major axis range from
      	3~\si{\astronomicalunit}~to~2.5~\si{\astronomicalunit}. 
      	As semi-major axis decreases, perihelia first gravitate towards 
      	$l_{peri,\, Mars} +90$\si{\degree}, followed by a stronger attraction towards 
      	roughly $l_{peri,\, Mars} -80$\si{\degree}
      	once semi-major axis passes Mars's aphelion distance.
      	Migrating further, the particles' apse line alignment relaxes again.
      	}
        \label{fig:PERI_DRIFT_population}
	\end{figure}

	In the previous section we noticed crescent-shaped density features generated by Mars, which could not
	be connected to particles trapped in its mean-motion resonances.
	We therefore looked into the orbital evolution of these particles more closely, 
	as they migrated the planet's orbital region.
	Figure~\ref{fig:XY_PERI_DRIFT_exemplary} illustrates, how the particles' apsidal precession resulting from the secular 
	interaction with Mars generates this density pattern: 
	It displays the orbit of an exemplary particle at two different points in 
	its evolution -- before and after traversing Mars's orbit.  
	Neither did this particle get trapped in a mean-motion resonance, nor did it suffer a close 
	encounter with Mars.
	As the particle orbit drops inside the orbit of Mars for the first time, it starts to experience a steady negative 
	apsidal drift that is maintained until it revolves entirely within the orbit of Mars.
	Due to the drift, the particle's trajectory falls off faster on the side of its orbit that the perihelion is moving into and
    more slowly on the opposite side, apparent in the varying gap width between the two orbits.
    The particle's migration is effectively decelerated in one region and accelerated in the other.
	Thus, the generation of the mirrored crescent-shaped pattern is consequential, if the apse lines
	of all particles tend to align in a preferred direction while traversing Mars orbit.
	This preferred alignment of apse lines is shown in Fig.~\ref{fig:PERI_DRIFT_population}.
	Here, the distribution of ecliptic longitude of particle perihelia projected onto the ecliptic 
	$(l_{peri})$ is given over their semi-major axis.
	The ecl. long. of Mars's perihelion $(l_{peri,\, Mars})$, which is constant throughout
    the simulation due to the absence of other planets, is indicated by the red line.
    We note, that $l_{peri}$ starts off distributed uniformly in a region 
    beyond 2.5~\si{\astronomicalunit}. 
    As semi-major axis decreases, however, particle perihelia tend to gravitate towards an ecl. long. 
    $90\si{\degree}$ behind $l_{peri,\, Mars}$.
    Then, as their semi-major axis passes Mars's aphelion distance, the apsidal drift is reversed as well as
    amplified. 
    Perihelia now drift towards an ecl. long. around 80\si{\degree} ahead of $l_{peri,\, Mars}$, 
    reaching a relative frequency of a factor of 3 over uniform distribution as the semi-major axis passes 
    Mars's perihelion distance.
    Migrating further, the particle's apse line alignment relaxes again.
	This confirms a directed apsidal precession induced by Mars as the cause for the crescent-shaped features.
	\begin{figure*}[]
		\centering
		\begin{tabular}{ccc}
		\hspace{-2mm}\includegraphics[width=60mm,trim={3mm 0 4mm 0},clip]{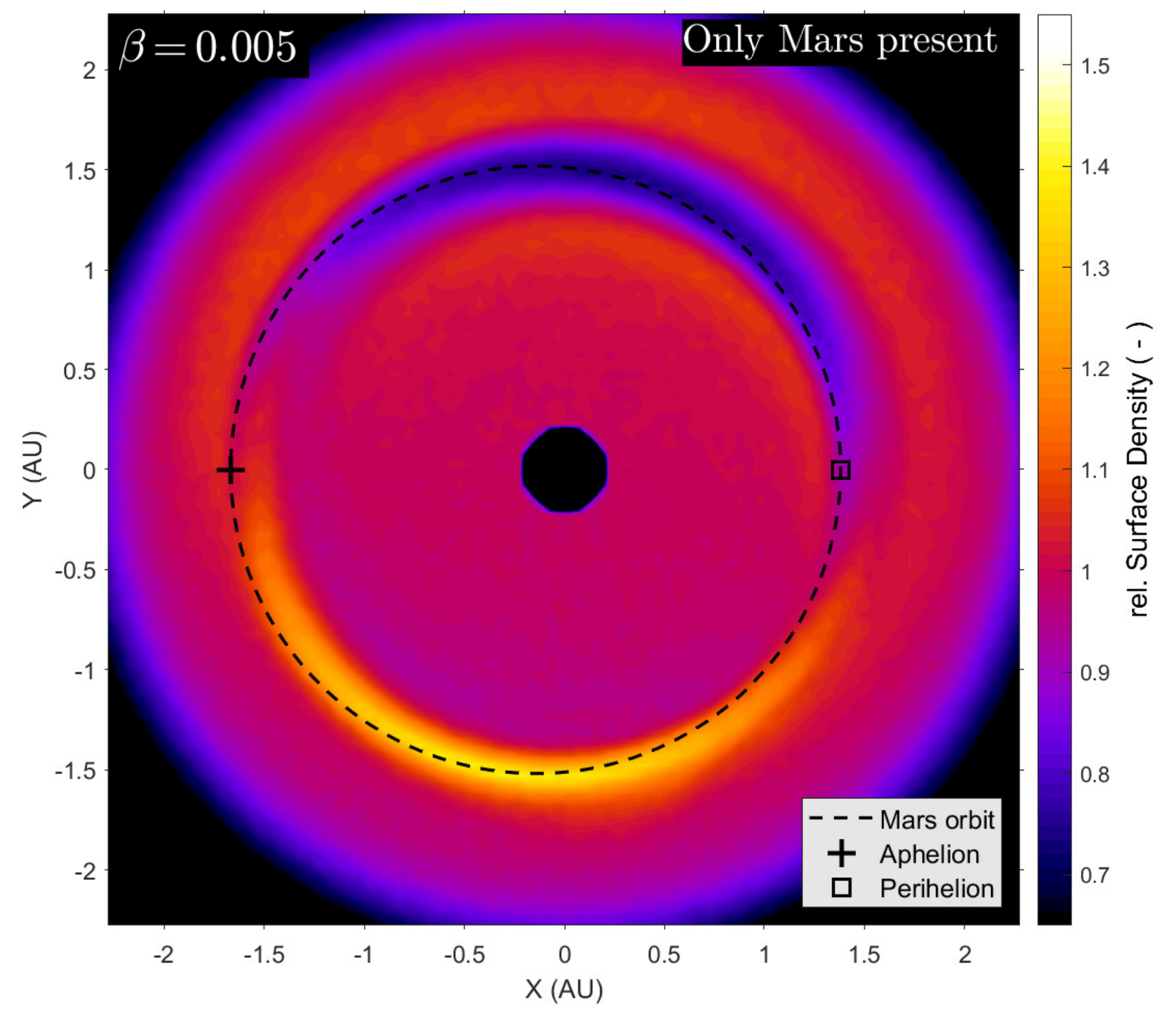} &
		\hspace{-2mm}\includegraphics[width=60mm,trim={3mm 0 4mm 0},clip]{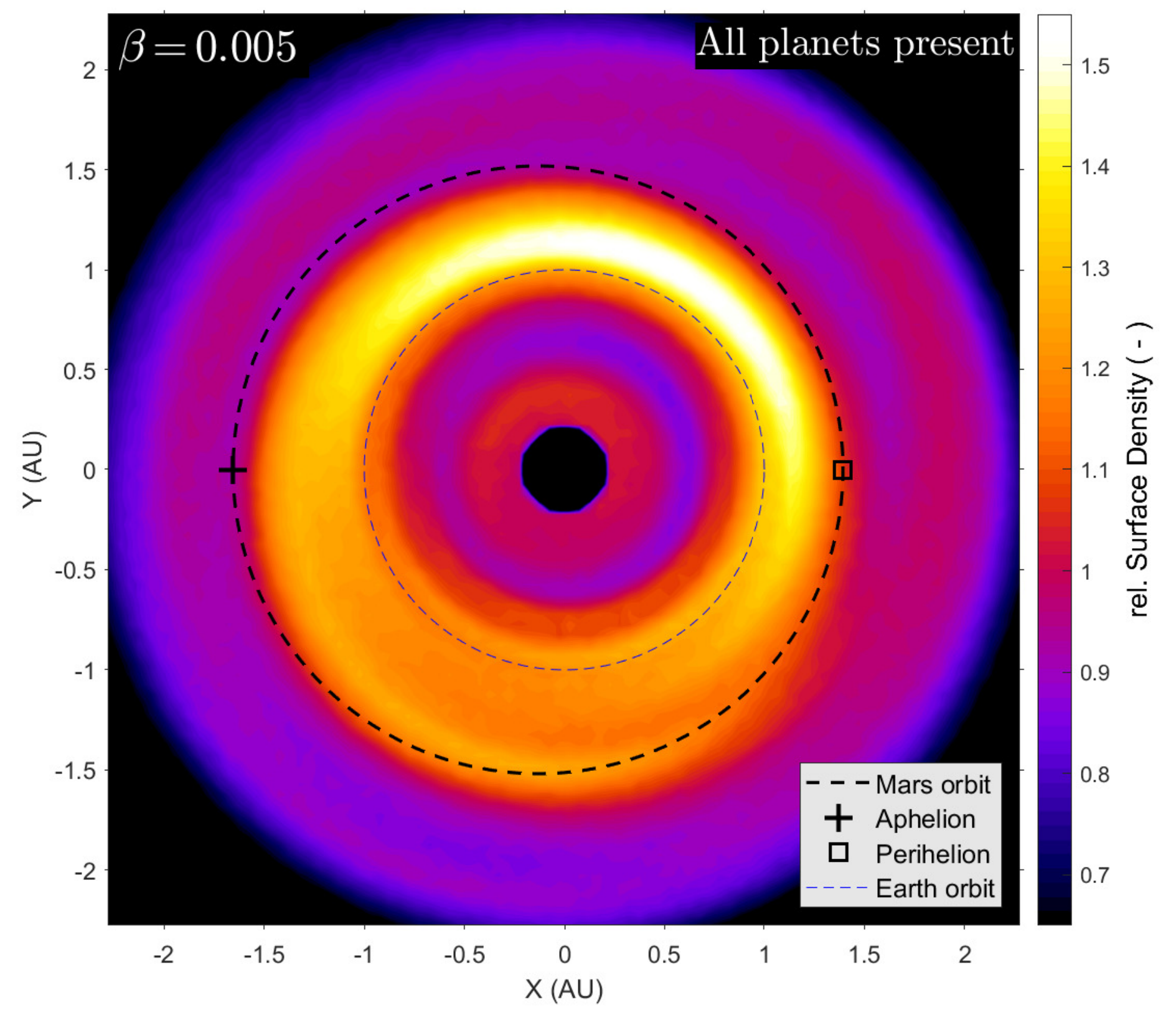} &
		\hspace{-2mm}\includegraphics[width=60mm,trim={3mm 0 4mm 0},clip]{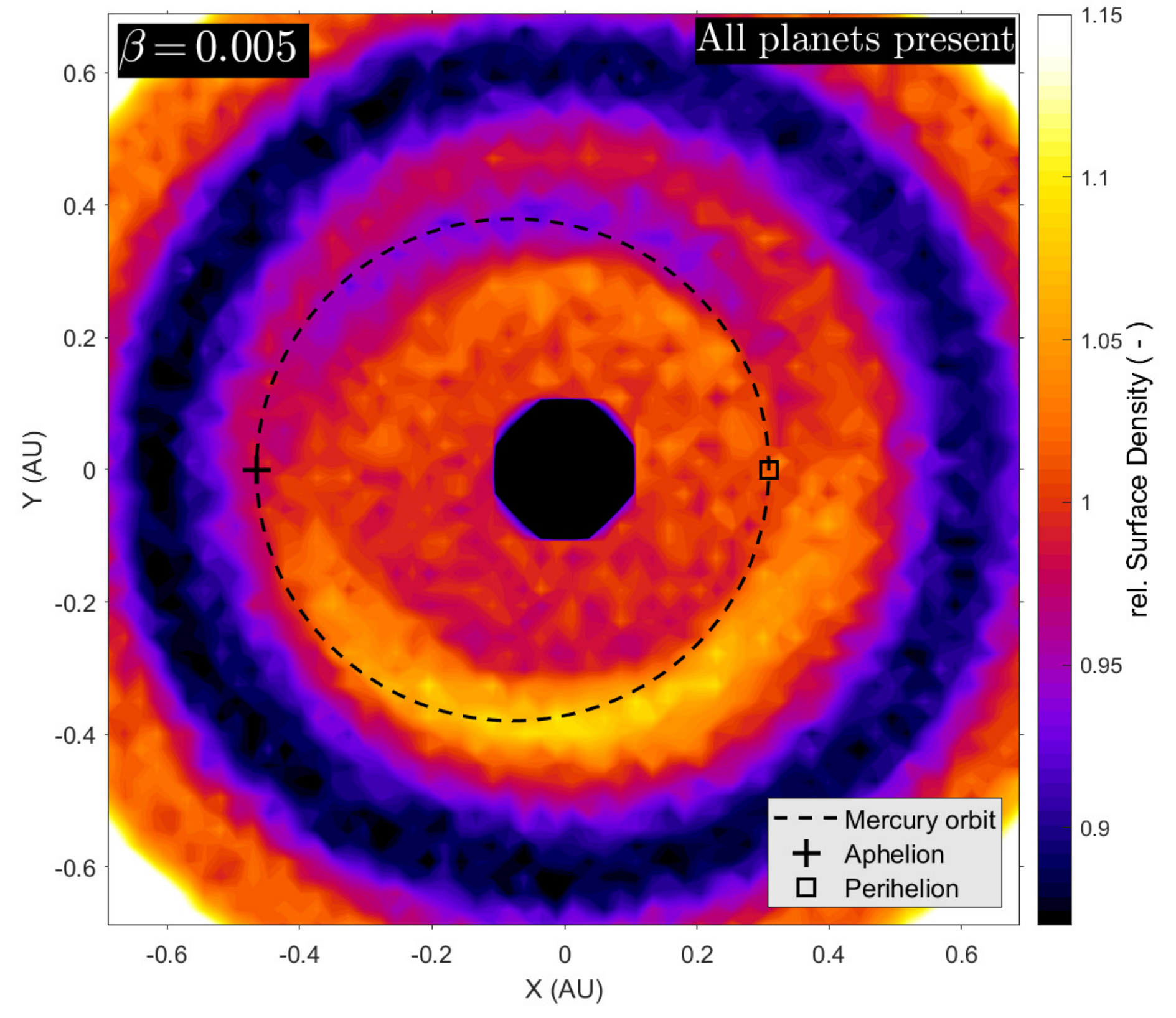} \\
		\hspace{-4mm}(a) & \hspace{-4mm}(b) & \hspace{-4mm}(c)
		\end{tabular}
      	\caption{Density distribution emerging when recording particle positions in a frame co-rotating
      	with the planet's line of apsides ($\beta=0.005\hspace{.2mm}$). 
      	(a): Apsidal alignment pattern that forms under the presence of Mars alone.
      	Also apparent is a disparity of the density inside the inner edges of the crescents.
      	(b): Density distribution under the presence of all planets, recorded in a frame co-rotating with Mars's apse line.
      	The colour scale is the same as in (a).
      	The presence of the apsidal alignment pattern generated by Mars is evident from the modulation of 
      	the contrast of Earth's resonant ring with respect to Mars's apse line.
      	The orbit of Mars was plotted with the mean eccentricity it had over the duration of the simulation. 
      	The plotted orbit of Earth is a circular 1~\si{\astronomicalunit} orbit.
      	(c):  Density distribution under the presence of all planets, recorded in a frame co-rotating with Mercury's apse line.
      	Mercury likewise generates an apsidal alignment pattern, forming just inside the depletion ring at the internal 
      	Venus MMRs.
      	The orbit of Mercury was plotted with the mean eccentricity it had over the duration of the simulation. 
      	}
        \label{fig:XY_ApseAlign}
	\end{figure*}
	
	With the indication that the apsidal alignment pattern is fixed with respect to the apse line of the perturbing planet,
	we can test whether this effect is occurring only under the presence of Mars alone or also under the presence
	of its planet neighbours.
	We revise the stacking method such that recording of particle positions takes place in a frame fixed 
	to the apse line of Mars, irrespective of the planetary phase.
	The resulting density distributions for $\beta=0.005$ are given in Fig.~\ref{fig:XY_ApseAlign}.
	For reference, Fig.~\ref{fig:XY_ApseAlign}~(a) again gives the distribution in a Martian single planet system, 
	but using 	the revised stacking method.
	Here, besides the distinct crescents, we also note the effect of the weaker yet reversed apsidal alignment
	occurring in particles revolving on the outside and inside of the orbital region inhabited by Mars.
	This weaker alignment produces an opposing density variation around the edges of the crescents, for instance 
	next to the inside of the depleted crescent an overdensity of 8\% is reached, while the inner edge of the 
	aggregating crescent exhibits a density reduction of 7\%.
	In Fig.~\ref{fig:XY_ApseAlign}~(b), all planets of the Solar System are reintroduced.
	While the apsidal alignment pattern generated by Mars is overlapped by the resonant ring of Earth, 
	it's presence is evident from the modulation of the contrast and width of Earth's ring with respect to Mars's 
	apse line.
	The aggregating crescent extends the width of Earth's ring, whereas the cavity at its inner edge
	undercuts the ring's contrast; the ring appears broad and diffuse.
	On the contrary, the depleted crescent curtails the extend of Earth's ring while increasing its density
	at the crescent's inner edge; the ring appears condensed.
	We can derive that the apsidal alignment induced by Mars not only occurs under the presence of its
	planet neighbours, but also that the density features it entails considerably alter the appearance
	of the Earth's resonant ring.
	Going to smaller particle sizes, the effect on Earth's ring diminishes in accordance with the noted
	weakening of the apsidal alignment pattern in the Martian single planet system.

	Mercury, likewise exhibiting a notable orbital eccentricity, is also able to generate the apsidal alignment 
	pattern at $\beta=0.005$, as is shown in Fig.~\ref{fig:XY_ApseAlign}~(c). 
	The appearance of these features is comparable to those of the Martian single planet system, 
	even though they only reach a fraction of their contrast (about 1/\nth{4}).
	It is noteworthy that this apsidal alignment pattern, albeit weak, is present notwithstanding Mercury's 
	inability to trap particles in its mean-motion resonances.
	
  \section{Discussion} \label{Ch4}
	The modelling results reported here show that the formation of resonant dust features
	associated with a planet is highly sensitive to the presence of other planets.
	While trapped in a resonance, particles experience a periodic stabilising pull 
	from the body they are resonating with,
	that counteracts their loss of energy due to PR drag.
	Contrary to that, neighbouring planets cause a rather chaotically occurring perturbation.
	In the case of external Venus MMRs for example 
	-- the most dominant being located just between the orbits of Venus and Earth -- 
	it is easily conceivable that perturbations of Earth become comparable in magnitude to 
	the stabilising influence of Venus, given their similar mass.
	These perturbances result in a lowered efficacy of resonances that is amplified 
	by the fact that resonating particles have their eccentricity gradually raised, increasing
	their likelihood of suffering a resonance ending interaction with a neighbouring planet.
	This interaction must not be misunderstood as a close encounter. 
	As our simulations showed, Jupiter is quite effective in weakening the resonant ring of 
	Earth, even though the particles constituting it never move beyond a heliocentric 
	distance of 1.8~\si{\astronomicalunit}.
	
	The impairment of resonances seen here becomes more effective with decreasing $\beta$.
	To explain this, we have to consider that particle concentration in resonances is 
	directly dependent on the relation of resonance dwell time to PR lifetime.
	A lower particle $\beta$ increases dwell times substantially, due to 
	the slower energy loss \citep{1989Natur.337..629J}.
	The probability of occurrence of a resonance ending gravitational perturbation, however, 
	must be considered independent of $\beta$ but rather increasing with dwell time.
	The impairment of resonant features is therefore more pronounced for slowly migrating particles,
	which require accordingly longer resonant dwell times to produce meaningful contrast.
	This may be one explanation why the impact of neighbouring planets becomes more crucial 
	with decreasing $\beta$.
	
	Another resonance impairing effect has to be considered relevant to the MMRs of Venus in particular.
	After traversing Earth's resonant region, particles having undergone lengthier trapping in Earth's external MMRs
	have their eccentricities substantially pumped. 
	Notwithstanding a partial re-circularisation during the subsequent migration to the resonant region of Venus,
	the raised eccentricities lower trapping probability
	and thus contribute to the diminished occupancy of Venus MMRs.
	To what extent the impairment of Venus resonances must be attributed to the limiting of dwell times 
	in resonances or to an overall reduced trapping probability cannot be determined 
	conclusively in this study and should be subject of more thorough investigation.
	
	Interestingly, the presence of other planets is not necessarily limited 
	to an impairment of resonances. 
	As detailed in appendix~\ref{ChFR}, there are certain MMRs of Venus whose efficacy 
	is maintained or even increased despite their close proximity to Earth.
	Although their contribution to the overall density distribution remains minuscule, 
	they hint at the intricate nature of resonant phenomena that may arise in multi-planetary systems.

	There are some limitations implied in our model, that one has to keep in mind when  
	trying to infer the actual density distribution of the zodiacal cloud from these results.
	We consider these simplifications acceptable since this work is mainly concerned with the general 
	implications of planet neighbours, rather than making precise density or brightness predictions. 
	For one, we focus on the dynamically cold cloud component, specifically that of asteroidal origin.
	In doing so, we note that the distribution of our initial test particles is a rather poor 
	reproduction of the distribution of main belt asteroids, disregarding the belt's inner grouping structure 
	that has been shaped by the gravitational interaction with Jupiter and Saturn, 
	as well as by the fragmentation of larger bodies.
	This grouping of orbital elements in the source population may introduce effects which are 
	unaccounted for in the present model.
	Further, to account for the contribution of dynamically hot dust, one has to reduce the overdensities
	simulated here according to the fraction of asteroidal dust in the zodiacal cloud.
	The share of asteroidal dust is presumably subjected to vary depending for instance on 
	heliocentric distance or distance to the ecliptic plane. 
	The overall asteroidal contribution, however, has been estimated in various studies
	concerning the composition of the zodiacal cloud, typically falling in the range
	of 1/\nth{5} to~/\nth{20} (22\% \citet{2013MNRAS.429.2894R};
	20\% \citet{SojaR_2019_IMEM2}; $\leq$10\% \citet{2017AJ....153..232U}; 
	$\leq$10\% \citet{Nesvorn__2010}; 6\% \citet{2015ApJ...813...87Y}).
	Moreover, the fraction of dynamically cold dust susceptible to MMRs may increase
	at lower heliocentric distances, as PR drag works to increasingly circularise dust particle orbits 
	as they approach the Sun.
	
	In addition, our model does not consider the collisional lifetimes of particles
	imposed by grain-grain collisions.
	Collisional lifetimes derived by \citet{1985Icar...62..244G} indicate 
	that the particle size range studied here ($D\lesssim$\hspace{.5mm}100~\si{\micro\meter}) 
	should not be particularly collisionally dominated.
	These lifetimes, however, have been the subject of much debate due to an incompatibility with the 
	distribution of orbital elements of dust grains in the millimetre size range observed by ground-based meteor
	radars \citep{Nesvorn__2010, Nesvorn__2011, Pokorn__2014, SojaR_2019_IMEM2}.
	The derived particle dynamics require considerable evolution under PR drag and thus collisional lifetimes
	up to orders of magnitude higher than those derived by \citet{1985Icar...62..244G}.
	On the other hand, \citet{Stark_2009} found that collision rates inside resonant
	rings can increase to a multiple of the collision rates outside the ring, 
	due to the higher grain density	as well as pumped eccentricities.
	They report the most notable resulting morphological change of the resonant ring 
	to be a decreased overdensity and sharpness of the inner ring edge, in addition to azimuthal smearing 
	and an overall	reduced contrast.
	To some extent this can be considered akin to the reduction of feature sharpness
	that we have simulated here. 
	Since the edge features are a result of long-lived particles approaching the equilibrium solutions
	of their resonances, a limiting of resonant dwell time whether through grain collisions
	or gravitational disruption should manifest itself in much the same way.
	At the lower end of our $\beta$-range, the consideration of collisional lifetimes might therefore
	even further the diffusion of the ring edges that we have found to occur due to the 
	presence of other planets.
		
	The results of this study are especially relevant to the search of a resonant ring 
	associated with Venus. 
	From imagery data of \textit{STEREO}, \citet{2017Icar..288..172J} derive an increase of 
	density occurring at two radial distances located just inside and just outside the 
	orbit of Venus ($r_{0}=0.71524$~\si{\astronomicalunit} and $r_{1}=0.73917$~\si{\astronomicalunit}), 
	adding up to an azimuthally varying total overdensity of typically 4--8\% reached at the second
	step relative to the density inside $r_0$ (with a maximum overdensity of $8.5\%$).
	Although our model produces an increase of density that occurs gradually at a heliocentric
	distance of 0.70~\si{\astronomicalunit} to 0.76~\si{\astronomicalunit} of around the same
	magnitude for particle sizes 10--50~\si{\micro\meter}, this overdensity has to be diluted by the
	featureless contribution of the dynamically hot cloud.
	Assuming a cometary component of 90\% and notwithstanding an even higher dilution when 
	factoring in the contribution of dust outside this resonance prone particle size range, 
	this would yield an actual density increase of only $\lesssim\hspace{-1.1mm}1\%$ -- 
	much less than what is required to explain the \textit{STEREO} observations.
	\citet{2017Icar..288..172J} further state that the azimuthal structure of the observed ring
	might be indicative of a major contribution of dust in the 2:3 resonance of Venus.
	However we do not find this specific resonance to significantly contribute to the ring of
	Venus, given its proximity to Earth and the resulting impairment (the semi-major axis of the 
	2:3V resonance is located at around 0.94~\si{\astronomicalunit} depending on particle $\beta$).
	
	Similarly, \citet{2007A&A...472..335L} postulate an overdensity of 10\% for a ring just outside 
	the orbit of Venus to account for the increase in brightness measured by the zodiacal light
	photometer aboard \textit{Helios}. 
	Again this can not be matched by our model after considering the contribution of the smooth
	background cloud.
	Without deviating from a dominant featureless background contribution,
	we therefore find migrating dust trapped in Venus resonances of the size range studied here
	inadequate of producing the observed ring, confirming results of previous modellers
	\citep{2012LPICo1667.6201J,2019ApJ...873L..16P}.
	Our results show that the weakness of the thus obtained ring of Venus
	has to be attributed to large extent to the gravitational interference of Earth, 
	suppressing external Venus MMRs.
	\citet{2019ApJ...873L..16P} hypothesise that a so far undiscovered population of dust
	producing Venus co-orbital asteroids might release particles directly into the 1:1 resonance
	of Venus, thus bypassing a low trapping probability. 
	Their analysis shows that a dust ring produced this way is potentially able to match both
	the observations of \textit{Helios} and \textit{STEREO}.
	Their hypothesis might offer a more plausible explanation than the trapping of
	migrating dust in external resonances alone. 
	The latter, as we conclude, would likely require a featureless background contribution at Venus of
	less than 50\%.
	This is contrary to the typically assumed 90\% \citep{2019ApJ...873L..16P}, which 
	stem from the cometary dominated compositional constraints on the zodiacal cloud 
	put forward by various studies (\citet{Nesvorn__2010}; \citet{2017AJ....153..232U}; \citet{SojaR_2019_IMEM2}).
	\citet{Nesvorn__2010}, however, also show that the orbits of small meteoroids 
	($D\lesssim$\hspace{.5mm}100~\si{\micro\meter}) released by Jupiter-family comets~(JFCs)
	become significantly circularised by PR drag before reaching $a\!\approx\!1$~\si{\astronomicalunit}, albeit
	the broad distribution of inclination remaining largely unchanged.
	Conceivably, this may make a fraction of them susceptible to resonances of Earth and Venus thus increasing
	the contribution of the feature bearing disc, without breaking compositional constraints.
	Nevertheless \citet{2019ApJ...873L..16P} showed, that the signal produced by resonating JFC dust 
	near the orbit of Venus is still insufficient to meaningfully account for the observed density variation.
	It should be noted, however, that a featureless background of 90\% also reduces the contrast of the 
	Earth's resonant ring modelled here to a few percentage density-wise (in our model the most efficient 
	ring forming particle size of $\beta=0.01$ generated an Earth ring that reached an azimuthally 
	averaged surface overdensity of around 40\%).
	Even though observations from locations near Earth cannot conclusively determine structure and contrast
	of the Earth's resonant ring, \citet{1998ApJ...508...44K} have fitted a parametrised geometric ring model to the 
	zodiacal light foreground brightness derived from \textit{COBE/DIRBE} data.
	While the fitted model finds a density of 16\% over the smooth background at ring centre,  
	a comparison of contrast is problematic as their geometric parameters (radius of peak density 
	at~1.03~\si{\astronomicalunit} (fitted), radial FWHM of 0.06~\si{\astronomicalunit} (fixed))
	arguably lead to a much more confined ring, than what was modelled in this work. 
	
	In this context, these results are also relevant to studies attempting to derive the 
	asteroidal dust component of the zodiacal cloud from relating the strength of the leading--trailing 
	asymmetry in all-sky infrared surveys to the leading--trailing asymmetry of the resonant ring 
	in single planet simulations \citep[e.g.][]{1994Natur.369..719D,2017AJ....153..232U}.
	The presence of all planets of the Solar System causes a reduction of ring contrast 
	and leading--trailing asymmetry, which may lead to an underestimation of the dynamically cold
	contribution	when comparing observations to simulations considering only Earth.
	
	The presence of a resonant ring of Mars has been theorised, yet a search for its trailing
	enhancement was unable to prove its existence \citep{KUCHNER200044}.
	The infrared study using \textit{COBE/DIRBE} data conducted by \citet{KUCHNER200044} determined
	for the trailing enhancement an upper limit fractional overdensity of 18\% of that of Earth's.
	They state that a detection would have been successful if the surface area of dust in the wake 
	scaled simply as the mass of the planet times the Poynting–Robertson time scale.
	Adding to that, \citet{2011MNRAS.413..554M} have predicted Mars to be capable of trapping
	$\lesssim25\%$ the dust trapped by Earth, using a semi-analytic approach to analyse resonance 
	capture in single planet systems. 
	However, our results suggest that the formation of a meaningful resonant enhancement 
	associated with Mars is severely impaired by the perturbations of Jupiter. 
	This is consistent with the observational findings by \citet{KUCHNER200044}.
	Nevertheless we find that even though Mars is unable to produce distinct resonant features,
	its inherent orbital eccentricity induces an alignment of apse lines in particles traversing 
	its orbital region.
	This in turn, generates a crescent-shaped density pattern that is fixed 
	with respect to Mars's line of apsides.
	These apsidal alignment features showed to be less affected by the presence of other
	planets and formed even at particle sizes not susceptible to Mars's MMRs, 
	although a more relevant impact was reached only at the larger end of our 
	particle size range.
	The quasi-stationarity of these Martian dust features would have further frustrated searches 
	for structures revolving with Mars such as that of \citet{KUCHNER200044}.
	Their signal might nonetheless be discernible in the infrared foreground brightness data obtained by \textit{COBE/DIRBE}.
	Merging into Earth's resonant ring, Mars's apsidal alignment features would manifest in a modulation of the Earth's 
	ring in contrast and radial extent, as our model suggests. 
	This might justify re-examination of \textit{COBE/DIRBE} data focussing on the variability of 
	Earth's ring with respect to Mars's line of apsides.
	
	Mercury neither showed meaningful trapping of migrating dust in the modelled $\beta$-range
	nor did it seem to affect the efficacy of resonances of its neighbour Venus.
	This must be attributed to its small mass, as well as proximity
	to the Sun ($a_{Mercury} = 0.54\,\times\,a_{Venus}$).
	At larger particle sizes, however, it produced a weak apsidal alignment pattern similar 
	in appearance to that of Mars.
	We included Mercury in this study due to the recent findings by \citet{Stenborg_2018}, that 
	indicate the existence of a circumsolar dust ring near Mercury's orbit.
	Using optical data obtained by the \textit{Parker Solar Probe}, \citet{Stenborg_2018} derive
	a maximum excess of dust density of 3--5\%.
	Whether the double-crescent feature seen here can account for the observed signal is questionable
	and would require a much more thorough investigation, which is outside of the scope of this work.
	However based on these results, we do not consider the existence of an MMR-driven dust ring at Mercury likely.
	
	The effects described here must also be considered when trying to connect observations of 
	feature-bearing dust discs around other stars to the existence of exoplanets. 
	As described by \citet{Liou99} in the case of Edgeworth-Kuiper belt dust, a 
	distinct resonant ring can only form at the outermost planet of a contiguous 
	group of planets (such as those of the outer Solar System).
	Our study further showed, that the gravitational interference of even relatively small or 
	distant outer neighbouring planets can severely reduce resonant ring contrast and may thus 
	lead to an underestimation of planet mass derived from dust feature strength alone, 
	if not preventing the ring signal's observability entirely.
	The apsidal alignment of dust particles migrating the orbital region of an eccentric planet
	might also generate observable density features in exo-zodiacal discs, and thus should
	receive more attention in future modelling works.
	
  \section{Summary} \label{Ch5}
In this paper we have presented an approach to model resonant dust features 
in the inner Solar System that arise under the presence of all eight planets.
We then looked specifically into how these dust structures depend
on the multi-planet configuration.
In doing so, we found that planetary neighbours can have a significant impact on the contrast
as well as the shape of emerging circumsolar resonant rings.
The gravitational interference of other planets proves to be effective in lowering
the efficacy of trapping, external mean-motion resonances and displacing, internal mean-motion
resonances alike. 

For the formation of resonant rings most essential external MMRs are impaired by the presence
of an outer planet neighbour, leading to an overall reduced ring contrast, in addition to a 
dilution of the ring edges. 
Especially noteworthy is the suppression of the inner ring edge by the presence of an outer planet.
Forming as a result of a culmination of dust particle perihelia due to long lasting entrapment, 
the distinct inner edge is squashed as particles are kicked out of resonance by gravitational 
perturbances long before approaching the limiting eccentricity of the resonance.
While resonating, eccentricity pumping leads the particles closer and closer to the 
orbits of neighbouring planets, making a resonance ending gravitational interference
more and more likely.
Consequently, we can consider closely neighbouring planets to introduce a natural limit
on the resonant dwell times of external MMRs.
Furthermore, the impact of neighbouring planets becomes more decisive with increasing 
particles size.
Trapping of smaller, quickly migrating dust particles is much more limited by PR drag rather
than gravitational interference. 

In the inner Solar System, these gravitational interactions have consequences for 
the ability to form resonant features of Venus and Earth, as well as Mars. 
In the case of Venus, the presence of Earth causes a severe impairment of Venus'
external MMRs, which leads to the weakening of its resonant ring to only a fraction
of its potential contrast (i.e. in the absence of any neighbour planets).
Although the qualitative appearance of the modelled Venus ring is to some degree
consistent with the findings of \textit{Helios} \citep{2007A&A...472..335L} and \textit{STEREO}
\citep{2017Icar..288..172J} -- a rather sudden change of density
at the orbit of Venus -- its intensity is too low to adequately match the observations.
Our model suggest that without considering other theories of origin for the Venusian ring 
\citep[e.g.][]{2019ApJ...873L..16P} a contribution of the smooth background cloud at the 
orbit of Venus of less than 50\% would be required. 

The intensity of the Earth's resonant ring is likewise damped by the presence of Mars and Jupiter,
albeit not as drastically as that of Venus.
The influence of its outer neighbours is more notable at the larger modelled particle sizes, with
a reduction of overdensity in the order of 50\% at a particle diameter of 50~\si{\micro\meter}.
Mars alone would be capable of forming faint resonant features at larger particle sizes.
Our results indicate, however, that the presence of Jupiter prevents them from reaching a 
meaningful contrast.
Nevertheless we found Mars to produce crescent-shaped density features with considerable
contrast, driven by a directed apsidal precession of particles migrating its orbital region.
Mercury, likewise exhibiting substantial orbital eccentricity, generates a weaker
version of this apsidal alignment pattern.
A meaningful interaction of Mercury with the migrating dust cloud through mean-motion resonances,
however, could not be shown.

In summary, this work has shown that the impact of planet neighbours can be decisive in
the formation of resonant rings. 
Moreover, even the presence of small outer neighbours (such as Mars), that are not capable of 
producing their own distinct resonant ring, might very well be able to diminish the 
intensity of a resonant ring of an inner planet.
These findings might be particularly relevant to surveys of feature-bearing, and
thus potentially exoplanet harbouring debris discs around distant stars.

\begin{acknowledgements}
MS gratefully acknowledges support from the Japan Society for the Promotion of Science 
for part of this work, which was initiated under the JSPS Summer Program 2017.
We thank Douglas P. Hamilton and Eberhard Gr\"un for insightful discussions, as well as the reviewer 
Petr Pokorný for his helpful comments and suggestions.

\end{acknowledgements}

	\bibliographystyle{aa}  
	\bibliography{references} 
	
	\begin{appendix}

\onecolumn
\section{The Nominal Solar System}

	\begin{figure*}[h]
	\centering
	\vspace{-50mm}
	\begin{tabular}{ccc}
	
	\includegraphics[width=59mm,trim={3mm 2mm 3mm 1mm},clip]{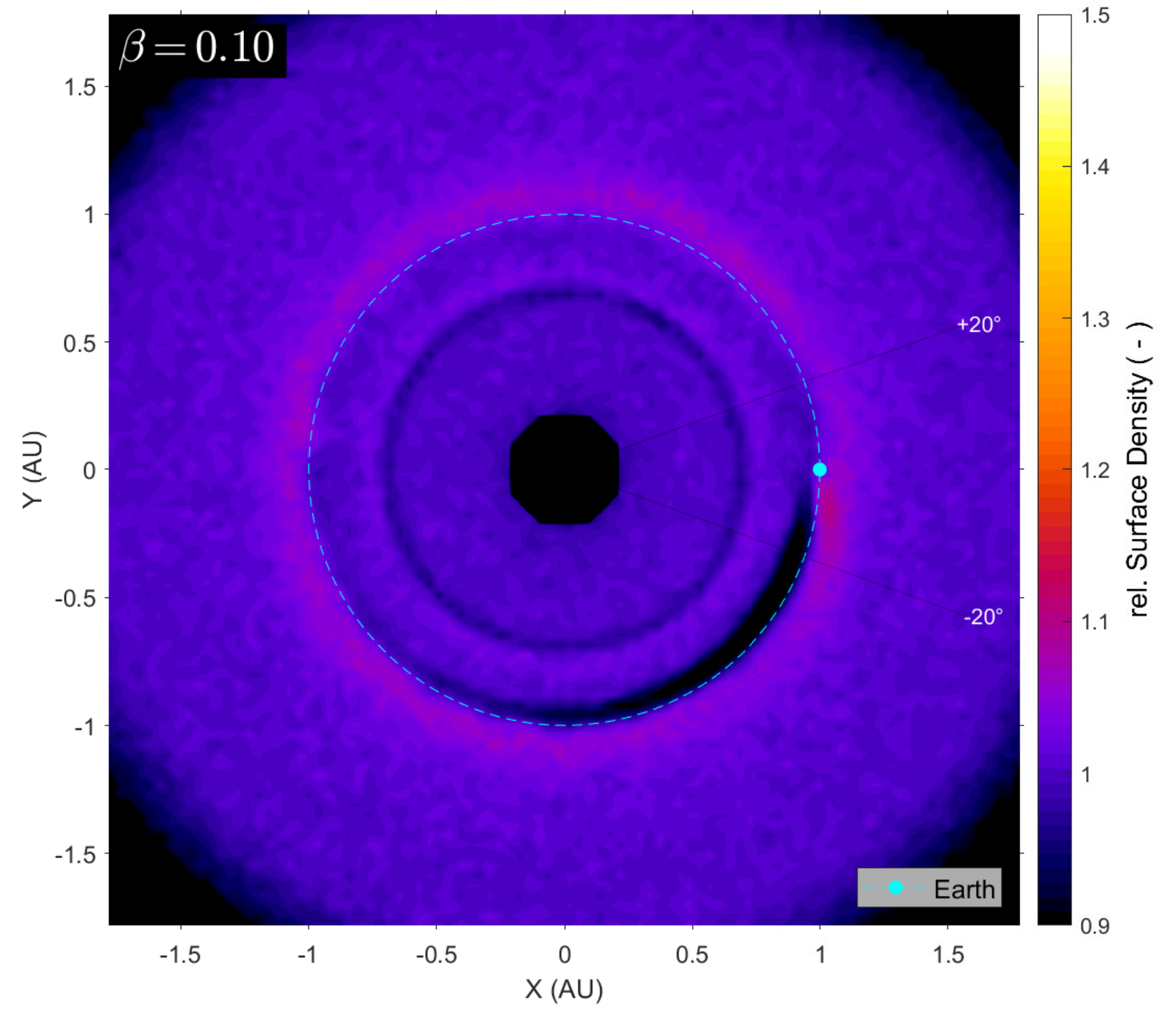} &
	\hspace{-1mm}\includegraphics[width=59mm,trim={3mm 2mm 3mm 1mm},clip]{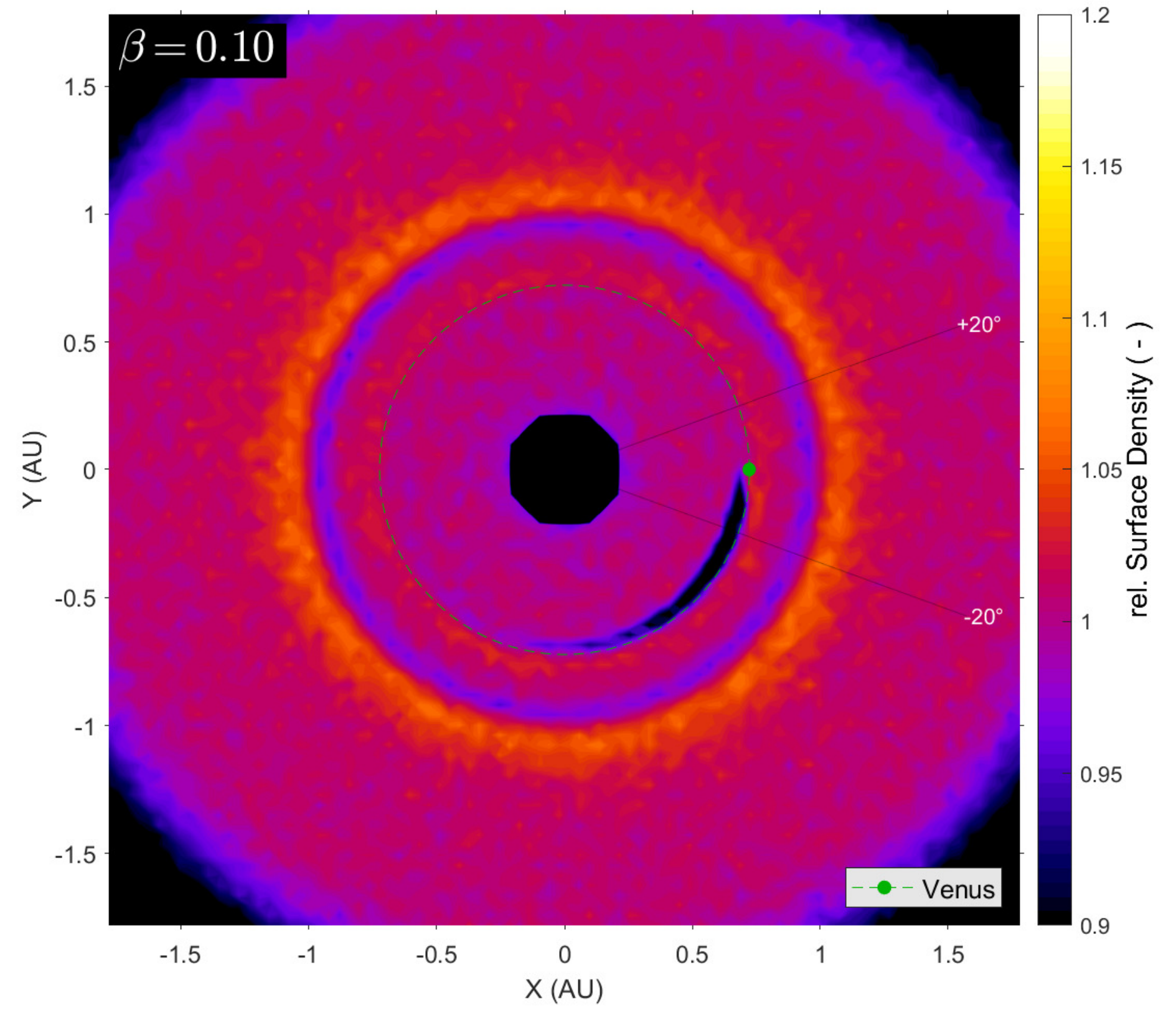} &
	\hspace{-1mm}\raisebox{28.05mm}{\includegraphics[width=59mm,trim={3mm 11.8mm 3mm 1mm},clip]{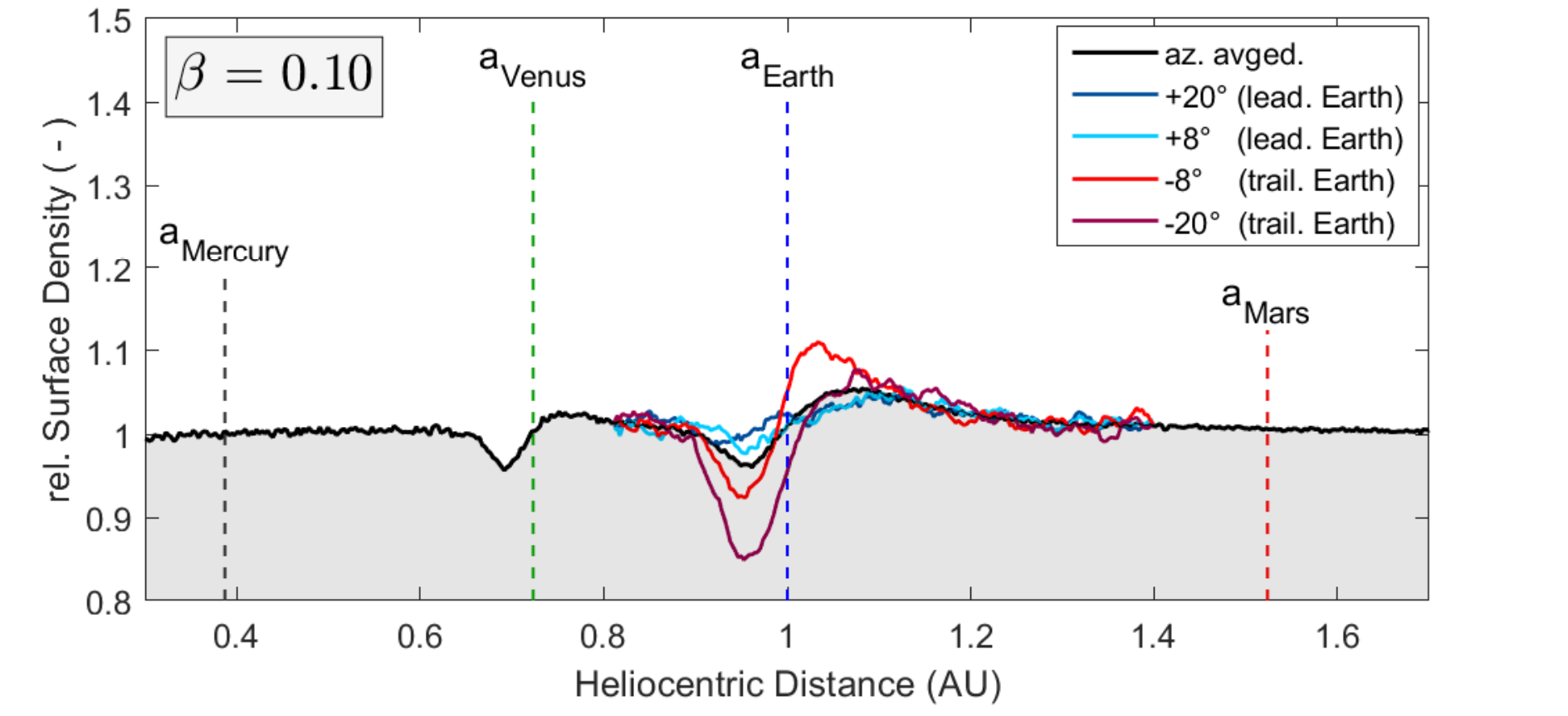}}
	\hspace{-59.9mm}\raisebox{1.7mm}{\includegraphics[width=59mm,trim={3mm 6.3mm 3mm 1mm},clip]{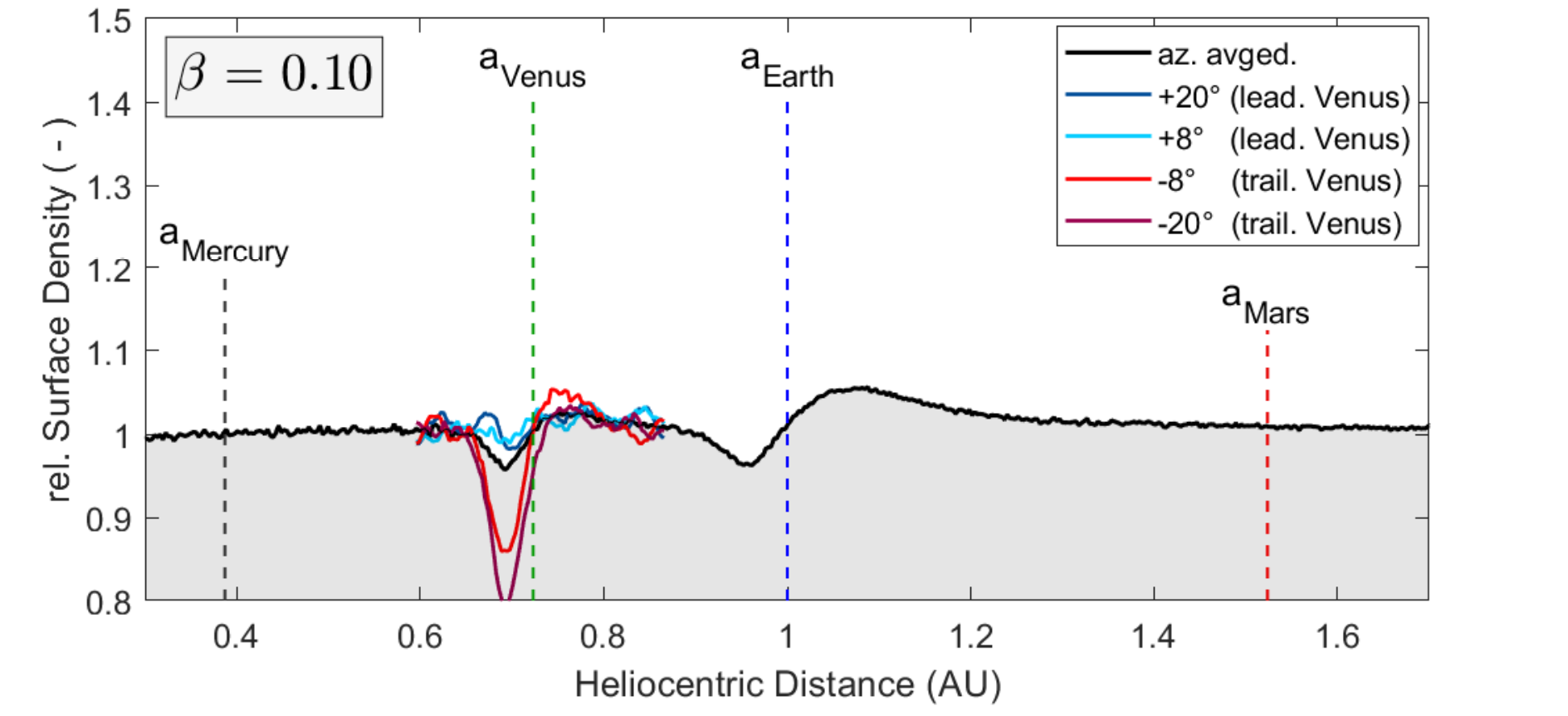}} \\[-1.6ex]
	
	\includegraphics[width=59mm,trim={3mm 2mm 3mm 1mm},clip]{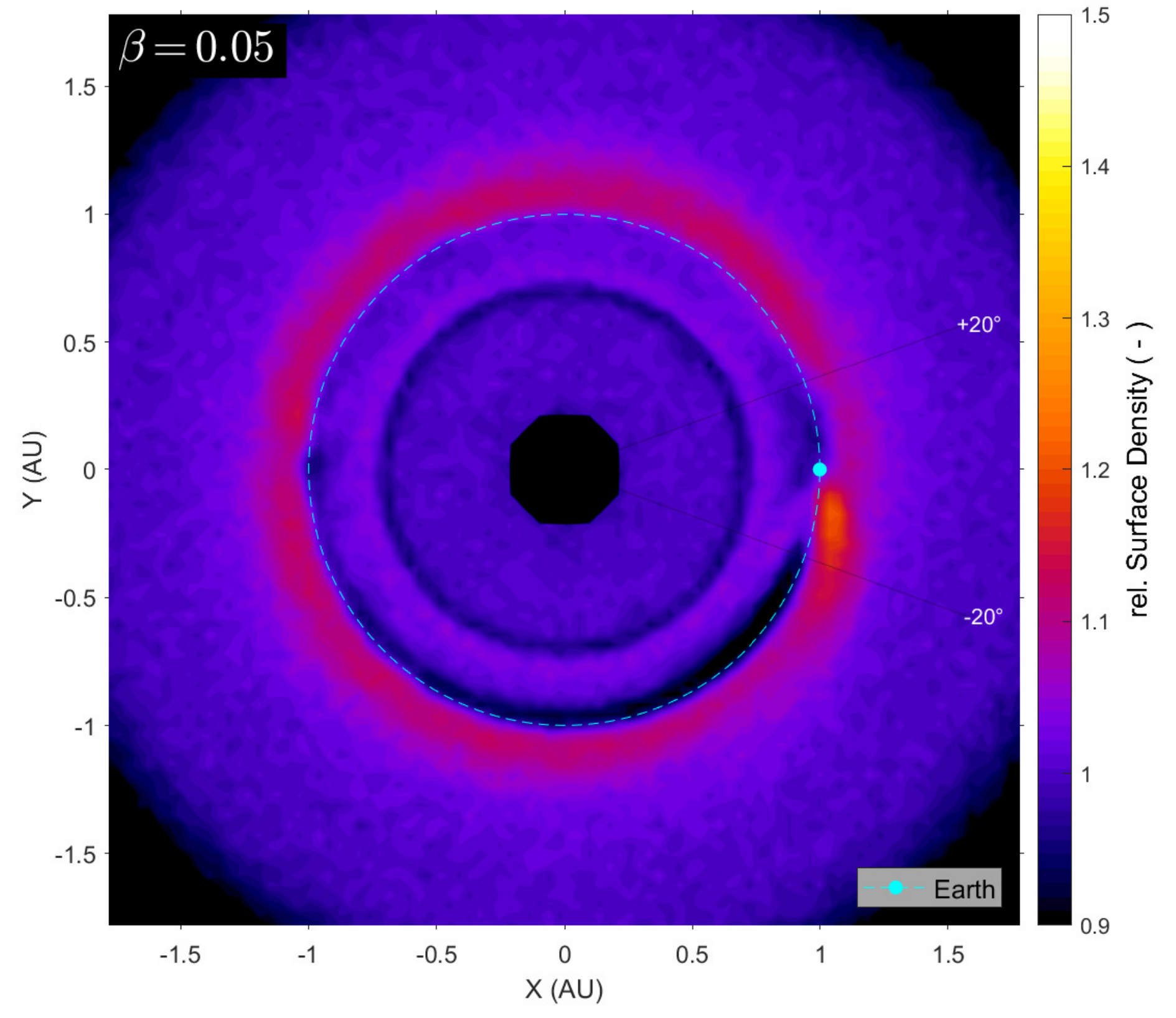} &
	\hspace{-1mm}\includegraphics[width=59mm,trim={3mm 2mm 3mm 1mm},clip]{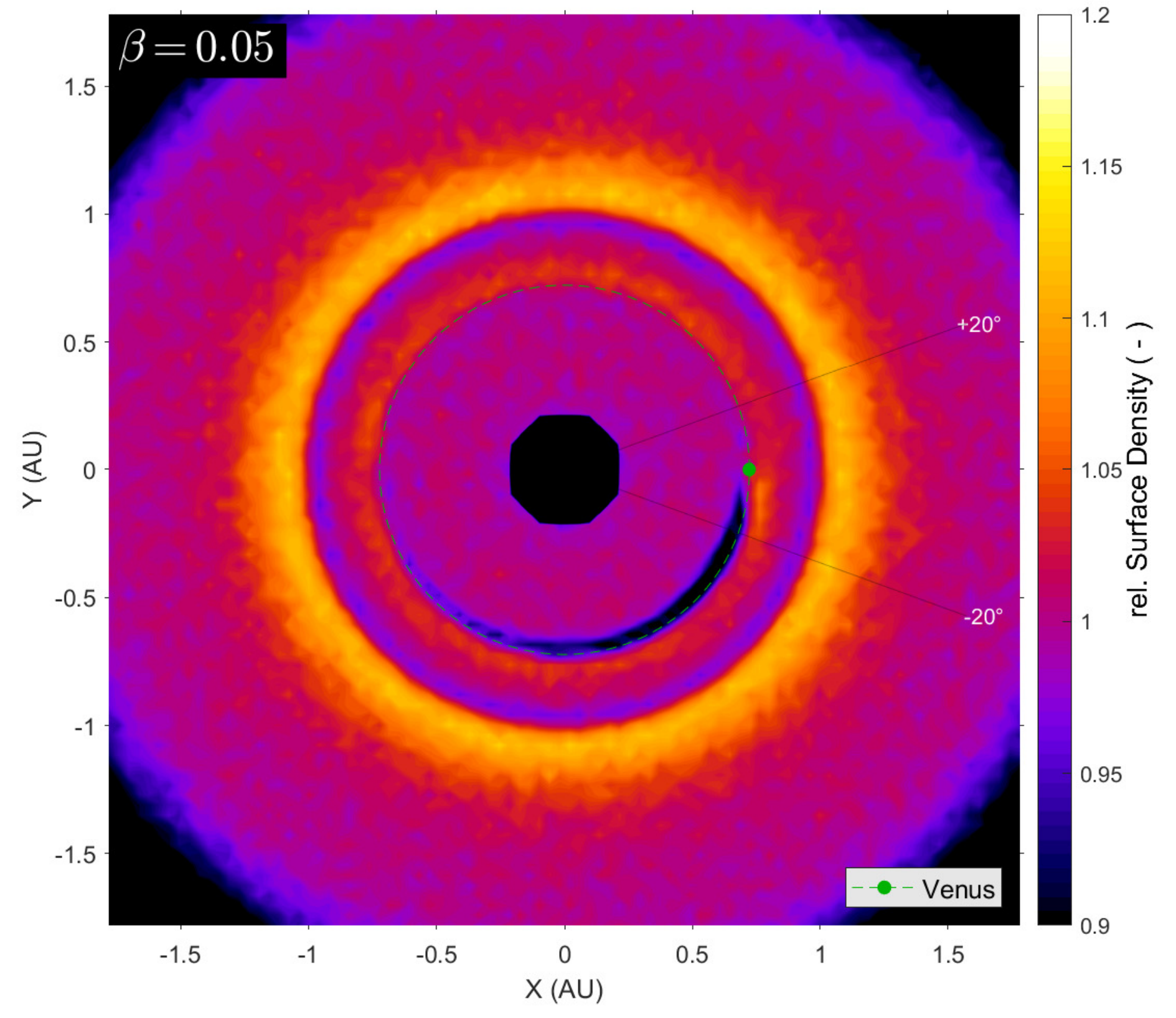} & 
	\hspace{-1mm}\raisebox{28.05mm}{\includegraphics[width=59mm,trim={3mm 11.8mm 3mm 1mm},clip]{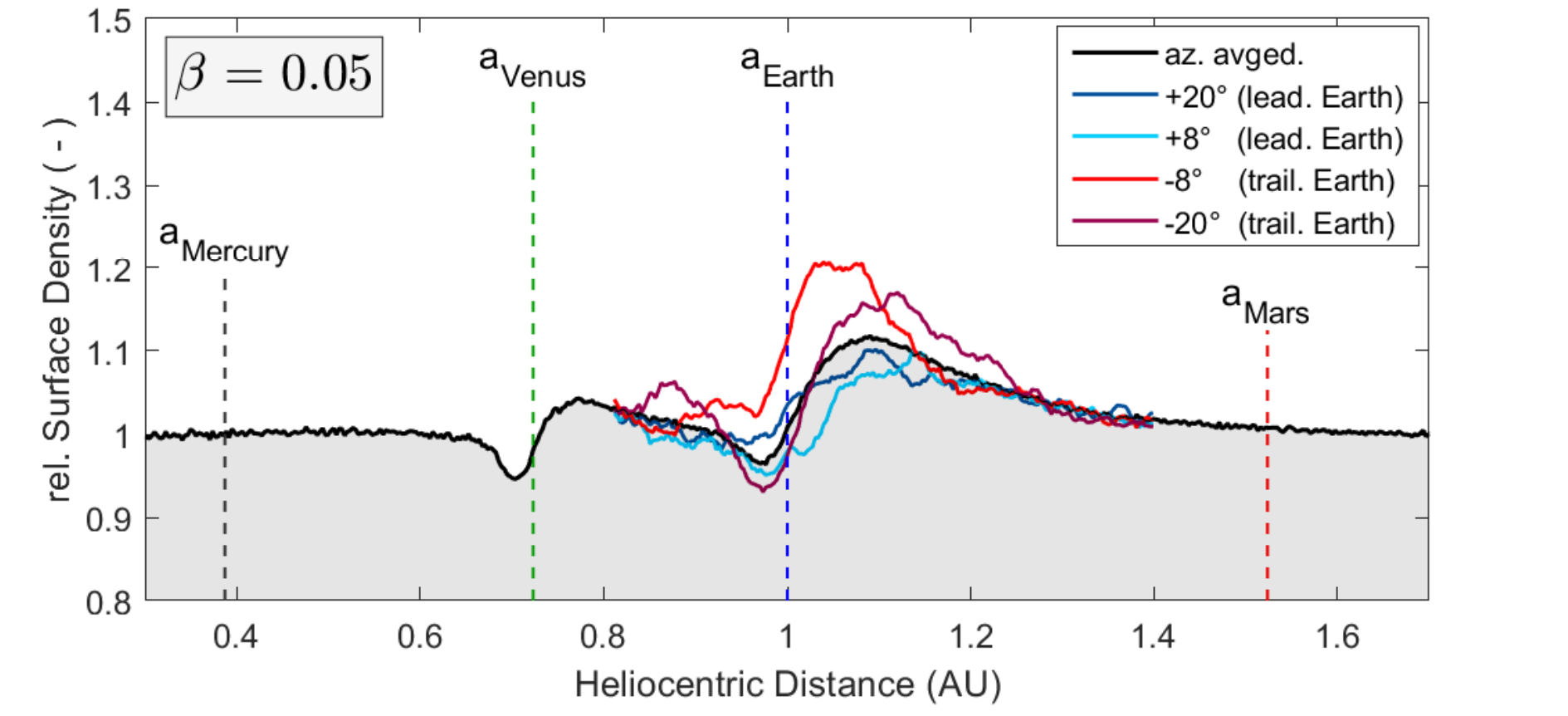}}
	\hspace{-59.9mm}\raisebox{1.7mm}{\includegraphics[width=59mm,trim={3mm 6.3mm 3mm 1mm},clip]{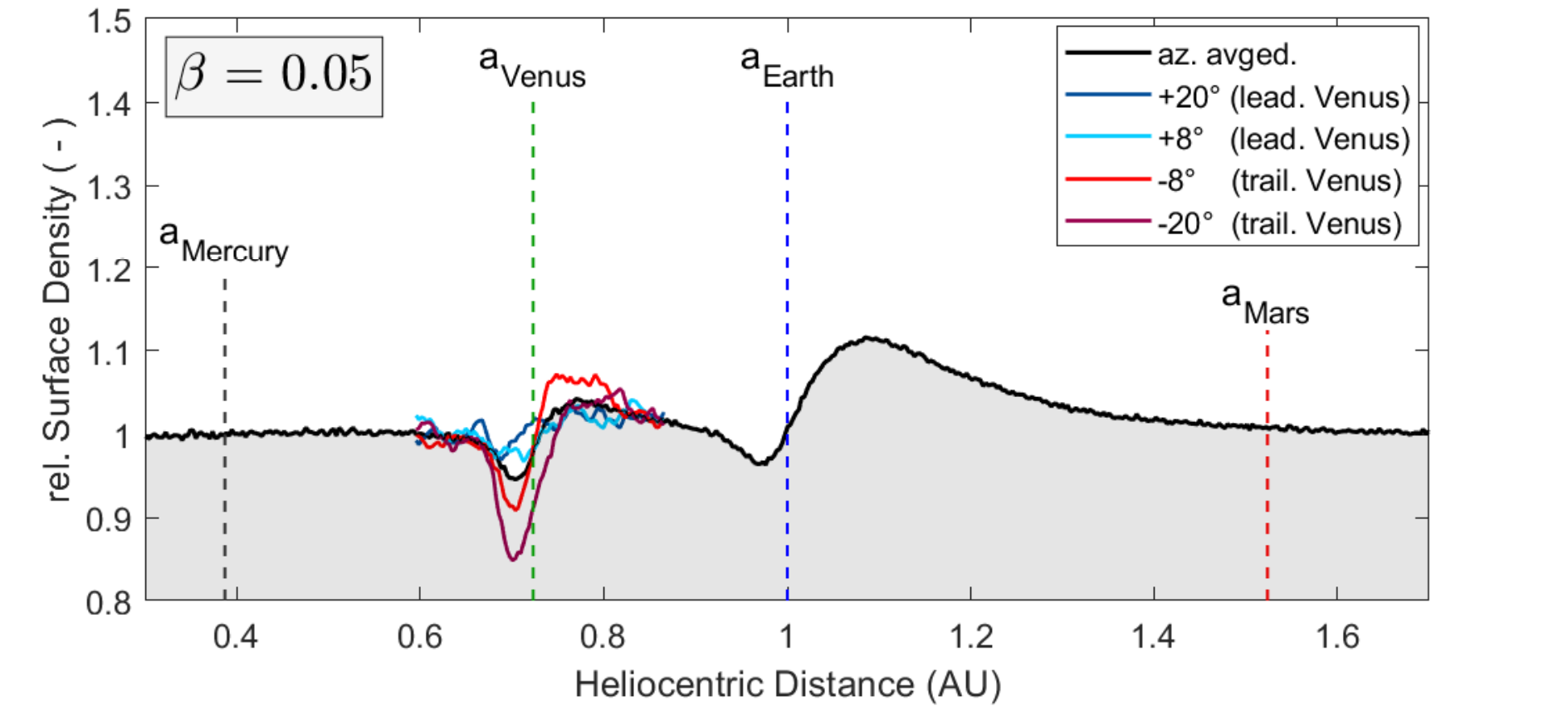}} \\[-1.6ex]
	
	\includegraphics[width=59mm,trim={3mm 2mm 3mm 1mm},clip]{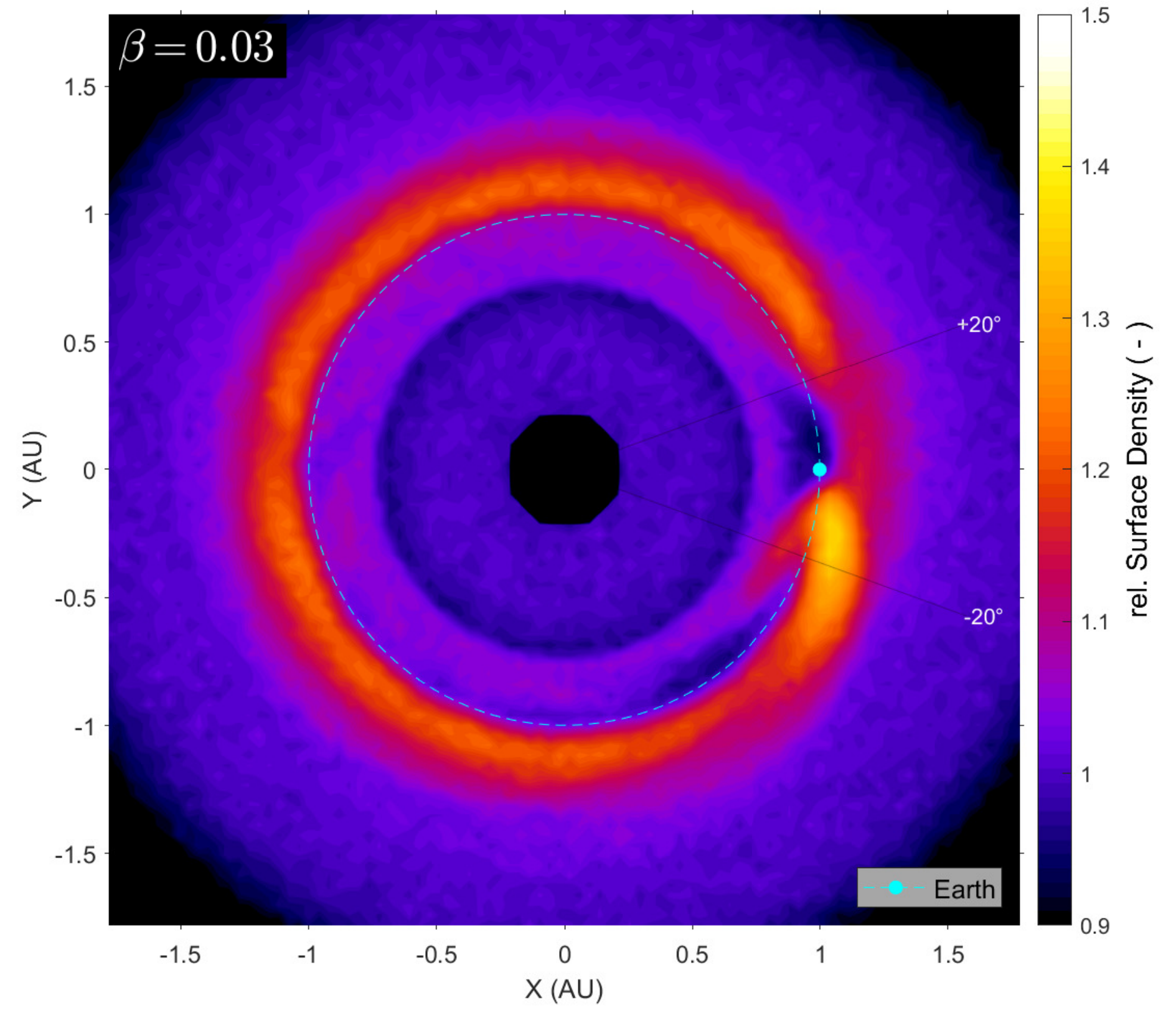} &
	\hspace{-1mm}\includegraphics[width=59mm,trim={3mm 2mm 3mm 1mm},clip]{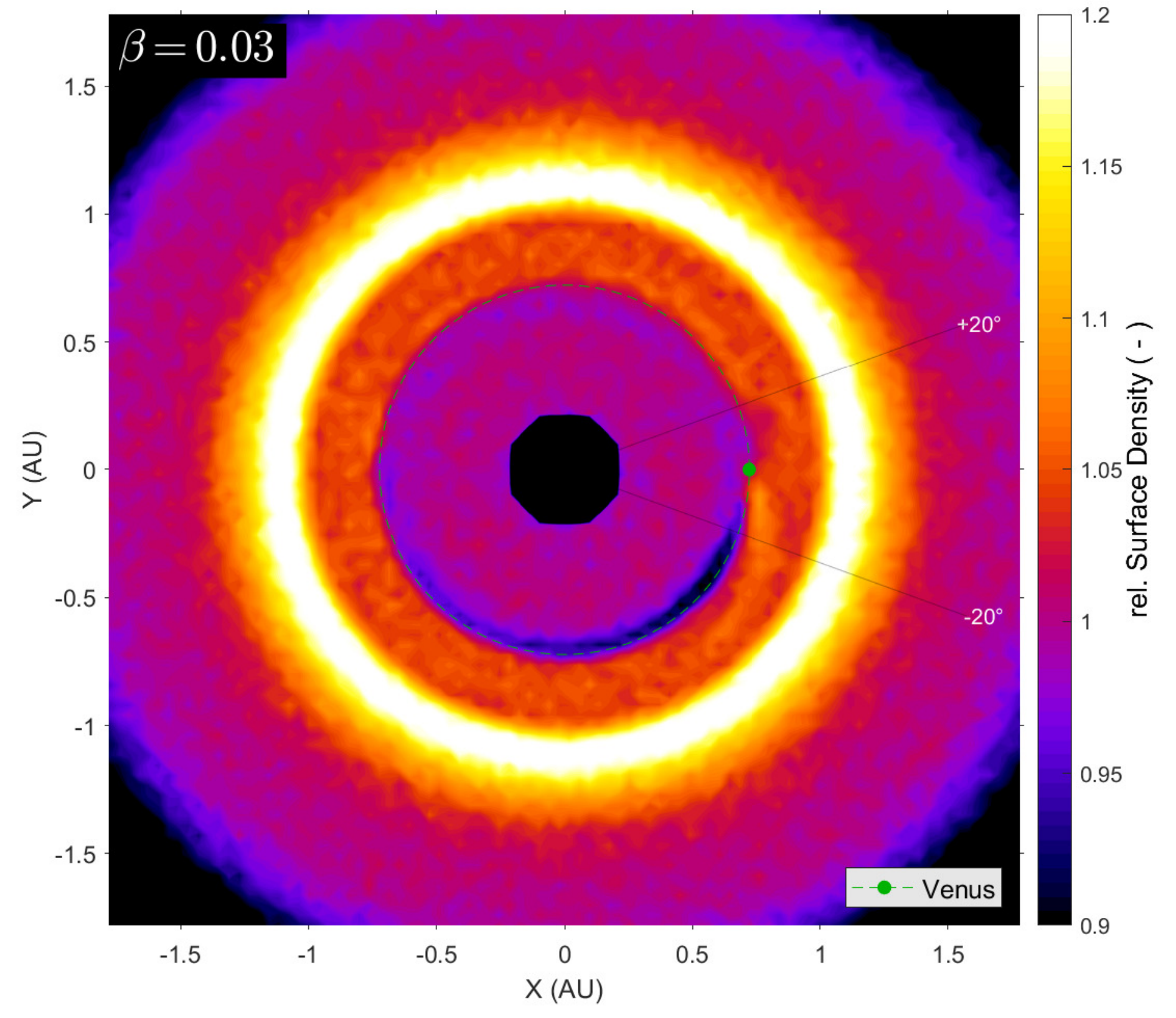} & 
	\hspace{-1mm}\raisebox{28.05mm}{\includegraphics[width=59mm,trim={3mm 11.8mm 3mm 1mm},clip]{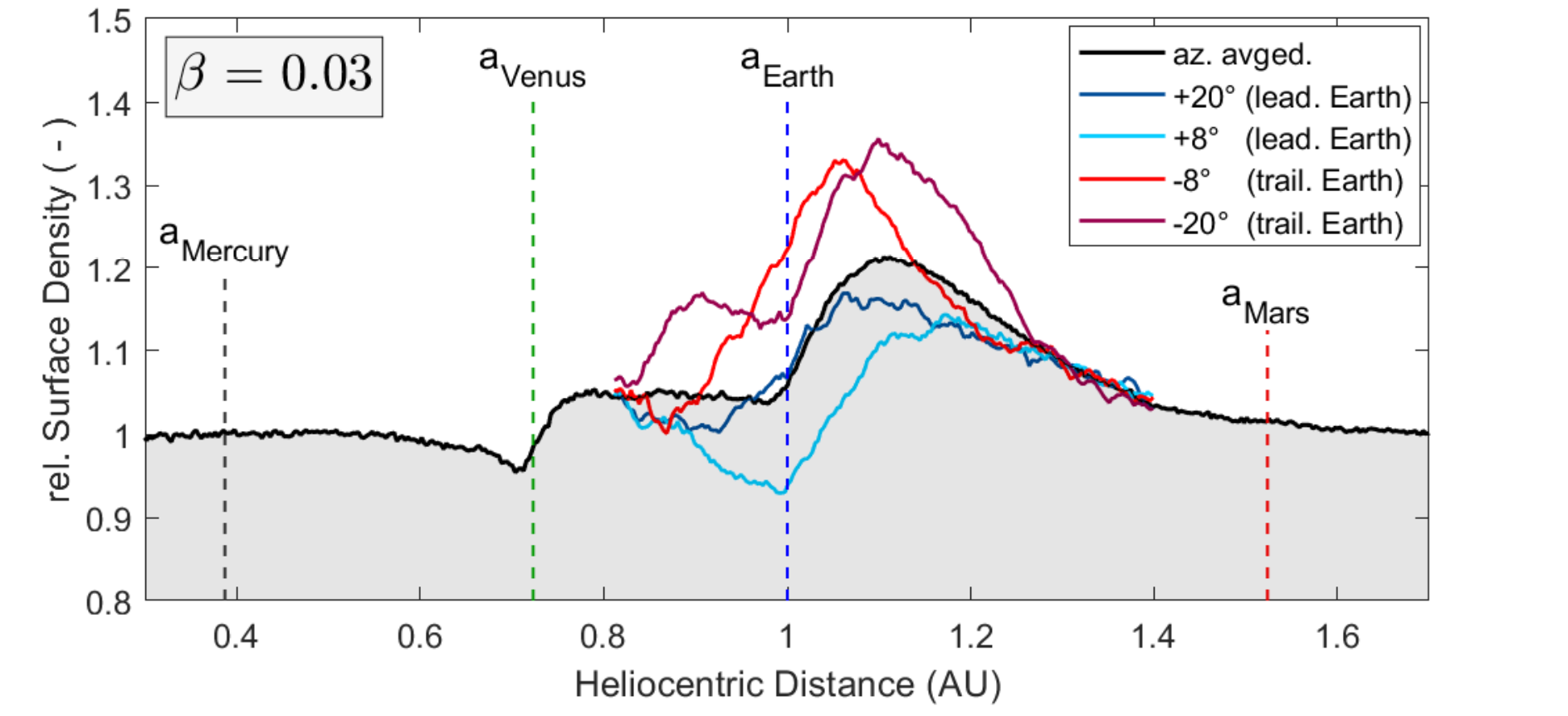}}
	\hspace{-59.9mm}\raisebox{0mm}{\includegraphics[width=59mm,trim={3mm 1.2mm 3mm 1mm},clip]{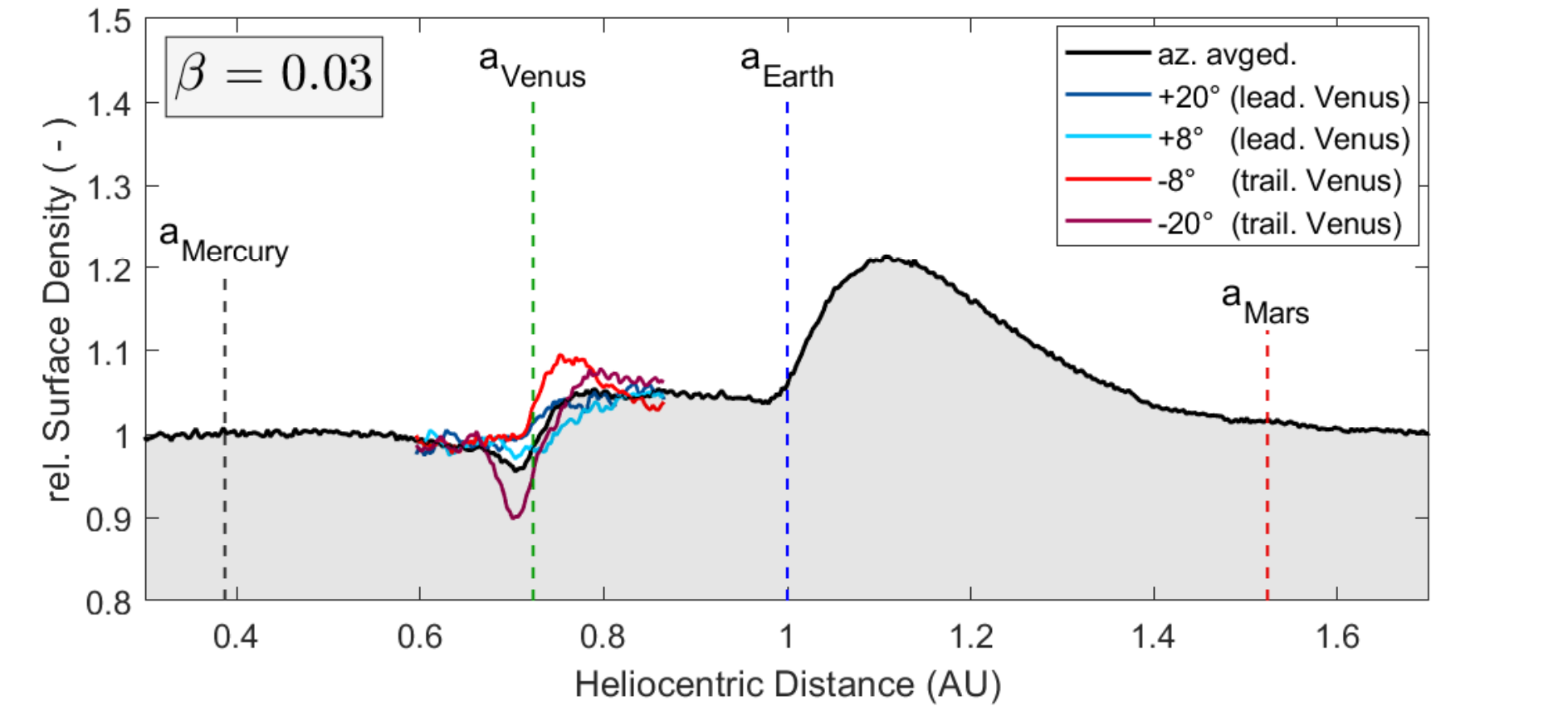}} 
	
	\end{tabular}
	\caption{Density distribution under the presence of all planets for 
			$\beta=$~0.1, 0.05, 0.03 ($D=5~\si{\micro\meter}, 10~\si{\micro\meter}, 17~\si{\micro\meter}$) 
			from top to bottom, recorded in 
			a frame co-rotating with Earth \textit{(left)} and Venus \textit{(middle)}. 
			2D-histogram bin size is 0.02~\si{\astronomicalunit}~$\times$~0.02~\si{\astronomicalunit}.
			\textit{Right:} Radial profiles of the obtained surface density for the Earth and Venus ring. 
			Radial profiles are given at cuts leading and trailing Earth and Venus, 
			as well as for the azimuthal average. 
			Vertical dashed lines indicate the location of the semi major axes of the planets.}
	\label{fig:A_XY_AP_1}
	\end{figure*}
	\pagebreak

	\begin{figure*}[h]
	\vspace*{40mm}
	\centering
	\begin{tabular}{ccc}
	\includegraphics[width=59mm,trim={3mm 2mm 3mm 1mm},clip]{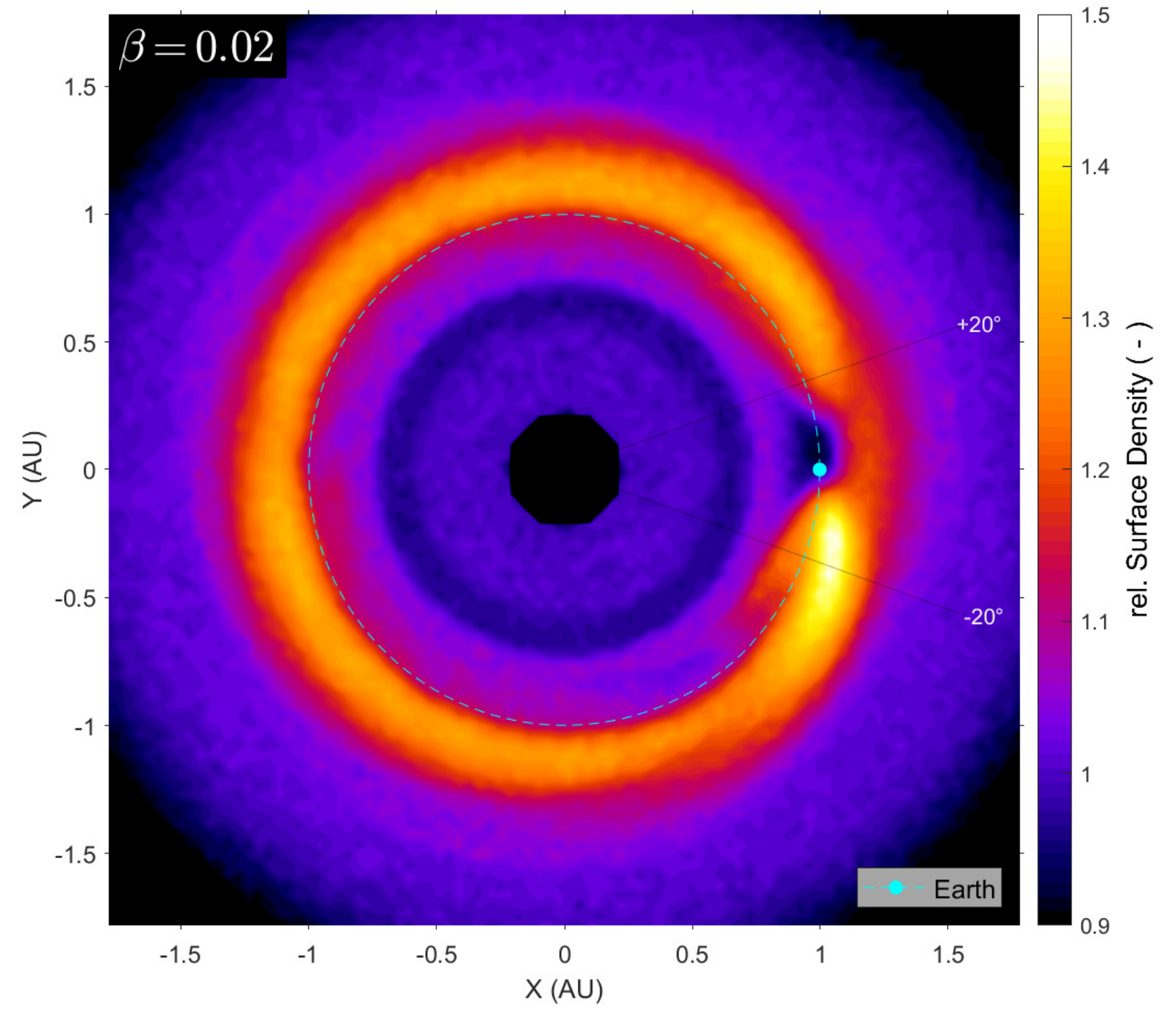} &
	\hspace{-1mm}\includegraphics[width=59mm,trim={3mm 2mm 3mm 1mm},clip]{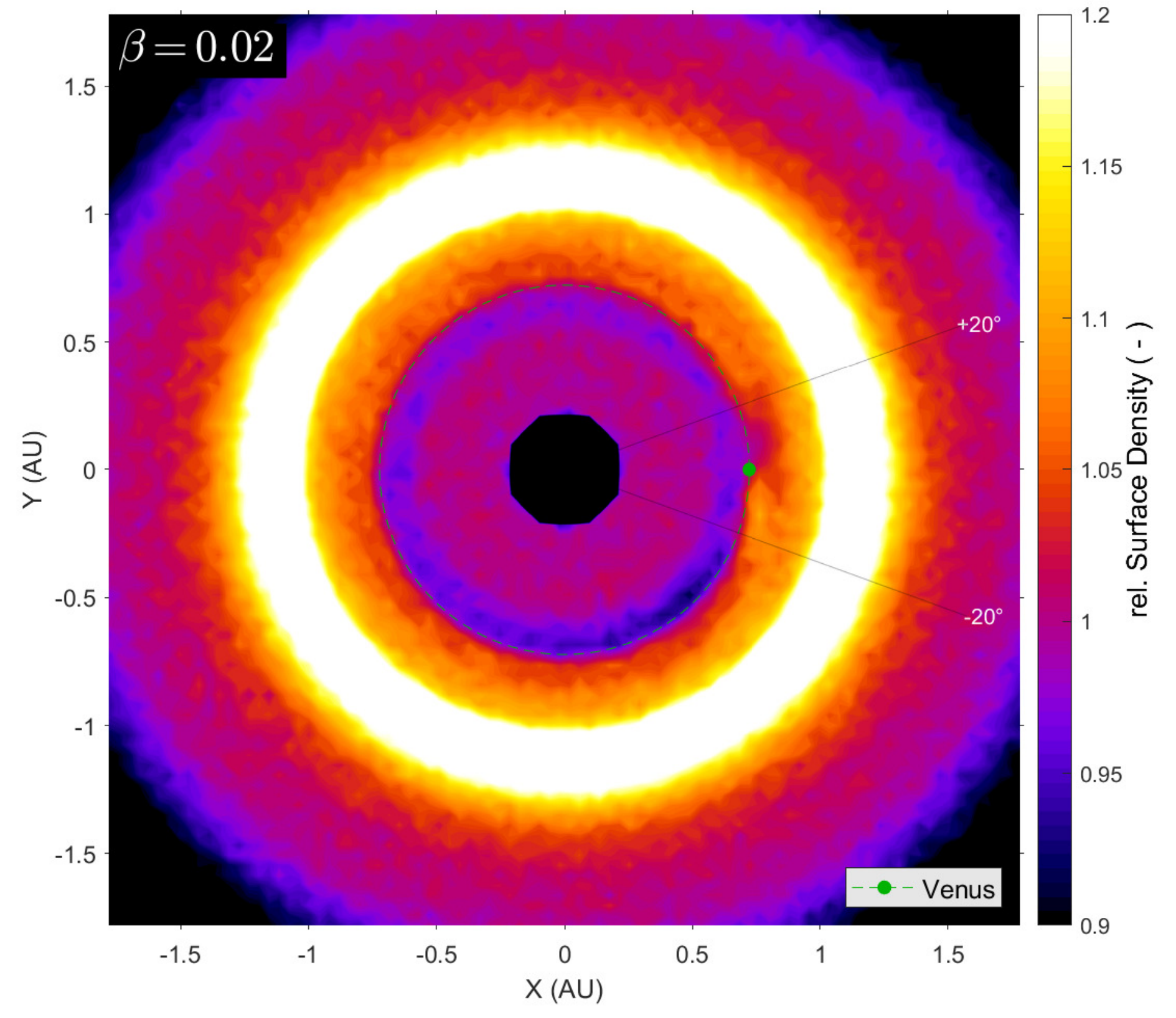} & 
	\hspace{-1mm}\raisebox{28.05mm}{\includegraphics[width=59mm,trim={3mm 11.8mm 3mm 1mm},clip]{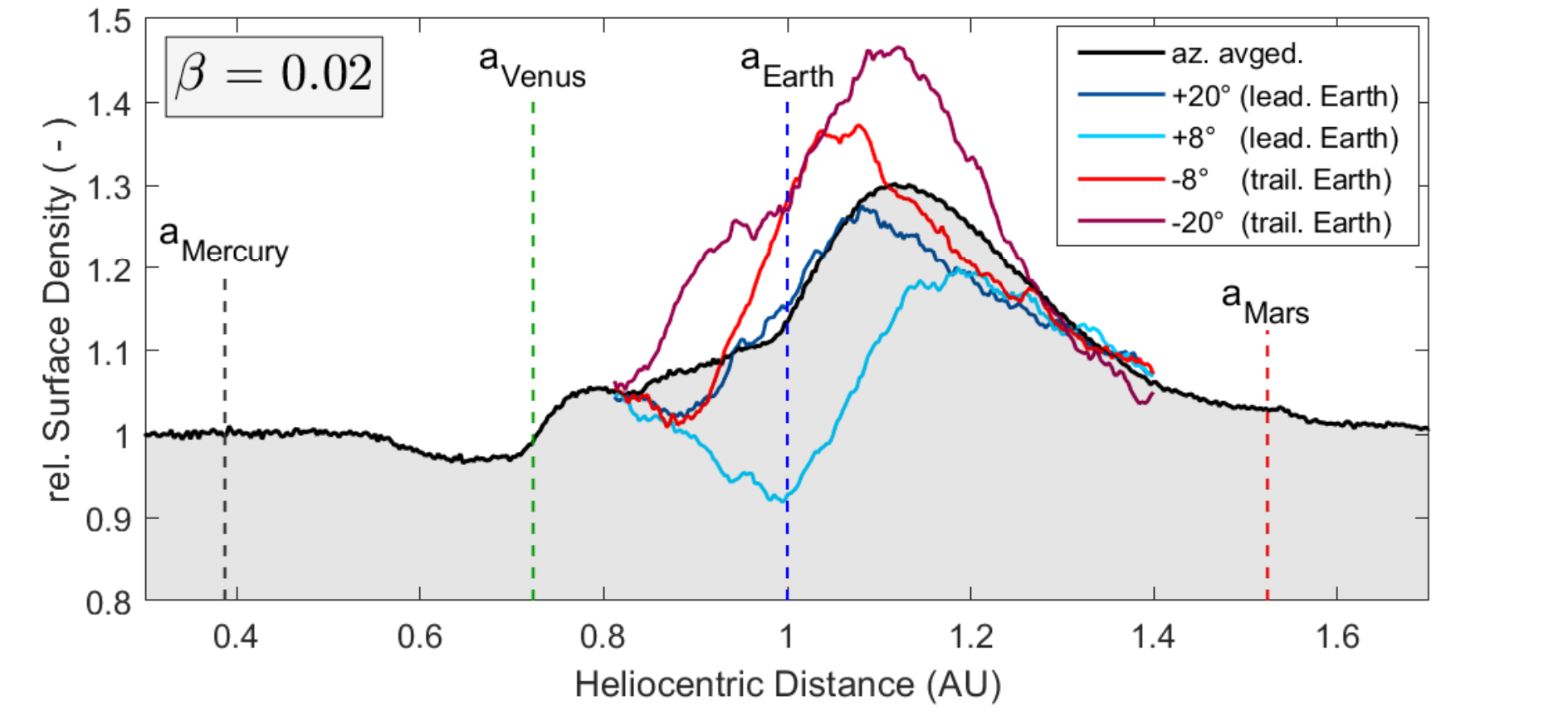}}
	\hspace{-59.9mm}\raisebox{1.7mm}{\includegraphics[width=59mm,trim={3mm 6.3mm 3mm 1mm},clip]{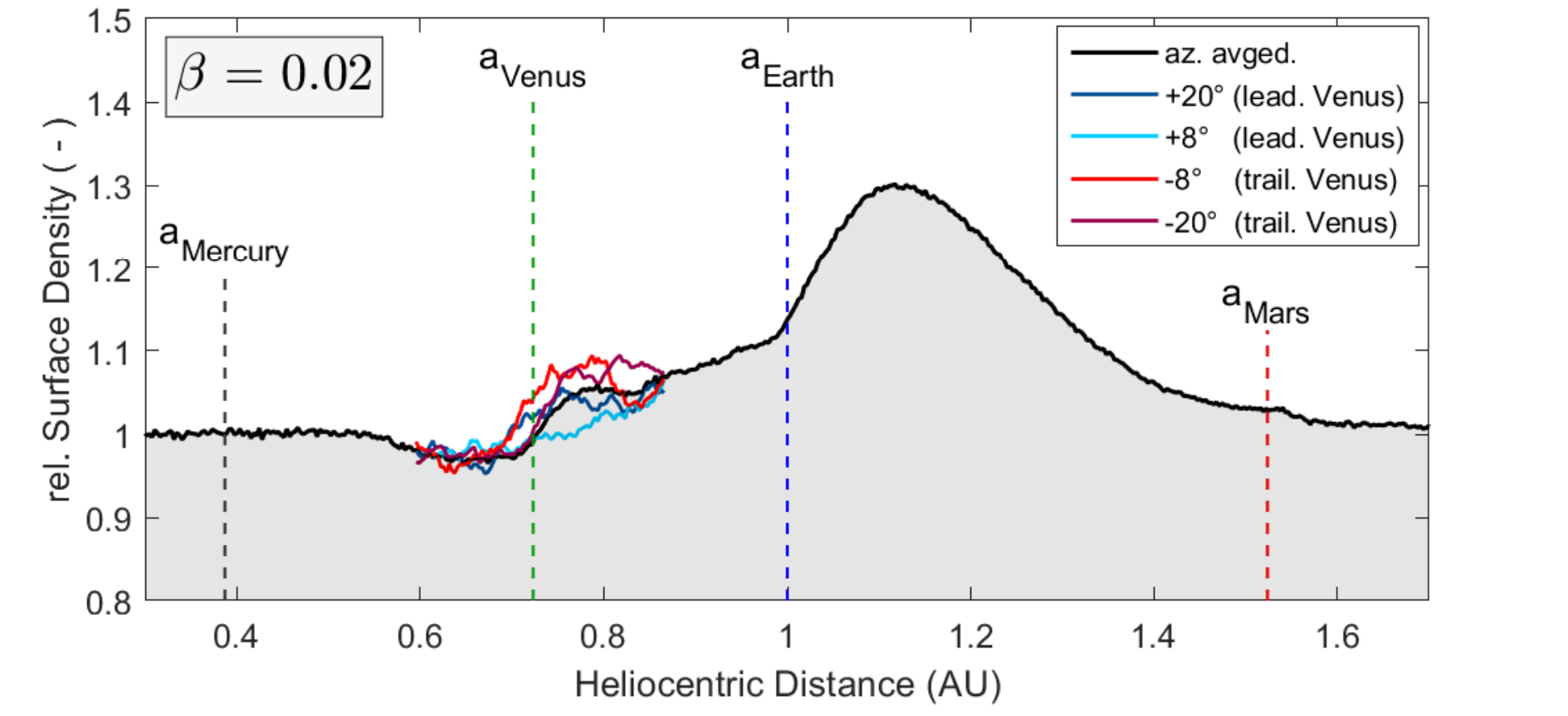}} \\[-1.6ex]
	
	\includegraphics[width=59mm,trim={3mm 2mm 3mm 1mm},clip]{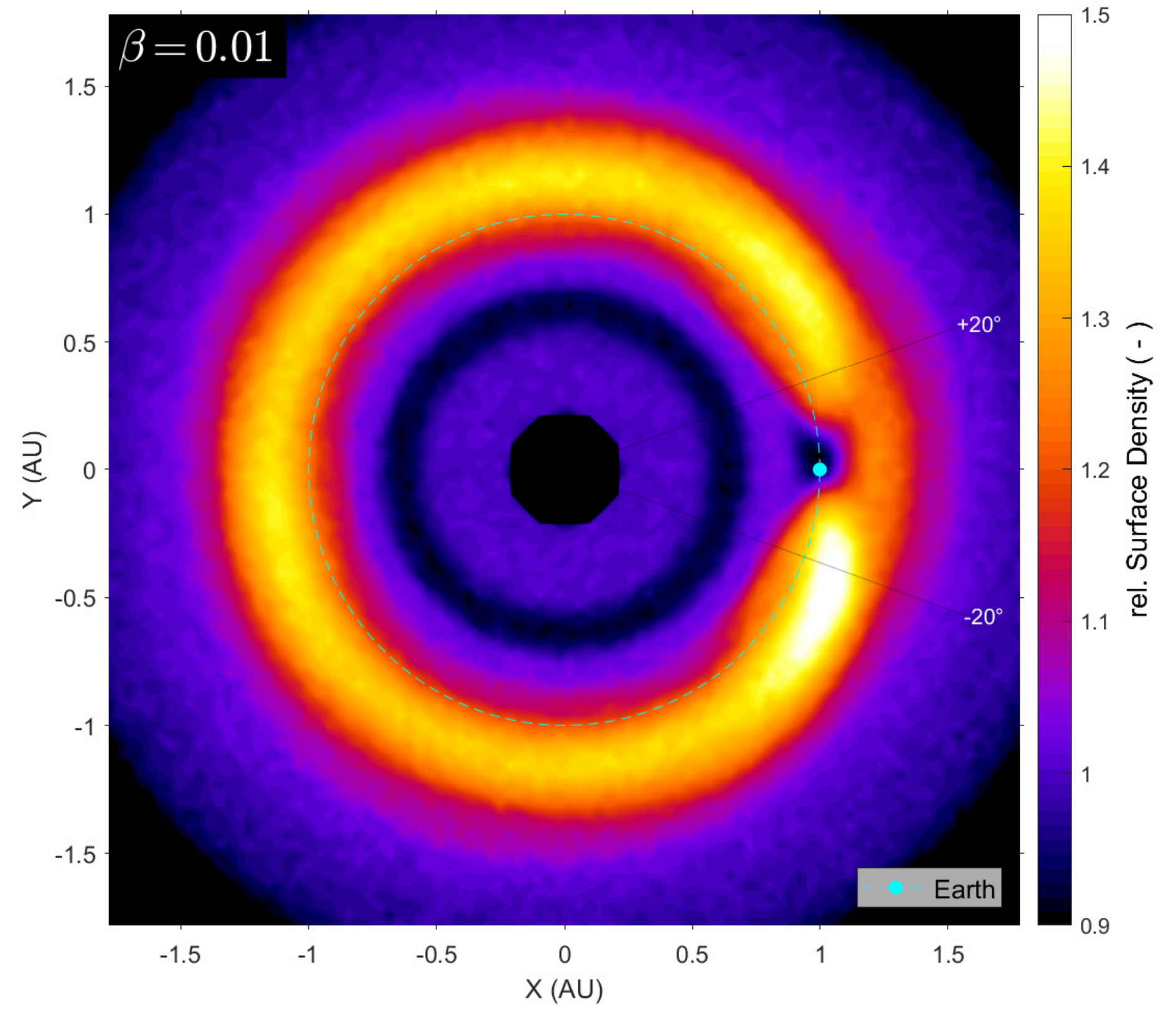} &
	\hspace{-1mm}\includegraphics[width=59mm,trim={3mm 2mm 3mm 1mm},clip]{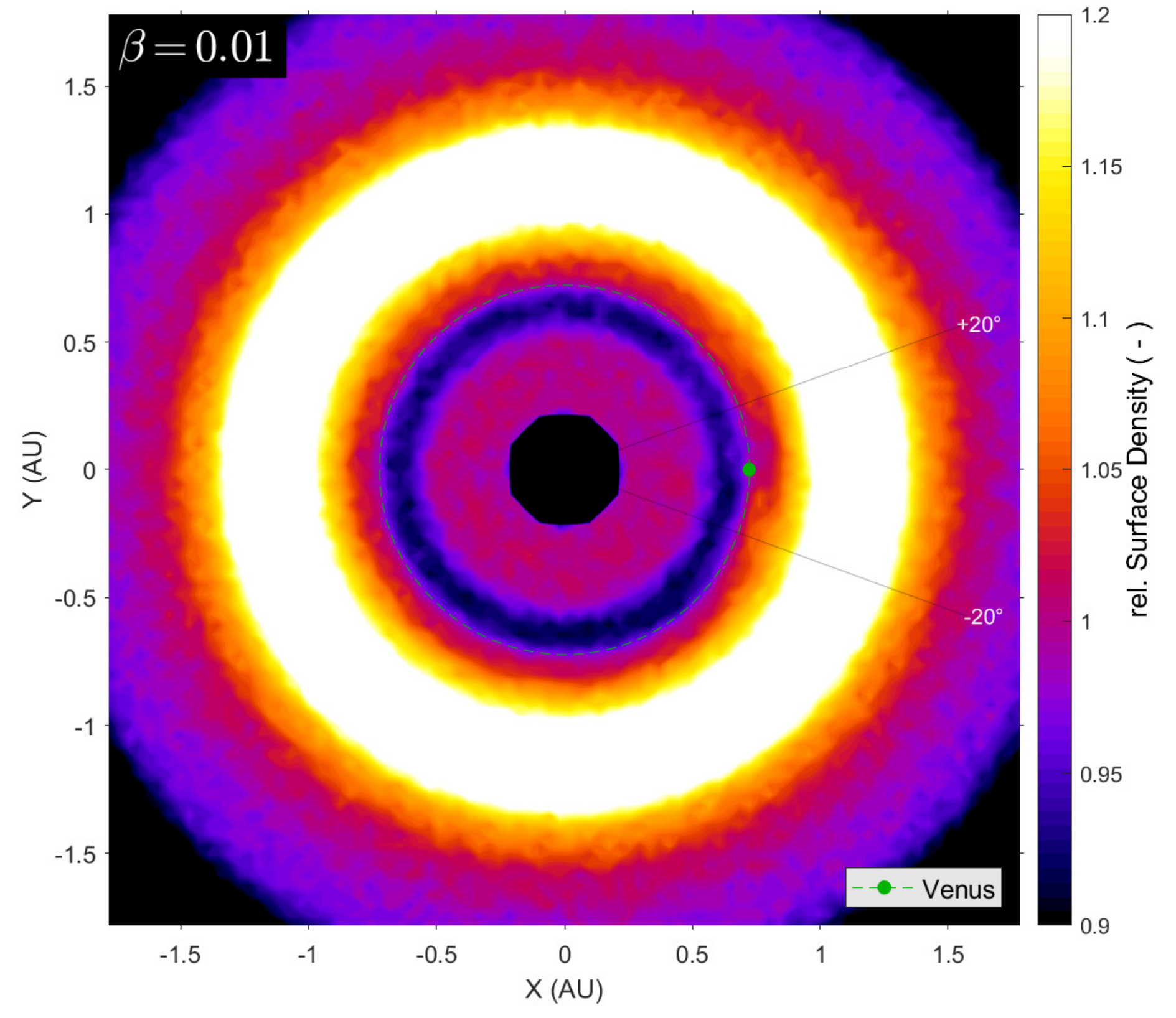} & 
	\hspace{-1mm}\raisebox{28.05mm}{\includegraphics[width=59mm,trim={3mm 11.8mm 3mm 1mm},clip]{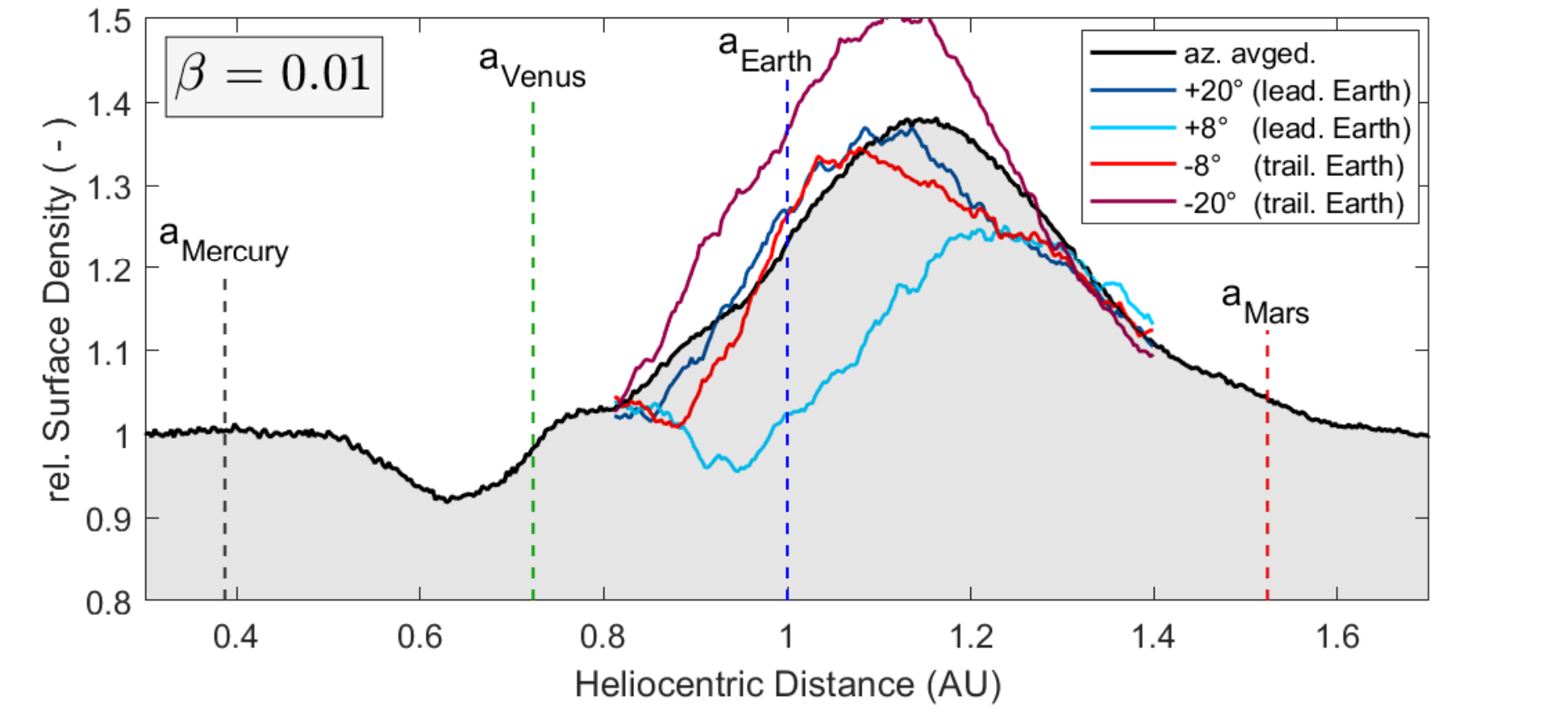}}
	\hspace{-59.9mm}\raisebox{1.7mm}{\includegraphics[width=59mm,trim={3mm 6.3mm 3mm 1mm},clip]{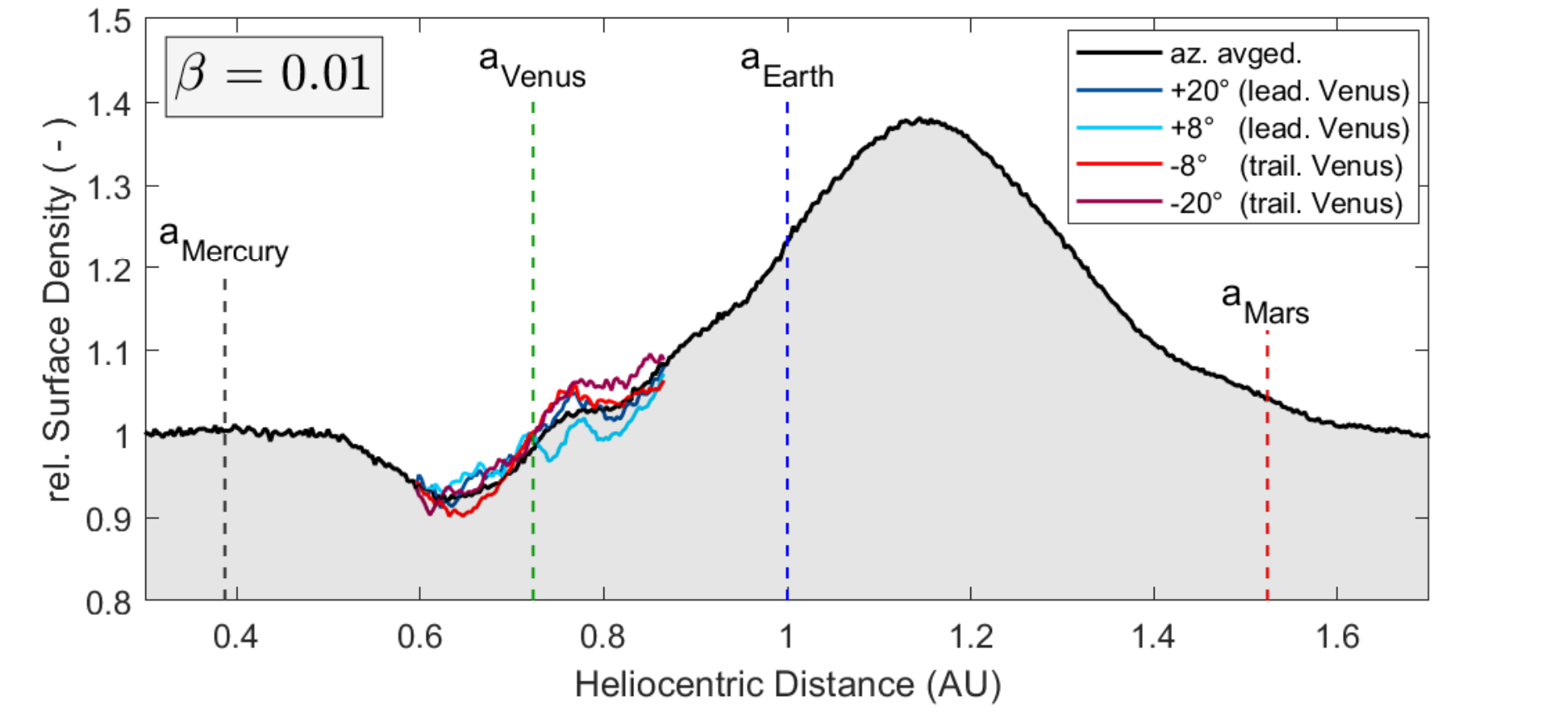}} \\[-1.6ex]
	
	\includegraphics[width=59mm,trim={3mm 2mm 3mm 1mm},clip]{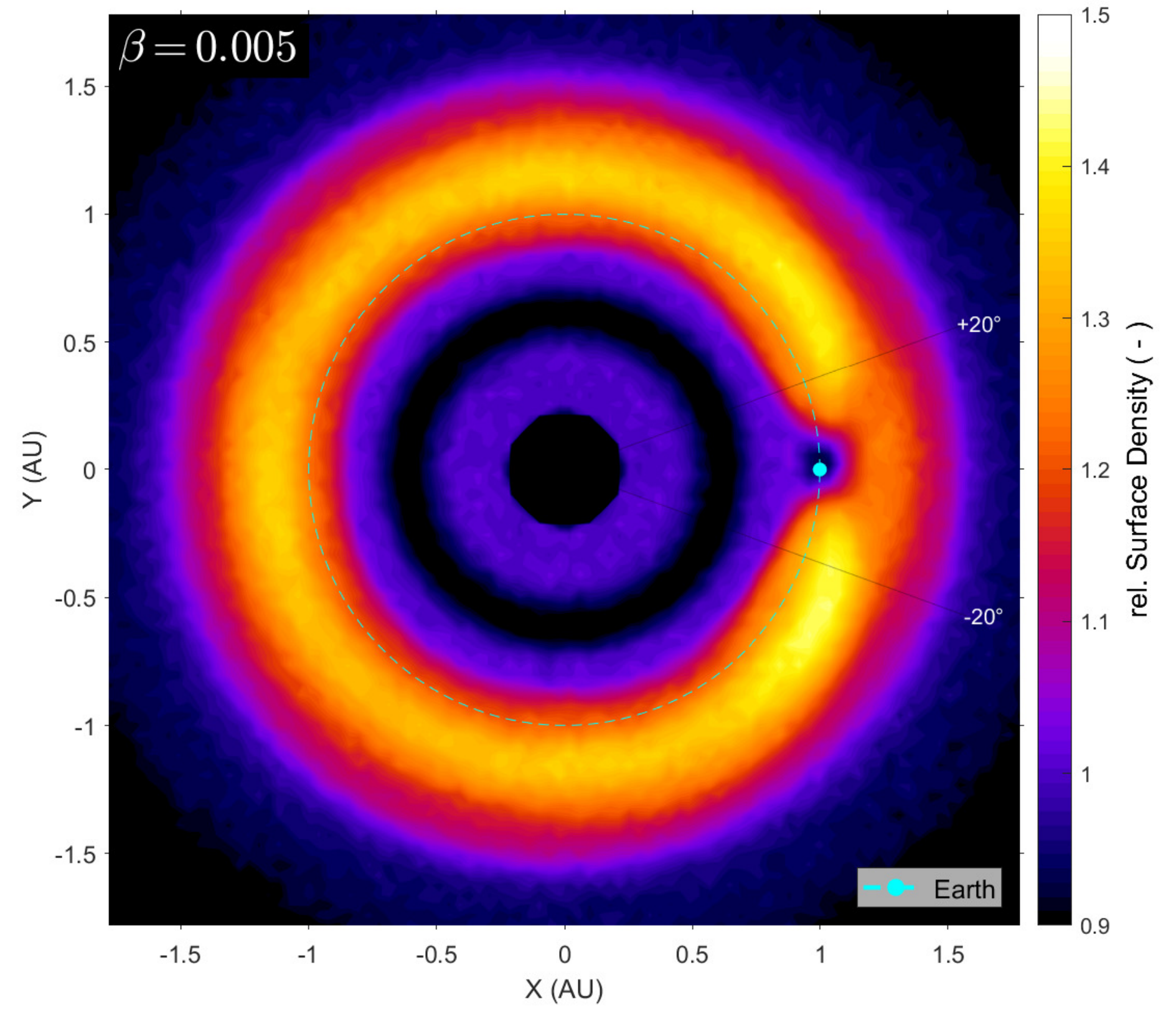} &
	\hspace{-1mm}\includegraphics[width=59mm,trim={3mm 2mm 3mm 1mm},clip]{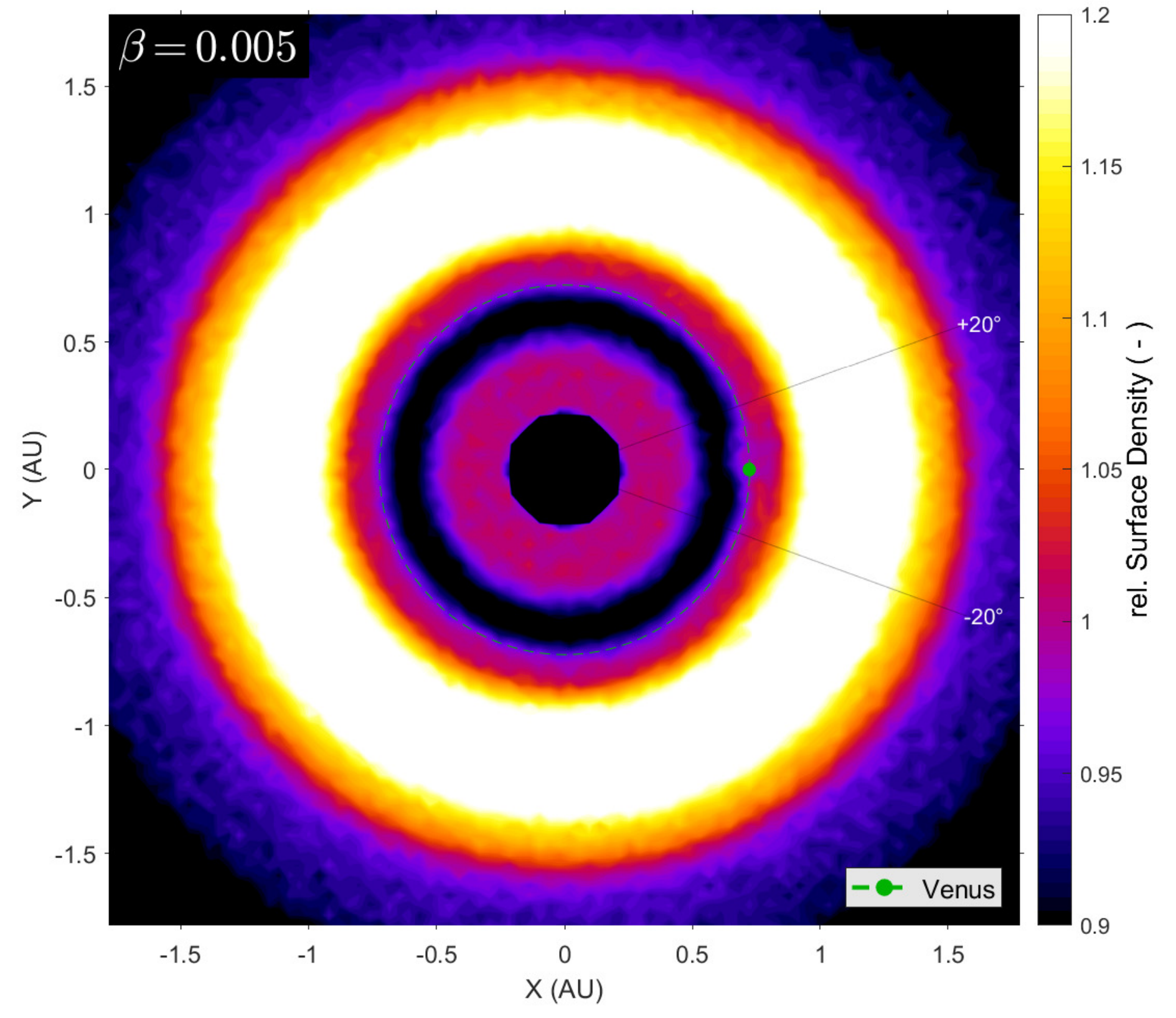} & 
	\hspace{-1mm}\raisebox{28.05mm}{\includegraphics[width=59mm,trim={3mm 11.8mm 3mm 1mm},clip]{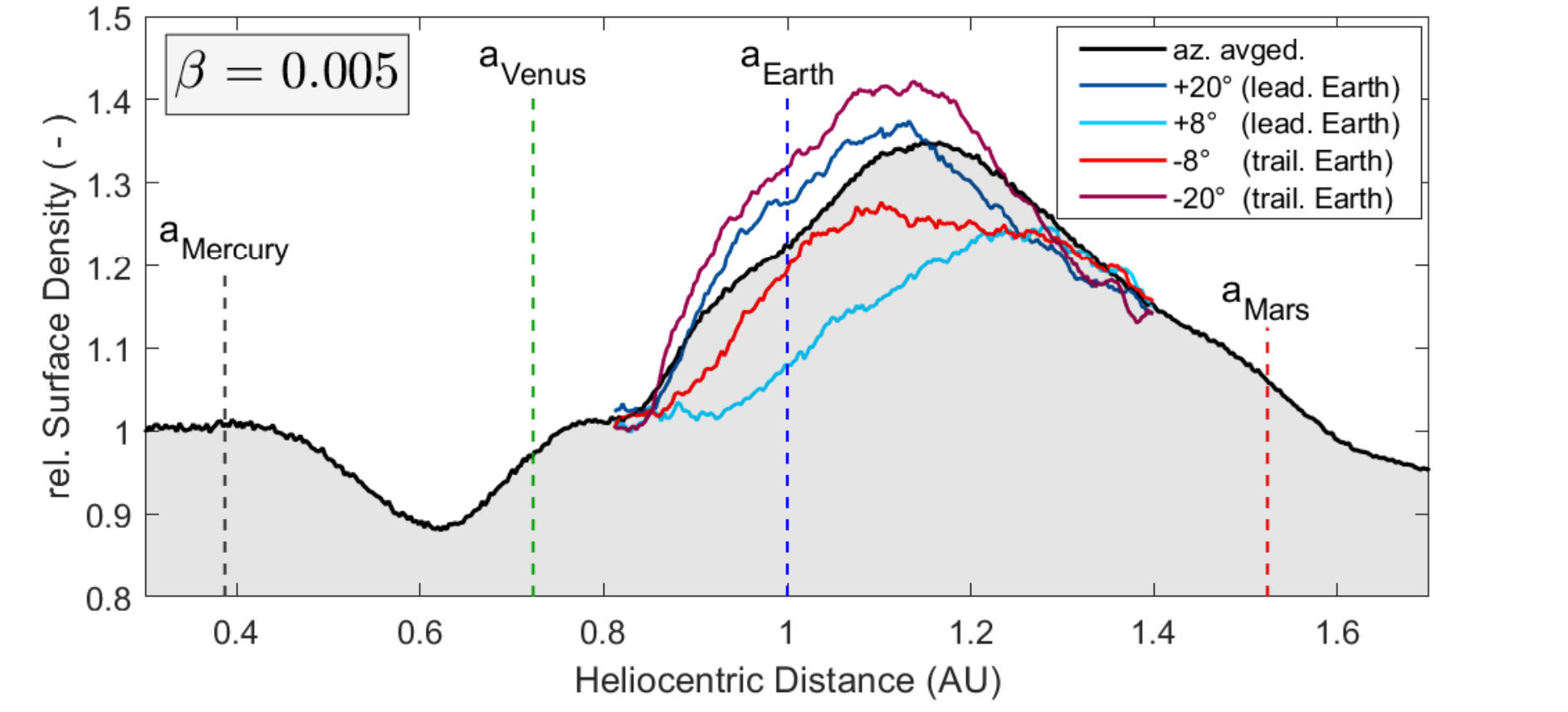}}
	\hspace{-59.9mm}\raisebox{0mm}{\includegraphics[width=59mm,trim={3mm 1.2mm 3mm 1mm},clip]{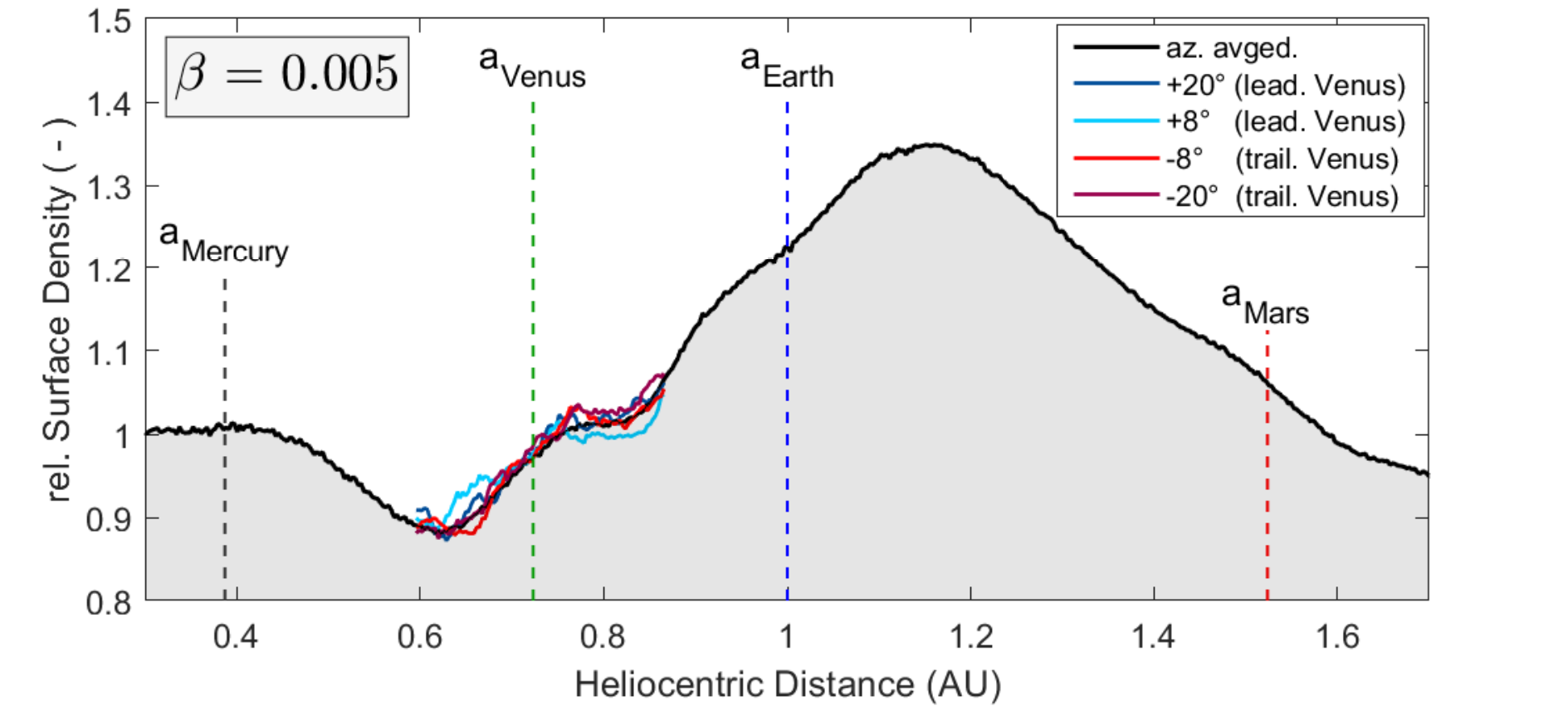}}
	
	\end{tabular}
	\caption{Density distribution under the presence of all planets for 
			$\beta=$~0.02, 0.01, 0.005 ($D=25~\si{\micro\meter}, 50~\si{\micro\meter}, 100~\si{\micro\meter}$) 
			from top to bottom, recorded in 
			a frame co-rotating with Earth \textit{(left)} and Venus \textit{(middle)}. 
			2D-histogram bin size is 0.02~\si{\astronomicalunit}~$\times$~0.02~\si{\astronomicalunit}.
			\textit{Right:} Radial profiles of the obtained surface density for the Earth and Venus ring. 
			Radial profiles are given at cuts leading and trailing Earth and Venus, 
			as well as for the azimuthal average. 
			Vertical dashed lines indicate the location of the semi-major axes of the planets.}
	\label{fig:A_XY_AP_2}
	\end{figure*}

	\begin{figure*}[!htbp]
		\vspace*{1mm}
		\centering
		\includegraphics[width=0.8\textwidth,trim={0 5mm 0 1.9mm},clip]{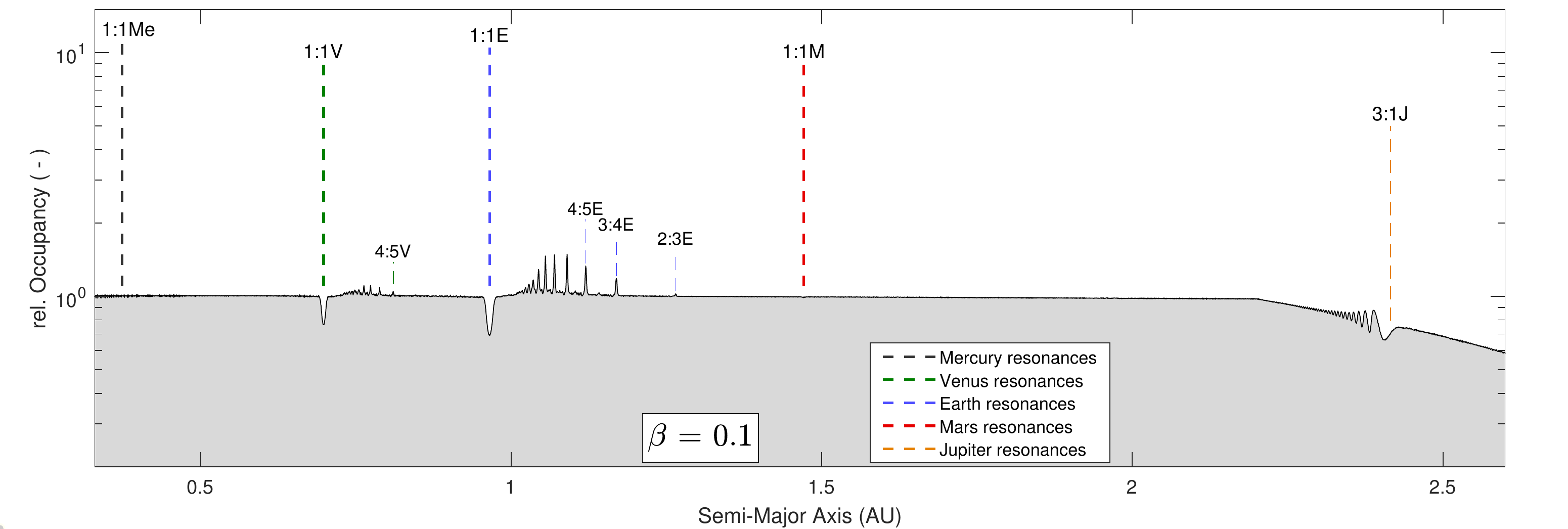}\\
		\includegraphics[width=0.8\textwidth,trim={0 5mm 0 1.9mm},clip]{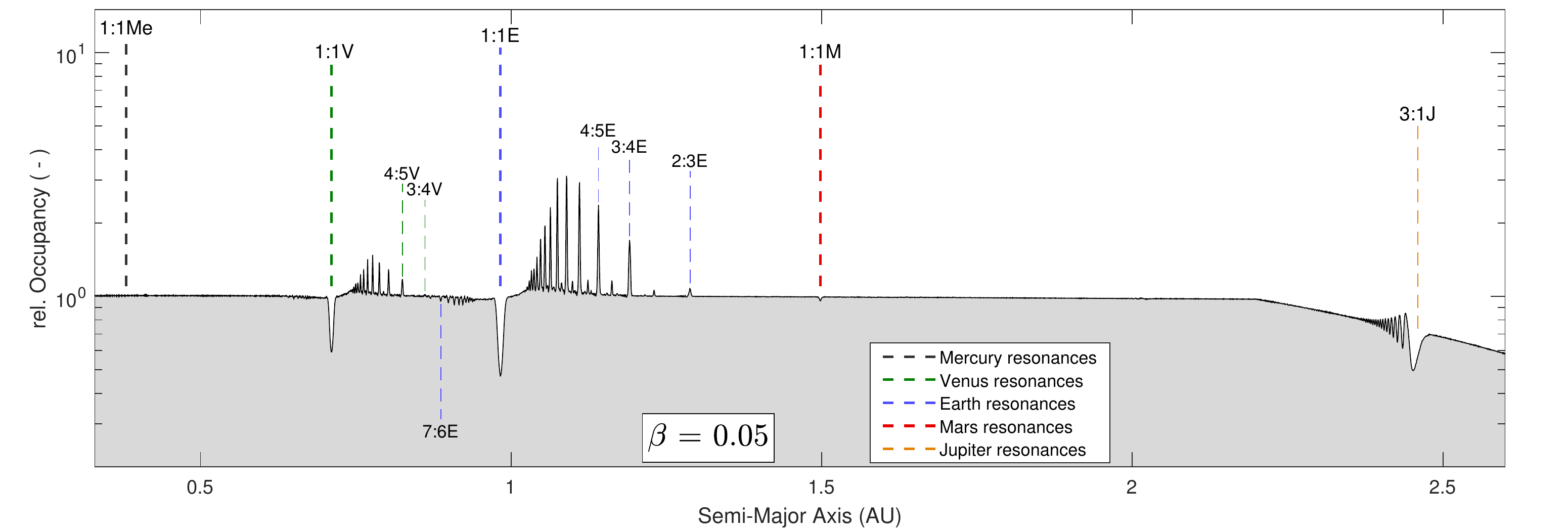}\\
		\includegraphics[width=0.8\textwidth,trim={0 5mm 0 1.9mm},clip]{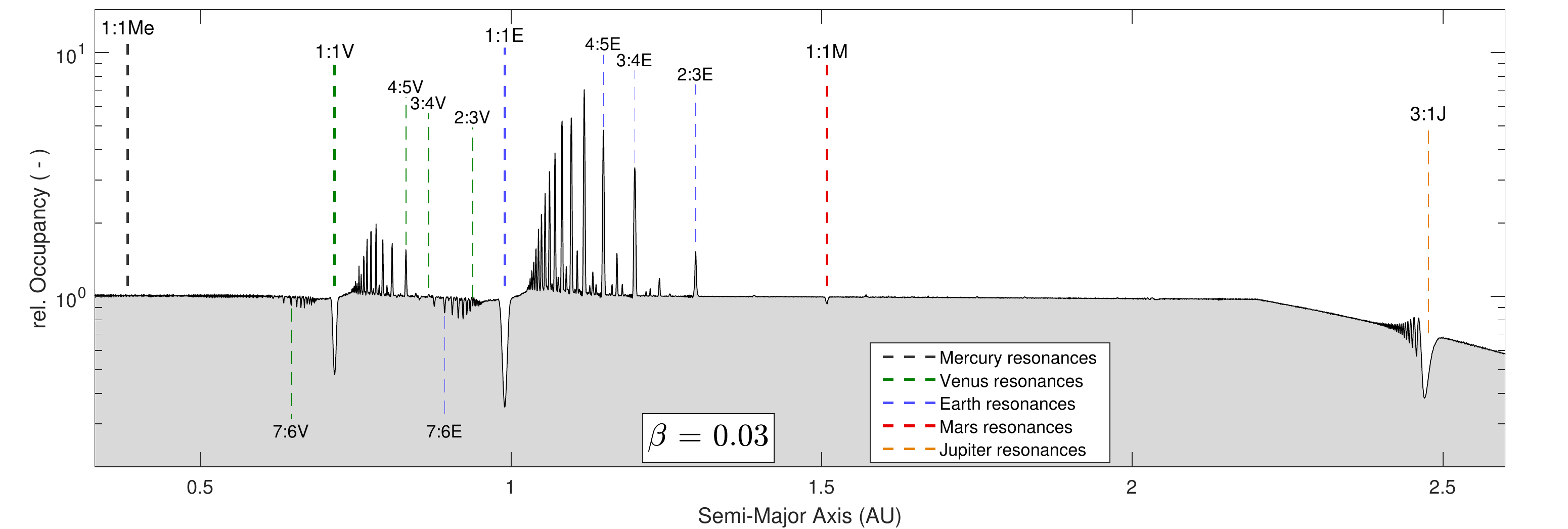}\\
		\includegraphics[width=0.8\textwidth,trim={0 5mm 0 1.9mm},clip]{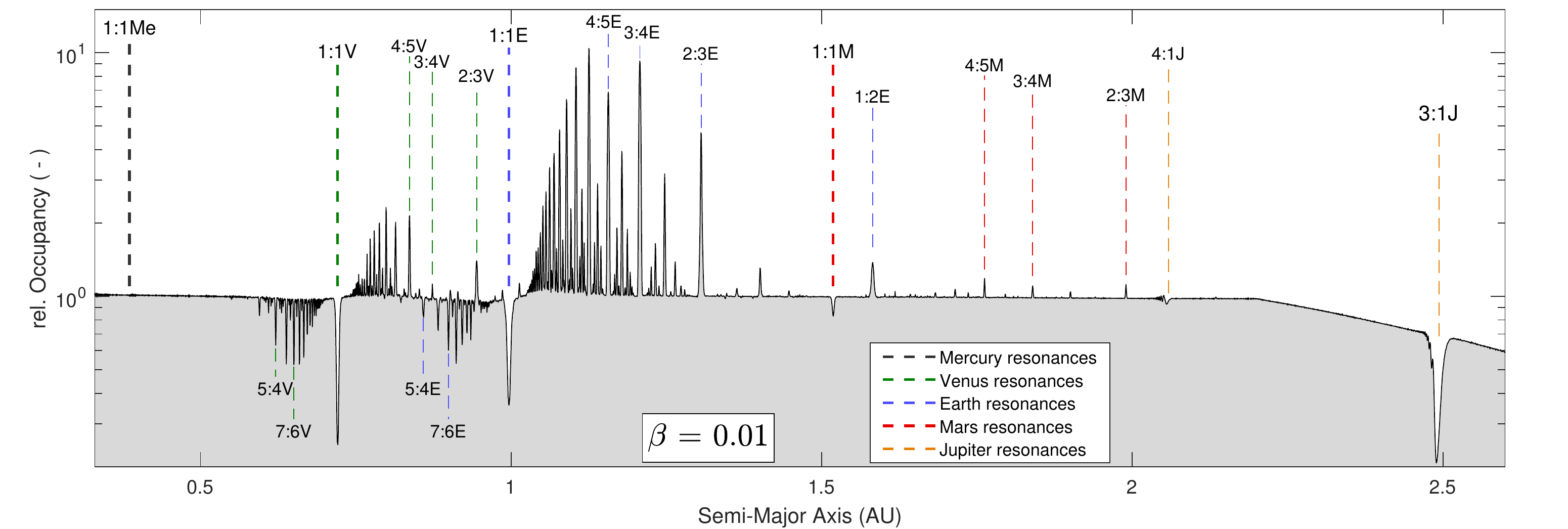}\\
		\hspace{0.9mm}\includegraphics[width=0.8\textwidth,trim={0 0mm 0 1.9mm},clip]{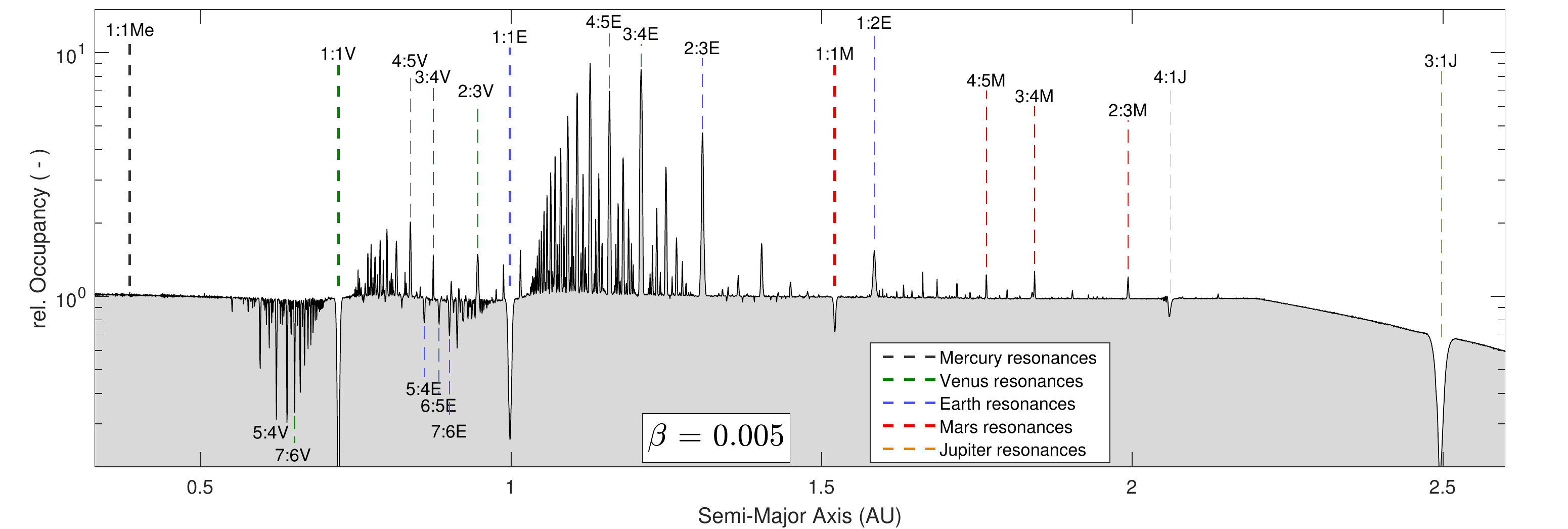}
	    \caption{Semi-major axis histogram under the presence of all planets for 
      	$\beta=$~0.1, 0.05, 0.03, 0.01, 0.005 ($D=5~\si{\micro\meter}, 10~\si{\micro\meter}, 17~\si{\micro\meter}, 
      	50~\si{\micro\meter}, 100~\si{\micro\meter}$) from top to bottom.
      	Occupancy is the number of particles per 0.0005~\si{\astronomicalunit} bin divided 
      	by solar distance of that bin and then normalised to background count 
      	(i.e. under the absence of resonances).
      	Vertical dashed lines indicate the locations of several first order as well as the 
      	co-orbital MMRs (including a shifting factor to account for radiation pressure).}
	        \label{fig:A_SMA_AP}
	\end{figure*}
	
	\twocolumn

\section{Facilitation of Resonances} \label{ChFR}
		
	In section \ref{Ch32} we noted the occurrence of some resonances that seemed to be facilitated by
	the presence of the respective neighbour planet.
	Specifically, these were the 5:8V and 3:5V resonance at $\beta=0.01$ that showed more trapping of 
	particles under the presence of Venus and Earth together than under the absence of either one
	planet (Fig.~\ref{fig:SMA_NENV}, \textit{bottom}).
	Although the occupancy of these Venus MMRs remained relatively low with respect to other external MMRs, 
	their occurrence is unexpected given their locations' proximity to the semi-major axis of Earth 
	with 0.988~\si{\astronomicalunit} and 1.015~\si{\astronomicalunit}.
	As we have argued, the vicinity to Earth and the resulting frequent gravitational perturbations 
	should impair the trapping efficacy of these MMRs severely but instead, 
	it is maintained or even amplified.
	Here we therefore give some consideration to this phenomenon.
	
	\begin{figure*}[]
		\includegraphics[width=\textwidth]{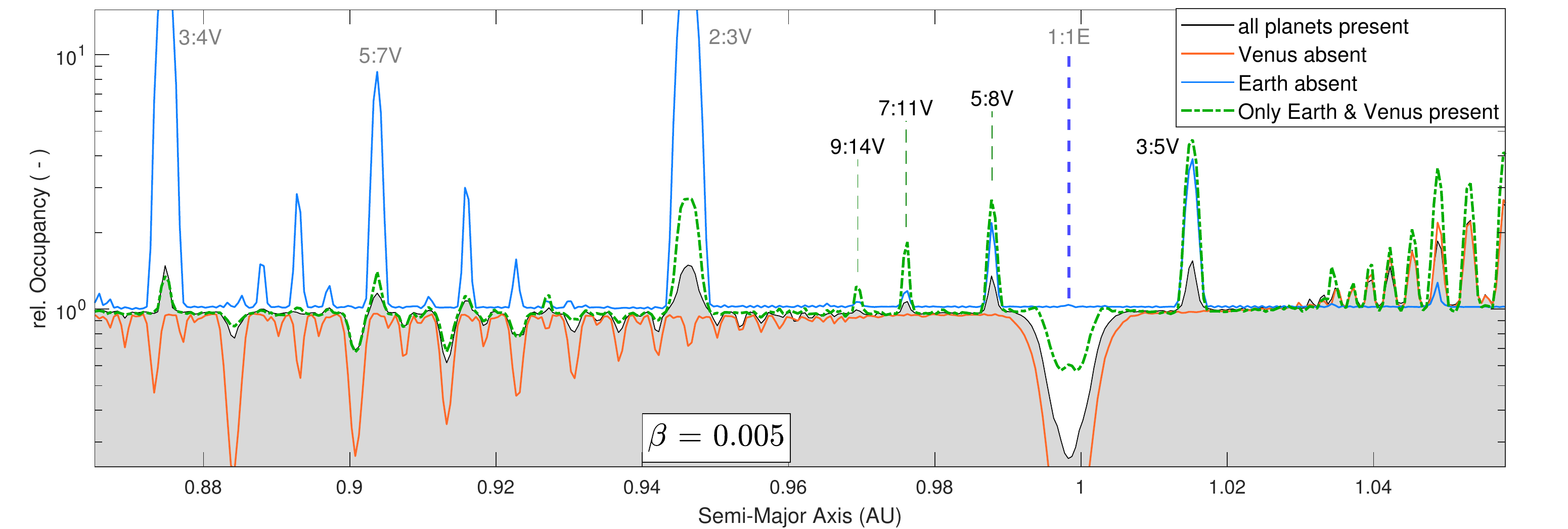}\\\\
      	\caption{Semi-major axis distribution in the vicinity of Earth under 
      	varying planetary configurations for $\beta=0.005$.
      	Vertical dashed lines show the locations of some of the MMRs of the planets.
      	The impairment of Venus resonances seems to be 
      	ineffective for those located in the chaotic region of Earth 
      	(9:14V, 7:11V, 5:8V, 3:5V).}
        \label{fig:co-inciding_SMA}
	\end{figure*}
	
	Figure~\ref{fig:co-inciding_SMA} shows the semi-major axis distribution in the vicinity 
	of these resonances at $\beta=0.005$.
	At this larger particle size, the occupancy at these resonances (5:8V \& 3:5V) is even higher, 
	reaching concentrations similar to the 3:4V and 2:3V resonance in the case of all planets present.  
	Even though the 5:8V and 3:5V MMR now also showing trapping under the absence of Earth, 
	it is apparent that the introduction of Earth does not have an impairing effect on them.
	This can be seen especially in the case of Earth and Venus being the only two considered planets.
	Here, the 3:5V resonance reaches the highest concentration of particles of any of the Venus MMRs.
	In addition we see the occurrence of even more Venus resonances in the vicinity of Earth, 
	namely 9:14V and 7:11V.
	On the other hand we still see the impairing effect of Earth, 
	drastically diminishing trapping at the more distant 3:4V, 7:5V, and 2:3V MMRs.
	It seems that the impairment caused by Earth is ineffective in the region closest to it.
	In fact, these MMRs all lie within the chaotic region of Earth resonances, 
	in which the libration widths of adjacent resonances overlap. 
	As found by \citet{wisdom1980resonance} and \citet{murray1999solar}, 
	the extent of this region for particle eccentricities $<0.15$ is given by:
	$$\Delta a_{overlap} \approx 1.3 \mu_{2}^{2/7} a' ,$$
	where $\mu_{2}$ is the fractional mass of the perturbing body in the three-body system, 
	and $a'$ its semi-major axis.
	In case of the Earth this results in a resonance overlapping region spanning from 
	0.9656~\si{\astronomicalunit} to 1.0344~\si{\astronomicalunit}.
	This should exclude the possibility of these resonances actually being resonances of Earth
	instead of being resonances of Venus.
	
\begin{figure*}[]
	\centering
	\begin{tabular}{cc}
	\includegraphics[width=88mm,trim={3mm 1.8mm 3mm 0mm},clip]{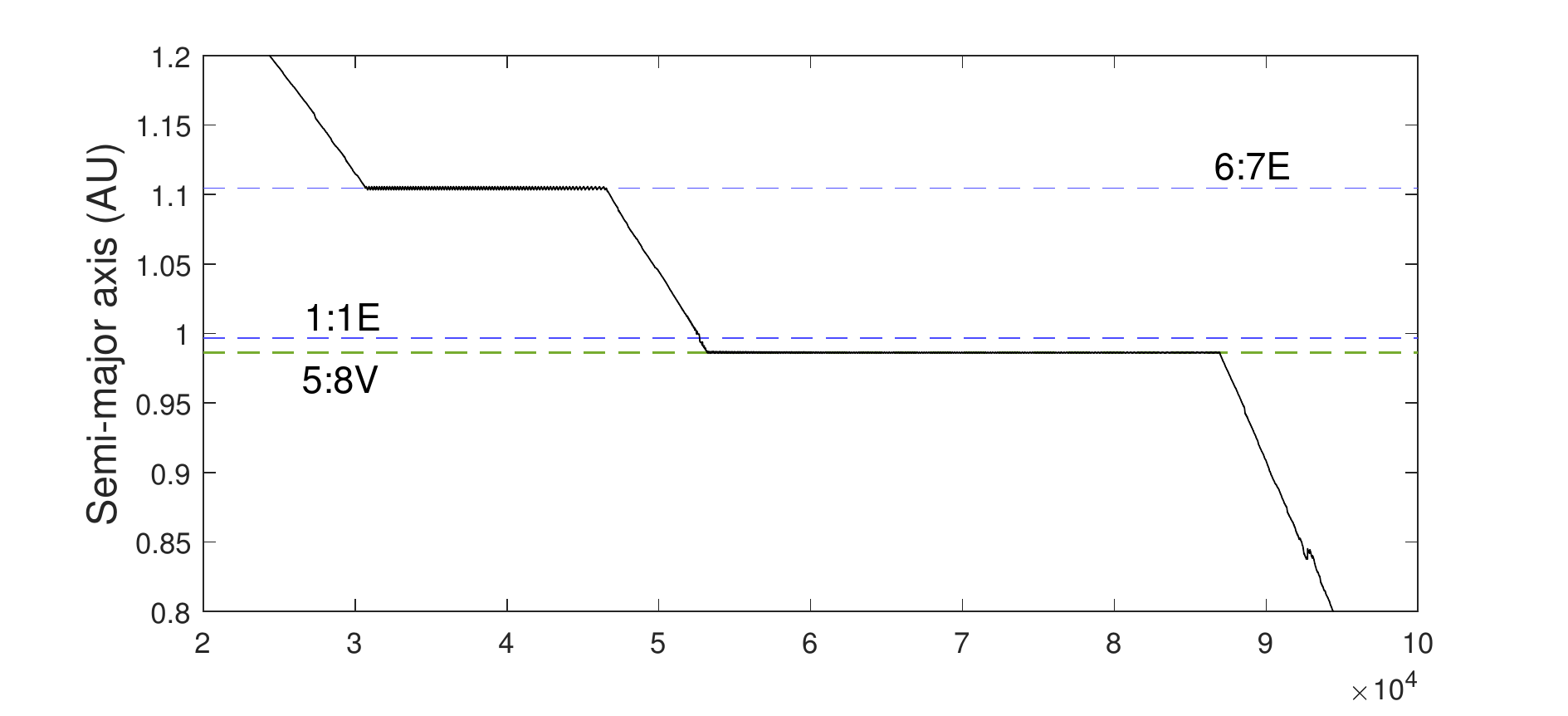} &
	\includegraphics[width=88mm,trim={3mm 1.8mm 3mm 0mm},clip]{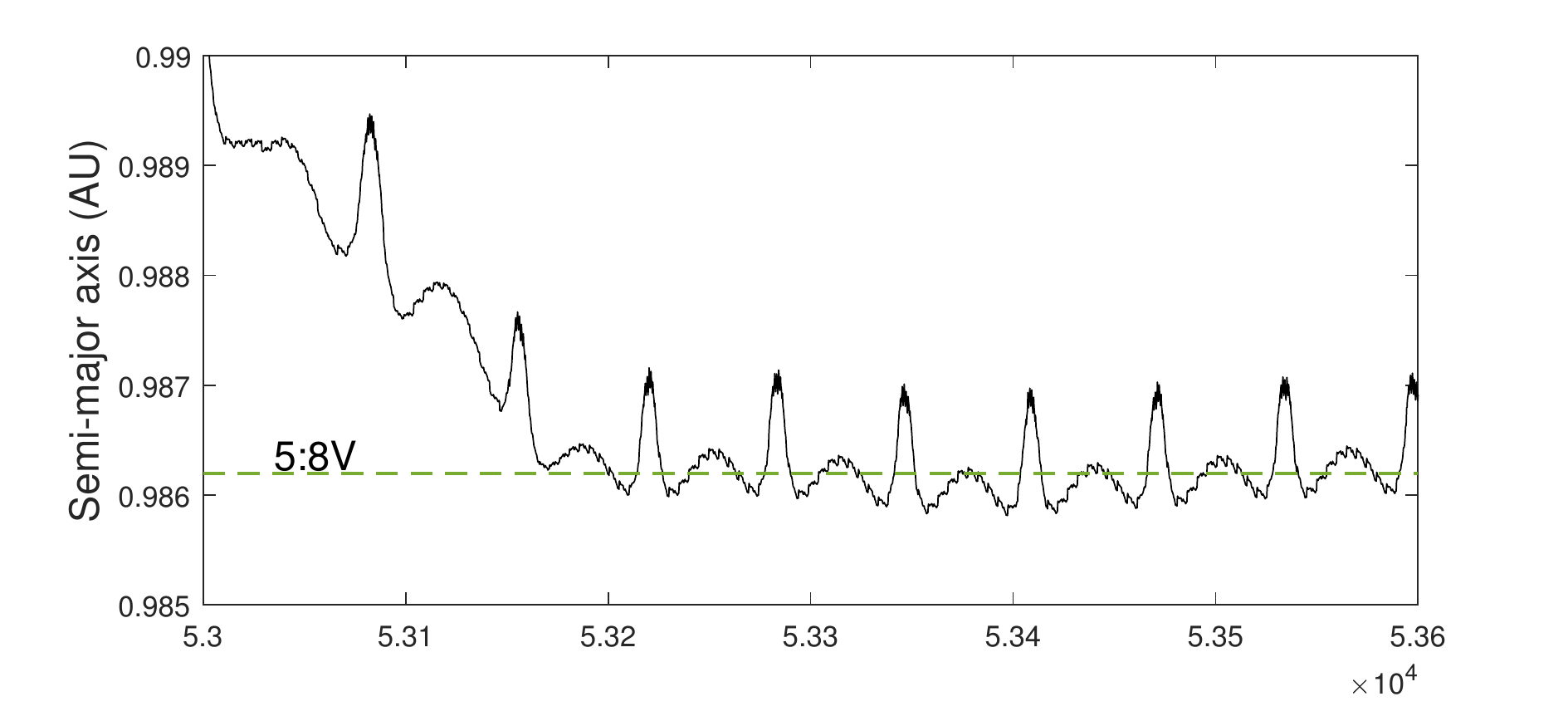} \\
	\includegraphics[width=88mm,trim={3mm 1.8mm 3mm 5.8mm},clip]{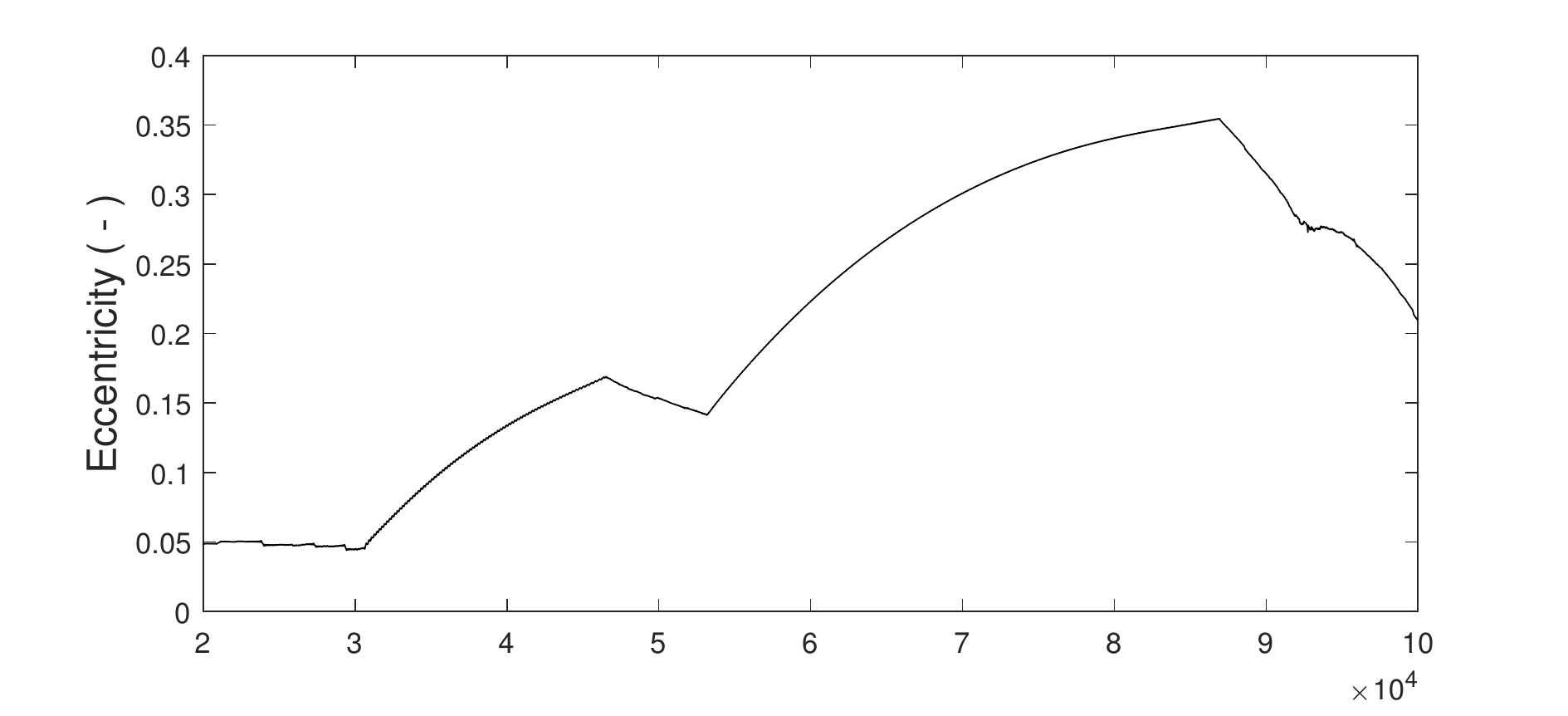} &
	\includegraphics[width=88mm,trim={3mm 1.8mm 3mm 5.8mm},clip]{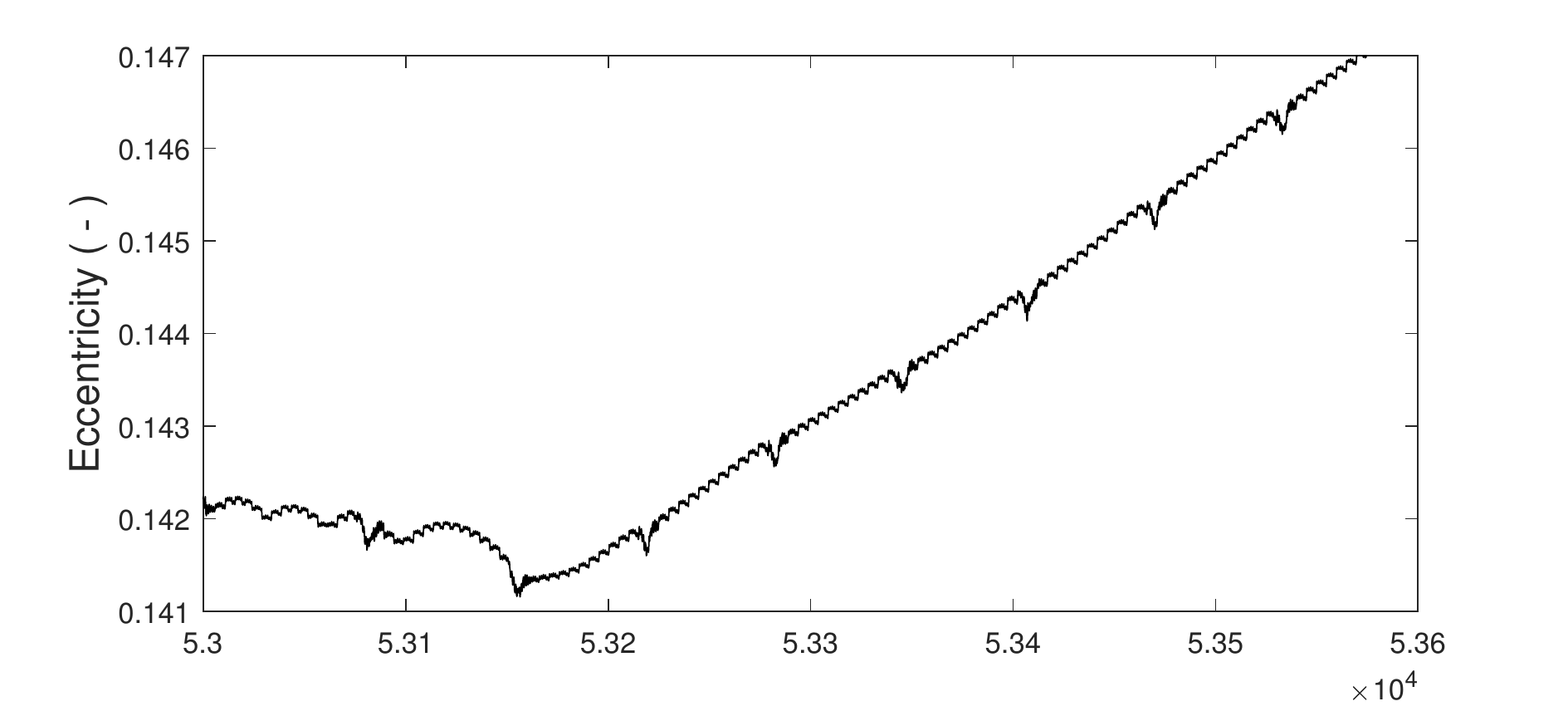} \\
	\includegraphics[width=88mm,trim={3mm 0mm 3mm 5.8mm},clip]{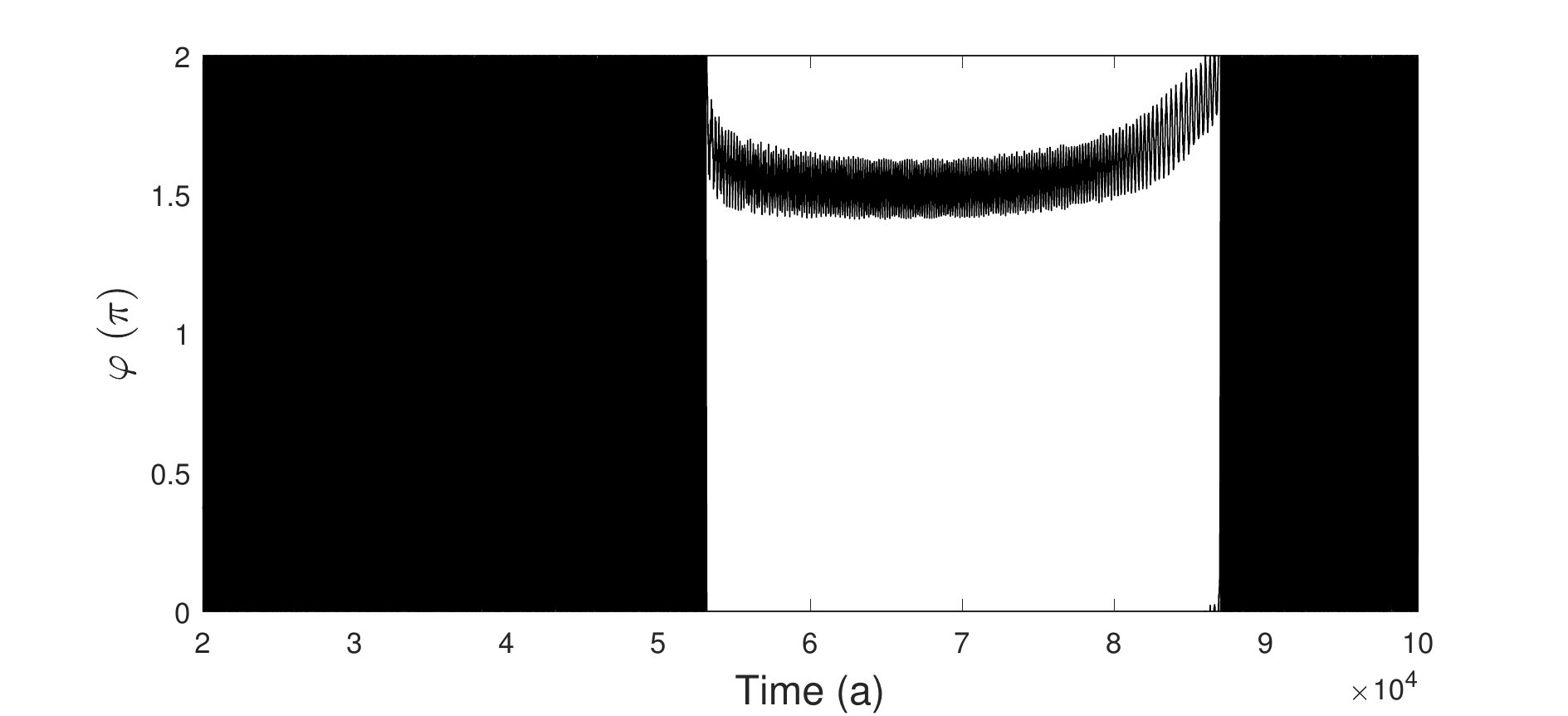} &
	\includegraphics[width=88mm,trim={3mm 0mm 3mm 5.8mm},clip]{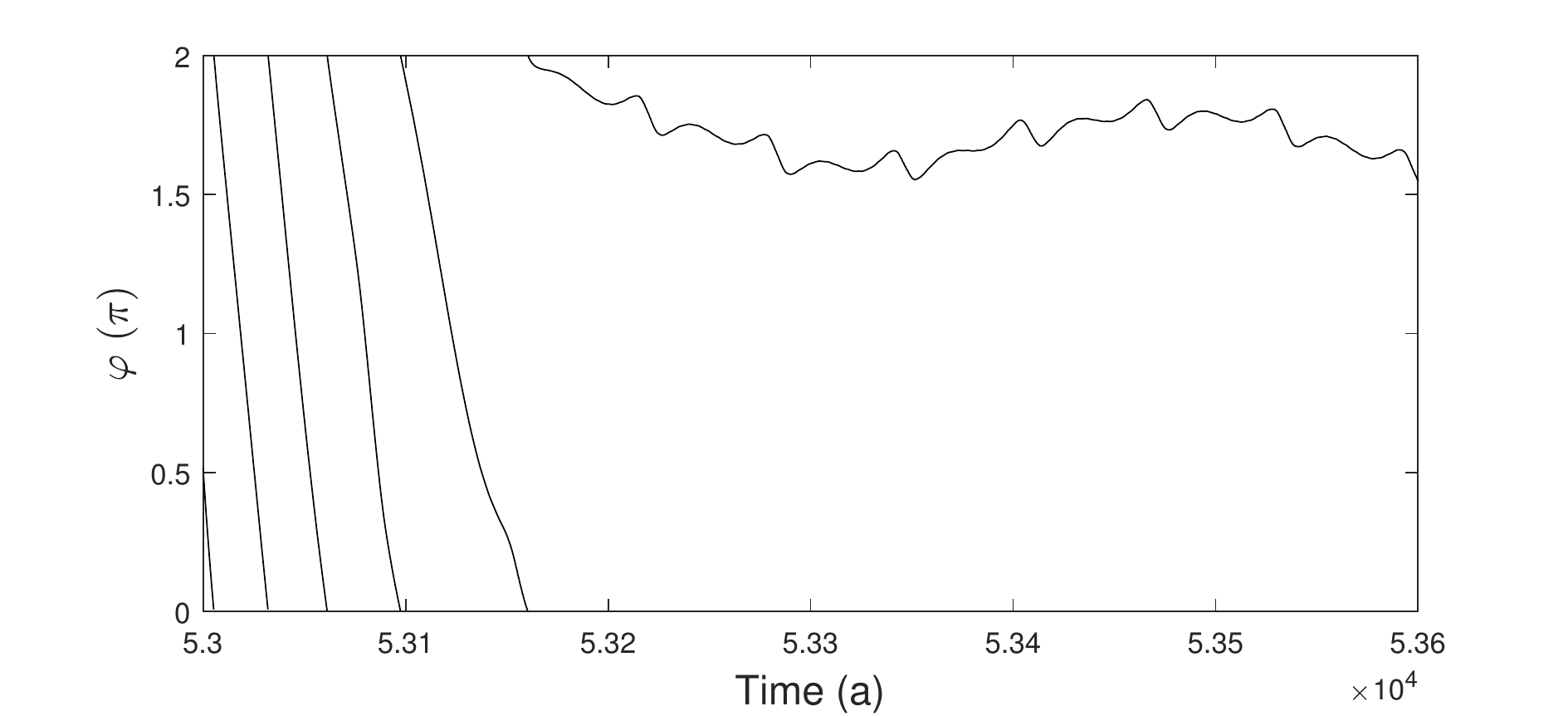} 

	\end{tabular}
	\caption{Evolution of semi-major axis \textit{(top)}, eccentricity \textit{(middle)}, 
			and resonant argument \textit{(bottom)} of a particle
			($\beta=0.01$) getting trapped in the 5:8V resonance. 
			\textit{Left:} long-term evolution.
			\textit{Right:} evolution as the particle is trapped.
			Only Venus and Earth were present in this simulation.
			The particle's inclination remained relatively constant around 14\si{\degree}
			throughout its life.
			The resonant argument was calculated as: 
			$\varphi= (p+q)\lambda_{particle} - p\lambda_{Venus} - q \varpi_{particle}$,
			with $p=5$ and $q=3$.
			}
	\label{fig:co-inciding_V58}
	\end{figure*}

	We considered that the occurrence of these resonances might have to do with the fact 
	that the orbital periods of Earth and Venus themselves commensurate to each other with a 
	ratio very close to 13:8
	\citep[$T_{Earth}/T_{Venus}=\frac{13}{8}\times 1.00032$,][]{chapman1986recurrent}, 
	with the idea that MMRs of Venus then also relate to MMRs of Earth through the same ratio,
	thus merely ensuring particles resonating with Venus to not suffer chaotic interaction with Earth.
	However, simulations with Earth and Venus relatively revolving more distant from the 13:8 ratio
	did not lead to a disappearance of these Venus resonances close-by the semi-major
	axis of Earth.
	Another possible explanation could be that these high-order resonances favour the trapping
	of particles with higher initial eccentricities. 
	Having just traversed Earth's external MMR sequence, particles in the region around 
	1~\si{\astronomicalunit} may still exhibit raised eccentricities as a result of eccentricity pumping.
	Accordingly, Fig.~\ref{fig:co-inciding_V58} shows the evolution of semi-major axis, eccentricity,
	as well as resonant argument 
	of a particle becoming trapped in the 5:8V resonance after having had its eccentricity pumped 
	in the 6:7E resonance. 
	This progression appears to be typical for particles getting trapped in these Earth-adjacent 
	Venus MMRs.
	The libration of the resonant argument confirms the particle's entrapment in the 5:8V MMR.
	
	A more sophisticated analysis will be necessary to reveal 
	the underlying dynamics, which should be subject of further investigation. 
	Even though the impact on the overall dust density of these phenomena may be trivial, 
	they further attest to the particularities of resonant dust discs in multi-planet systems. 
	
\end{appendix}

\end{document}